\documentclass[12pt]{article}
\pdfoutput=1
\usepackage{putex}
\usepackage{graphicx}
\usepackage{caption}
\usepackage{subcaption}
\usepackage{epstopdf}
\usepackage{enumerate}
\usepackage{cite}
\usepackage{tensor}
\usepackage{slashed}
\usepackage[aligntableaux=center, boxsize=1em]{ytableau}
\usepackage[utf8]{inputenc}
\usepackage[
      colorlinks=true,
      linkcolor=blue,
      urlcolor=blue,
      filecolor=black,
      citecolor=red,
      ]{hyperref}
\usepackage{float}

\renewcommand{\arraystretch}{1.5}

\newcommand{\grp}[1]{\mathrm{#1}}

\newcommand{\abs}[1]{\left\lvert #1 \right\rvert}

\newcommand {\be} {\begin {equation}}
\newcommand {\ee} {\end {equation}}

\newcommand {\bes} {\begin {equation*}}
\newcommand {\ees} {\end {equation*}}

\newcommand{\es}[2] {\begin{equation} \label{#1} \begin{split} #2 \end{split} \end{equation}}

\newcommand{\CP}{\mathbb{CP}}
\newcommand{\Z}{\mathbb{Z}}

\newcommand{\R}{\mathbb{R}}

\newcommand{\cO}{{\cal O}}

\def\Tr{\mop{Tr}}

\newcommand{\beq}{\begin{equation}}
\newcommand{\eeq}{\end{equation}}

\newcommand{\ov}{\over}
\newcommand{\Det}{{\rm Det}}
\def\le{\left}
\def\ri{\right}

\numberwithin{equation}{section}

\begin{document}

\preprint{PUPT-2537}

\institution{PU}{Joseph Henry Laboratories, Princeton University, Princeton, NJ 08544, USA}
\institution{Simons}{Simons Center for Geometry and Physics, SUNY, Stony Brook, NY 11794, USA}
\institution{PCTS}{Princeton Center for Theoretical Science, Princeton University, Princeton, NJ 08544, USA}

\title{Monopole Operators in $U(1)$ Chern-Simons-Matter Theories}

\authors{Shai M.~Chester,\worksat{\PU} Luca V.~Iliesiu,\worksat{\PU} M\'ark Mezei,\worksat{\Simons}$^{,}$\worksat{\PCTS} and Silviu S.~Pufu\worksat{\PU}}

\abstract{
We study monopole operators at the infrared fixed points of $U(1)$ Chern-Simons-matter theories  (QED$_3$, scalar QED$_3$, ${\cal N} =1$ SQED$_3$, and ${\cal N} = 2$ SQED$_3$) with $N$ matter flavors and Chern-Simons level $k$.  We work in the limit where both $N$ and $k$ are taken to be large with $\kappa = k/N$ fixed.  In this limit, we extract information about the low-lying spectrum of monopole operators from evaluating the $S^2 \times S^1$ partition function in the sector where the $S^2$ is threaded by magnetic flux $4 \pi q$.   At leading order in $N$, we find a large number of monopole operators with equal scaling dimensions and a wide range of spins and flavor symmetry irreducible representations.  In two simple cases, we deduce how the degeneracy in the scaling dimensions is broken by the $1/N$ corrections.    For QED$_3$ at $\kappa=0$, we provide conformal bootstrap evidence that this near-degeneracy is in fact maintained to small values of $N$.   For ${\cal N} = 2$ SQED$_3$, we find that the lowest dimension monopole operator is generically non-BPS.}
\date{\today}

\maketitle

\tableofcontents
%\newpage

\section{Introduction}

$U(1)$ gauge theories in three dimensions possess a topological global symmetry $U(1)_\text{top}$ \cite{Polyakov:1975rs}, whose associated conserved current and charge operators are given by
\es{topDef}{
j_\text{top}^\mu=\frac{1}{8\pi}\epsilon^{\mu\nu\rho}F_{\nu\rho}\,, \qquad q=\frac{1}{4\pi}\int_{\Sigma}F \,,
}
where $F_{\nu \rho} \equiv \partial_\nu A_\rho - \partial_\rho A_\nu$ is the gauge field strength, and $\Sigma$ is a closed two-dimensional surface.  The $U(1)_\text{top}$ symmetry is ``topological'' because the existence of operators charged under it is tied to the non-trivial topology of the gauge group, in particular to its nontrivial fundamental group $\pi_1(U(1))\cong \mathbb{Z}$\@. Such local, gauge invariant operators that carry non-zero $U(1)_\text{top}$ charge are called monopole operators \cite{Borokhov:2002ib}.  Due to Dirac quantization, their charge $q$ is quantized:  in the normalization of \eqref{topDef}, we have $q\in \mathbb{Z}/2$. 

In this paper, we will discuss monopole operators in $U(1)$ gauge theories at Chern-Simons level $k$ coupled to $N$ flavors of charged matter fields.  We take the charged matter to be either $N$ complex two-component fermions, $N$ complex scalars, or $N$ pairs of complex scalars and fermions preserving ${\cal N} =1$ or ${\cal N} = 2$ supersymmetry.    In the limit of large $N$ and/or $k$, these theories flow to interacting conformal field theories in the infrared \cite{Pisarski:1984dj,Appelquist:1988sr,Nash:1989xx}.\footnote{As $N$ is lowered, it is possible that this family of CFTs actually terminates at some critical value of $N$.  There is no consensus as to what this critical value of $N$ may be, as different approaches give different answers \cite{DiPietro:2015taa,Chester:2016ref,Giombi:2015haa,Karthik:2015sgq,Karthik:2016ppr,Appelquist:2004ib,Appelquist:1999hr,Strouthos:2008kc,Hands:2004bh,Hands:2002dv,Hands:2002qt,Pisarski:1984dj,Appelquist:1988sr}.  It is of course possible that there exists a non-trivial CFT for all $N>0$ \cite{Karthik:2016ppr,Karthik:2015sgq,Chester:2016ref}.}  As with any local operator in conformal field theories, the monopole operators in these theories are characterized by their scaling dimension, spin, and flavor symmetry representation.  Our goal will be to determine these quantum numbers for the monopole operators of low scaling dimension for any given $q$, in the limit of large $N$ and $k$ with fixed ratio $\kappa \equiv k/N$.

Our results generalize the current literature.  In the non-supersymmetric cases, the quantum numbers of the lowest monopole operators have been determined only when $k=0$ \cite{Murthy:1989ps, Borokhov:2002ib,Metlitski:2008dw,Pufu:2013vpa,Dyer:2013fja,Dyer:2015zha, Chester:2015wao}.\footnote{See also \cite{2013PhRvL.111m7202B,2015arXiv150205128K} for quantum Monte Carlo studies.}$^{,}$\footnote{Ref.~\cite{Borokhov:2002ib} gave the quantum numbers of the $q=1/2$ monopole also for non-vanishing $k$, but only in the range $\abs{k}\leq N/2$.}  In the supersymmetric examples with ${\cal N} = 2$ supersymmetry, the quantum numbers of all BPS monopole operators have been determined both using the large $N$ approximation \cite{Borokhov:2002cg} and supersymmetric localization \cite{Benini:2009qs,Benini:2011cma,Imamura:2011su,Kim:2009wb,Aharony:2015pla,Kapustin:2009kz}.  Our focus will not be on BPS operators, however, but instead on the monopole operators of the lowest dimension, regardless of whether they are supersymmetric;  generically they are not.

There are many reasons to be interested in monopole operators in the theories mentioned above.  For instance, it is known in many examples that when one of these theories arises as the continuum limit of a lattice theory, it is possible that certain monopole operators act as order parameters for symmetry-broken phases in second-order phase transitions beyond the Landau-Ginzburg paradigm \cite{Read:1989zz,Read:1990zza,SVBSF,SBSVF, Komargodski:2017dmc}.  Understanding the properties of monopole operators is important for characterizing the universality classes of these second order phase transitions.  Another motivation comes from the recently discussed web of non-superymmetric dualities \cite{Son:2015xqa,Aharony:2015mjs,Karch:2016sxi,Murugan:2016zal,Seiberg:2016gmd,Hsin:2016blu,Radicevic:2016wqn,Kachru:2016rui,Kachru:2016aon,Karch:2016aux,Metlitski:2016dht,Aharony:2016jvv,Benini:2017dus,Komargodski:2017keh}. Under the duality map, monopole operators sometimes get mapped to operators built from the elementary fields of the dual theory.   Comparing the scaling dimensions and quantum numbers of these operators across the duality could provide strong checks of these proposals.  Lastly, it was suggested in \cite{Chester:2016wrc} that the monopole operators provide a way to access conformal gauge theories using the conformal bootstrap program, hence gaining a better understanding of them in a larger class of theories could prove very fruitful for future studies.

As in \cite{Borokhov:2002ib,Borokhov:2002cg,Metlitski:2008dw,Pufu:2013vpa,Dyer:2013fja,Dyer:2015zha}, the method that we employ for determining the quantum numbers of the monopole operators relies on the state-operator correspondence.  In a CFT, the state-operator correspondence identifies the monopole operators of charge $q$ with the states in the Hilbert space on $S^2$ in the sector of $4 \pi q$ magnetic flux through the $S^2$ \cite{Borokhov:2002ib}. The scaling dimensions of the operators are given by the corresponding eigenvalues of the $S^2$ Hamiltonian.  In the limit of large $N$ and/or large $k$, the problem of determining the spectrum of the $S^2$ Hamiltonian simplifies because the theory becomes weakly coupled.  For fixed $q$, the ground state energy can be calculated using the saddle point approximation, as was done in \cite{Borokhov:2002ib,Borokhov:2002cg,Metlitski:2008dw,Pufu:2013vpa,Dyer:2013fja,Dyer:2015zha} when $k=0$.  

A new subtlety arises when $k\neq0$.  The Chern-Simons term induces a non-zero gauge charge proportional to $q$ for the naive $S^2\times\mathbb{R}$ vacuum.  To cancel this gauge charge, we must dress this vacuum state with charged matter modes, a procedure difficult to analyze in the usual path integral approach.  One can avoid this subtlety by computing, order by order in $1/N$, the free energy on $S^2\times S^1$, where the radius of $S^1$ is $\beta$.  (We will write $S^2 \times S^1_\beta$ for this space.)  This free energy should be interpreted as the thermal free energy of the theory placed on $S^2$ at temperature $T\equiv\frac1\beta$, and its small temperature limit captures the contributions of the low-lying physical states.  As we will explain, in this thermal computation, the dressing mentioned  above is enforced by the saddle point condition of the holonomy of the gauge field on $S^1_\beta$; the holonomy acts like a chemical potential for the matter fields.  

The picture we arrive at is as follows.  To leading order in $1/N$, the $S^2$ Hamiltonian in the sector of flux $4 \pi q$ generically has many degenerate low energy states whose energy scales as $N$ and that transform as a reducible representation of the symmetry group of the theory.  The leading order energy can be found using the saddle point approximation, and it is a non-trivial function of $\kappa = k/N$.  It matches the mode picture mentioned above, whereby one adds to the zero-point energy of the vacuum the energies of the modes required to cancel the gauge charge of the vacuum.\footnote{For states with quantum numbers of order $N$, this result receives large quantum corrections that we will not compute.}  
However, in cases where there are interactions between the matter fields that can be decoupled with Hubbard-Stratonovich fields, as is the case for instance in scalar QED$_3$, the energies of the modes depend on the saddle point values of these fields, which in turn also depend non-trivially on $q$ and $\kappa$.

Quite interestingly, while in general the answer for the scaling dimensions does not have a nice analytical expression, in scalar QED$_3$ we find that in the special case $\abs{\kappa} = \abs{q} + \frac 12$, the scaling dimensions of the monopole operators take the simple form
 \es{SpecialCase}{
  \abs{k} = \left( \abs{q} + \frac 12 \right) N : \qquad \qquad \Delta_q = \frac{2 \abs{q} (\abs{q}+1)(2\abs{q}+1)}{3} N + O(N^0) \,.
 }
In particular, for $q = 1/2$, we find that when $k = N$, we have $\Delta_{1/2} = N + O(N^0)$, which, when extrapolated to $N=1$ gives a scaling dimension $\Delta_{1/2} = 1$.  This is the exact scaling dimension in this case because we know that the monopole operator is dual to a free fermion \cite{Seiberg:2016gmd}. It would be interesting to see whether there is an explanation for the simple result \eqref{SpecialCase} more generally. 

Since generically the lowest physical states at leading order in $1/N$ transform as a reducible representation of the symmetry group, we expect that the degeneracy between the irreducible components should be lifted at higher orders in $1/N$.  We make this precise in a few cases where we argue that the energy splitting for states of spin $\ell$ is proportional to $\ell^2 / N$.  The argument we provide is rather indirect and comes from the evaluation of the thermal partition function to subleading order in $1/N$.

It would be interesting to know if the near-degeneracy between states in various irreducible global symmetry representations persists down to small values of $N$.  In Section~\ref{QED_Conformal_bootstrap}, we provide evidence that this may indeed be the case for QED$_3$ at $\kappa = 0$.  In this section, we apply the conformal bootstrap to this theory when $N=4$ to compute bounds on monopole scaling dimensions with the different spins and flavor representations that our thermal computation predicts. We find that monopoles in different representations have very similar bounds, which suggests that the near-degeneracy in scaling dimension inferred from the thermal computation holds even for small $N$.

The rest of this paper is organized as follows. In Section~\ref{PARTITION} we present our computation of free energies on $S^2 \times S^1_\beta$: we begin in Section~\ref{STRATEGY} by explaining the general structure of our answer, followed by actual computations of the free energy in QED$_3$ in Section~\ref{QED3All}, scalar QED$_3$ in Section~\ref{CPNLeading}, $\mathcal N=1$ SQED in Section~\ref{sec:sqed1} and $\mathcal N=2$ SQED in Section~\ref{sec:Neq2}. In Section \ref{MICROSTATES} we interpret the entropies obtained in the previous section in terms of a mode construction on $S^2\times S^1_\beta$.  Section~\ref{QED_Conformal_bootstrap} contains a conformal bootstrap analysis in the case of QED$_3$ at $k=0$ and $N=4$. Finally, in Section~\ref{conclusion}, we discuss future directions. Some of the technical details are relegated to the Appendices. In particular, in Appendix~\ref{regularization} we provide technical details on the zeta-function regularization procedure;  in Appendix~\ref{steep-desc}, we check in a particular case that the saddle point configuration we use is the physical one;  Appendix~\ref{Green} contains technical details on the subleading order computations in $1/N$;  and Appendix~\ref{SQEDAPPENDIX} contains more computation in ${\cal N} =2$ SQED, including a comparison of our large $N$ method to the supersymmetric localization results for the superconformal index of this theory.

\section{The $S^2 \times S_\beta^1$ partition function}
\label{PARTITION}

\subsection{General strategy and interpretation}
\label{STRATEGY}

As mentioned in the Introduction, in order to learn about monopole operators, we place the gauge theories of interest on $S^2 \times S^1_\beta$, where $S^1_\beta$ is interpreted as a thermal circle of circumference $\beta = 1/T$, and study the sector of magnetic flux $4 \pi q$ through the $S^2$.    For each of the four theories we study, we obtain a large $N$ expansion for the free energy $F_q$ of the form
 \es{FExpansion}{
  F_q = N F_q^{(0)} + F_q^{(1)} + \frac{1}{N} F_q^{(2)} + \ldots \,, 
 }
with $F_q^{(1)}$ possibly containing a $\log N$ dependence that we do not separate out for brevity.  Each term in \eqref{FExpansion} can be further expanded at large $\beta$, and this double expansion in $1/N$ and $1/\beta$ gives some information about the low-lying energy states and their degeneracies.

In particular, expanding $F_q^{(0)}$ and $F_q^{(1)}$ at large $\beta$, we find\footnote{Note that while taking $N \to \infty$ first and $\beta \to \infty$ afterwards is a well-defined procedure, if we neglect the exponentially small terms in the expression for $F_q^{(0)}$, then the sum $N F_q^{(0)} +F_q^{(1)}$ provides a good approximation to $F_q$ only if $\beta \gg \log N$.}
 \es{FCoeff}{
  F_q^{(0)} &= \Delta_q^{(0)} - \frac{1}{\beta} S_q^{(0)} + O(e^{-c \beta}) \,, \\
  F_q^{(1)} &= \Delta_q^{(1)} + \frac{1}{\beta} \left(\frac{1}{2} \log N + \frac{d}{2} \log \beta + O(\beta^0) \right) \,,
 }
where $c$ is a positive constant and $d$ is a positive integer.  In each case, we will calculate $\Delta_q^{(0)}$ and $S_q^{(0)}$, while for the non-supersymmetric cases we also calculate $d$ and $c$. We leave the evaluation of $\Delta_q^{(1)}$ for a future publication.

In order to interpret \eqref{FCoeff}, let us draw a distinction between the large $\beta$ behavior of the free energy of a system with a discrete versus a continuous spectrum.  For a system with a discrete spectrum, at large $\beta$ we have 
\es{FDiscrete}{
 F = E_0 - \frac{S_0}{\beta} + O(e^{-\beta(E_1 - E_0)}) \,,
}
where $E_0$ is the ground state energy, $e^{S_0}$ is its degeneracy, and $E_1$ is the energy of the first excited state.  For a system with a continuous spectrum starting at some energy $E_0$, for which the density of states near the bottom of the continuum behaves as ${\cal D}(E) \approx C (E - E_0)^\alpha$, for some constant $C$, we have
 \es{partCont}{
  Z = \int dE\, {\cal D}(E) e^{-\beta E} \approx C\Gamma(\alpha+1) \frac{e^{-\beta E_0}}{\beta^{\alpha + 1}} \,, \qquad \text{as $\beta \to \infty$} \,.
 }
This behavior of the partition function implies that the free energy behaves as 
 \es{FBehavior}{
  F = -\frac{1}{\beta} \log Z = E_0 + (\alpha + 1) \frac{\log \beta}{\beta} - \frac{\log (C \Gamma(\alpha+1))}{\beta} + O(1/\beta^2)\,, \qquad \text{as $\beta \to \infty$} \,.
 }
Thus, a way to distinguish between a system with a discrete spectrum and one with a continuous spectrum is that the free energy of the latter has a $\frac{\log \beta}{\beta}$ contribution that gives us information about the behavior of the density of states close to the bottom of the continuum. 

After this brief review, we can now interpret \eqref{FCoeff}.  To leading order in $1/N$, the partition function is dominated by $e^{NS_q^{(0)}}$ approximate ground states with approximate energy $N \Delta_q^{(0)}$, as can be seen from the expression for $F_q^{(0)}$ in \eqref{FCoeff}.  To subleading order in $1/N$, the interpretation of $F_q^{(1)}$ is as follows.  Because we first took $N \to \infty$ and then $\beta \to \infty$, the degeneracy of the $e^{NS_q^{(0)}}$ states is partially lifted into a large number of states of different energies;  these states can be approximated with a continuum whose density of states behaves as ${\cal D}(E) \propto \left[ E - (N \Delta_q^{(0)} + \Delta_q^{(1)}) \right]^{d-1}$ close to the bottom of the continuum.  Thus, the coefficient of the $\log \beta$ term in \eqref{FCoeff}, which we will compute, tells us how the low-lying energy levels are split.  In Section~\ref{MICROSTATES}, we will provide a more concrete perspective on this splitting.

In order for the interpretation above to hold, it is of course very important that we take $N \to \infty$ first, and $\beta \to \infty$ afterwards, because in the opposite limit there should not be any $\frac{\log \beta}{\beta}$ terms in the free energy.  Our field theories have a discrete spectrum, so at very large $\beta$ we expect a behavior of the form \eqref{FDiscrete}.  However, in the regime of large $\beta$ but with $\beta \ll N$, the continuum behavior \eqref{FBehavior} becomes possible.

\subsection{QED$_3$}
\label{QED3All}

Let us now proceed to concrete calculations.  We start with QED$_3$, where we aim to present more details than for subsequent theories.  

The Euclidean action for the QED$_3$ theory with $N$ two component complex fermions and ``bare'' Chern-Simons (CS) level $\hat k$ (which is different from the quantum-corrected CS level $k$ to be defined shortly) is 
 \es{QEDAction}{
  \mathcal{S} = \int d^3 x \, \sqrt{g} \left[\frac{1}{4e^2} F_{\mu\nu} F^{\mu\nu} - \sum_{i=1}^{N} \bar\psi_i(i \slashed \nabla +\slashed A) \psi^i \right]
   - \int d^3x\, \frac{i \hat k}{4 \pi}  \epsilon^{\mu\nu\rho} A_\mu\partial_\nu A_\rho \,,
 }
where $g$ is the determinant of the metric, $\psi^i$ are the $N$ fermion fields, $A_\mu$ is the (real-valued) $U(1)$ gauge field with field strength $F_{\mu\nu}$, and $e$ is the gauge coupling constant.  The appropriate large $N$ limit is taken with $\epsilon \equiv e^2 N$ held fixed.  In order to study the IR fixed point, we further take $\epsilon \to \infty$, thus dropping the gauge kinetic term in \eqref{QEDAction}.  Intuitively, this term is irrelevant at the IR fixed point, because $e^2$ has dimensions of mass.   

Following \cite{Seiberg:2016gmd}, we define the measure of the fermion path integral such that $N$ free fermions in a background gauge field have the partition function
\es{FreeFerm}{
Z[A]_\text{free fermions}=\abs{\Det(i \slashed \nabla +\slashed A)}^N\, e^{-{i\pi N\ov 2}\, \eta(A)}\,,
}
where the absolute value of the determinant is the regularized product of the absolute values of the eigenvalues, and the $ \eta(A)$ is the Atiyah-Patodi-Singer eta-invariant \cite{witten2016fermion}.  The formula \eqref{FreeFerm} is written assuming $A_\mu$ is real-valued.  Later on, when we use the saddle point approximation, we will have to relax the reality condition on $A$, and in this case we should extend \eqref{FreeFerm} to a holomorphic function of $A$.  With the definition \eqref{FreeFerm}, $Z[A]_\text{free fermions}$ is gauge invariant, hence gauge invariance of the full theory \eqref{QEDAction} requires that $\hat k\in\mathbb{Z}$. For our purposes, the phase in \eqref{FreeFerm} can be thought of as a level $-\frac N2$ CS term, which we combine with the bare CS level to define $k\equiv \hat k-N/2$. It is this effective $k$ that we use to label the family of QED$_3$ theories.

Let us take the space $S^2 \times S^1_\beta$ to be parameterized by $x=(\theta,\phi,\tau)$, with $\tau\in[-\beta/2,\beta/2)$ and metric
\es{metric}{
ds^2=d\theta^2+\sin^2\theta d\phi^2+d\tau^2 \,.
}
(We take the $S^2$ to be of unit radius.)  We are interested in studying the theory \eqref{QEDAction} in the sector of magnetic flux $4 \pi q$ through $S^2$, with $q \in \Z/2$.  The thermal free energy of this sector can be extracted from the partition function 
\es{FQEDdefine}{
Z_q = e^{-\beta F_q} 
=\int_{\int_{S^2} F = 4 \pi q} DA \exp\left[ N\,\tr\log |i\slashed \nabla+\slashed{A}|+N\frac{i\kappa}{4\pi}\int d^3x\epsilon^{\mu\nu\rho} A_\mu \partial_\nu A_\rho \right]\,,
}
where we performed the Gaussian path integral over the fermions and combined the bare CS level $
\hat k$ with the phase \eqref{FreeFerm} from the fermion functional determinant.  

\subsubsection{Leading free energy}

Eq.~\eqref{FQEDdefine} is an exact expression that is hard to evaluate in general.  However, in the limit of large $N$ and $k$ with ratio $\kappa = k/N$ fixed, one can evaluate it in a saddle point approximation whereby one replaces the integral over the gauge field $A_\mu$ by the saddle point value  of the integrand at $A_\mu={\cal A}_\mu$.  It is reasonable to expect that the gauge configuration that dominates in \eqref{FQEDdefine} is spherically symmetric and time-translation invariant---certainly, such a configuration is a saddle of the exponent in \eqref{FQEDdefine}.  The most general such background is
  \es{SaddleAnsatz}{
    \mathcal{A}_\tau &= -i \alpha \,, \qquad
    \mathcal{F}_{\theta \phi} \, d\theta \wedge d\phi = q \sin \theta d \theta \wedge d\phi \,, 
  }
  where $\alpha=i\beta^{-1}\int_{S_\beta^1}A$ is a constant independent of position.
 The saddle point configurations that we consider in this paper will correspond to real $\alpha$, which we refer to as the holonomy of the gauge field. Physically, real $\alpha$ corresponds to turning on a (real) chemical potential for the matter fields.  One should of course be careful when evaluating the functional determinant \eqref{FreeFerm} at real $\alpha$, because the absolute value in \eqref{FreeFerm} assumes purely imaginary $\alpha$, and it analytically continues non-trivially to complex $\alpha$.   The dimensions and flavor symmetry representations for operators with positive $q$ are related to those with negative $q$ via the charge conjugation symmetry, so without loss of generality, we consider $q\geq0$.

Evaluating the effective action in \eqref{FQEDdefine} on the saddle \eqref{SaddleAnsatz}, one finds $N \beta F_q^{(0)}(\alpha)$, with\footnote{To evaluate the CS term we extend $S^2\times S^1_\beta$ to the 4-manifold $S^2\times D^2_\beta$ and compute $\frac{1}{4\pi}\int_{S^2\times S^1_\beta} A\wedge F=\frac{1}{4\pi}\int_{S^2\times {D^2_\beta}} F\wedge F=\frac{1}{2\pi}\int_{D^2_\beta} F\times \int_{S^2}F=-2i \alpha \beta q$.}
\es{FQED}{
F_q^{(0)}(\alpha)=-{1\ov \beta}\, \tr\log|i\slashed \nabla+\slashed{\mathcal{A}}|-2\kappa q\alpha\,,
}
where the saddle point value $\alpha$ will be fixed later by the saddle point condition
\es{sadQED}{
\frac{\partial F_q^{(0)}(\alpha)}{\partial\alpha}=0\,.
}
After plugging in the value of $\alpha$ obeying \eqref{sadQED}, $F_q^{(0)}(\alpha)$ can be identified with the free energy coefficient $F_q^{(0)}$ appearing in \eqref{FExpansion}.  

To evaluate \eqref{FQED}, we must compute the spectrum of the operator $i\slashed \nabla+\slashed{\mathcal{A}}$ on $S^2\times S_\beta^1$.  The Dirac operator in the background \eqref{SaddleAnsatz} commutes with time translations and the total angular momentum $\vec{J}$, hence its eigenvalues are labeled by two quantum numbers $n,\, j$ \cite{Wu:1976ge, Wu:1977qk}:\footnote{We assume that the time dependence of the eigenfunctions is $\sim e^{-i\omega_n \tau}$.}
\es{QEDeigs}{
&j=q-1/2:\hspace{3.9cm} (\omega_n-i\alpha)\,,\\
&j\in \{q+1/2,\,q+3/2,\,\dots\}:\qquad \pm\sqrt{(\omega_n-i\alpha)^2+\lambda_j^2}\,,
}
and have degeneracy  $d_j=2j+1$ for each distinct eigenvalue.  Here, $\omega_n$ and $\lambda_j$ are the fermionic Matsubara frequencies and the energies of modes of the theory quantized on $S^2\times \R$, respectively:
\es{QEDeigs2}{
\omega_n&=\frac{(2 n+1)\pi}{\beta}\,, \qquad n\in \Z\,,\\
\lambda_j&=\sqrt{(j+1/2)^2-q^2}\,.
}
Using \eqref{QEDeigs}, the free energy \eqref{FQED} as a function of $\alpha$ is found to be
\es{FQED2}{
   F_q^{(0)}(\alpha) &=-2 \kappa q \alpha - \beta^{-1}\,\sum_{n\in \Z}\left[\sum_{j \geq   q+1/2} {d_j}  \log \left| (\omega_n  - i \alpha)^2 + \lambda_j^2 \right|+d_{q-1/2}\log\left|\omega_n-i\alpha\right|\right]\\
   &= -2  \kappa q \alpha - \beta^{-1} \left[\sum_{j \geq  q+1/2} {d_j}\log\left[ 2\left(\cosh(\beta\lambda_j)+\cosh(\beta\alpha)\right)\right]+d_{q-1/2}\log\left[2\cosh\left(\beta\alpha/2\right)\right]\right] \,.
 }
In going from the first to the second line in \eqref{FQED2} we performed the Matsubara sum assuming that $i\alpha$ is real, and then extended the result holomorphically to complex $i\alpha$. Note that under the large gauge transformation $\alpha\to\alpha+\frac{2\pi i}{\beta}$, the partition function of the fermions in the background \eqref{SaddleAnsatz},  $Z_q^{(0)}(\alpha)=\exp\left[-\beta N{F_q^{(0)}}(\alpha)\right]$,  is invariant as long as $\hat{k}= k+\frac{N}{2}\in\Z$, which is precisely the quantization condition.

\begin{figure}[t!]
\begin{center}
   \includegraphics[width=0.47\textwidth]{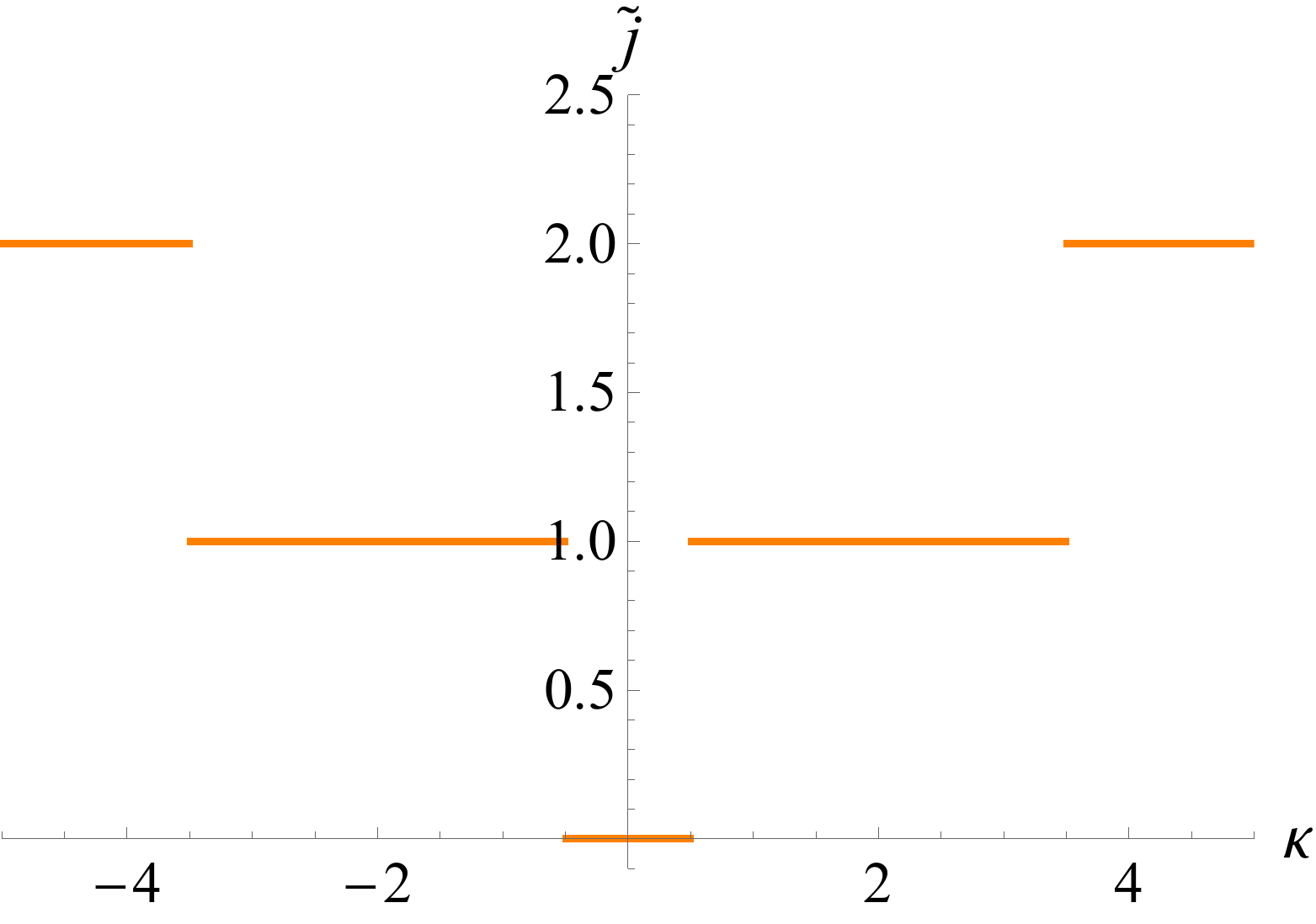}
   \includegraphics[width=0.47\textwidth]{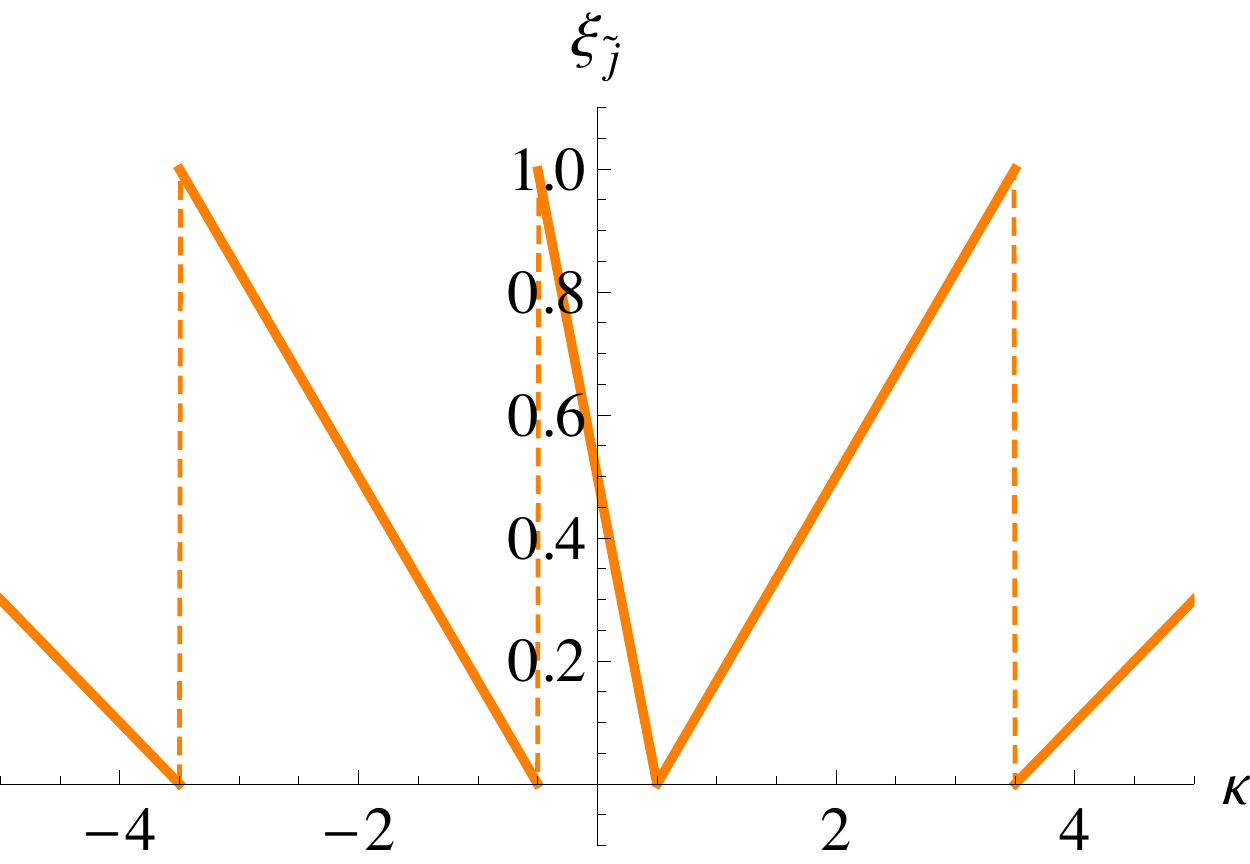}
\caption{\label{saddlesInfo}  The Landau level $\tilde j$ and the filling fraction $\xi_{\tilde j}$ for $q=1/2$ as a function of $\kappa=k/N$.
}
\end{center}
\end{figure}

Lastly, we should solve \eqref{sadQED} to find the saddle point value for $\alpha$.  While this can be done numerically for any $\beta$, we will only work at large $\beta$, where we can solve \eqref{sadQED} analytically.   For $q\neq0$, there are many solutions to \eqref{sadQED}.  They are $\alpha = \alpha^\pm_j(\kappa)$, being labeled by an index $j = q-1/2, q+1/2, \ldots$ (not to be confused with the summation variable in \eqref{FQED2}) and a choice of sign ($\pm$):
\es{QEDalphGen}{
\alpha_{j}^\pm (\kappa)&=\pm\left(\lambda_{j}+\beta^{-1}\log\frac{\xi_{j}}{1-\xi_{j}}\right)+O(e^{-(\lambda_{j+1}-\lambda_{j})\beta})\,,\\
\xi_{j}&\equiv\frac{1}{d_{ j}} \left(2q\le(\mp \kappa+\delta_{j,q-1/2}- \frac12\ri)-  \sum_{q-1/2<j'<  j} d_{j'}  \right)\,.
} 
Only one of these saddles corresponds to real $\alpha$ and real free energy, but precisely which one depends on $\kappa$. On physical grounds, we believe that this is the saddle through which we can deform the integration contour of the path integral. In Appendix~\ref{steep-desc}, in the $\kappa\to\infty$ limit we prove that this contour deformation is indeed possible, hence the saddle point with real $\alpha$ gives the correct answer.
 This physical saddle point has $j = \tilde j(\kappa)$ and the overall sign denoted by $\pm$ given by $-\sgn(\kappa-1/2)$, with 
\es{QEDexplicit}{
\tilde j(\kappa)=q+\frac12+\left\lfloor-\le(q+\frac12\ri)+\sqrt{\frac14+2q|\kappa|+q^2}\right\rfloor\,.
}
Plugging these into \eqref{QEDalphGen} gives:
\es{QEDalph}{
\alpha_{\tilde j}(\kappa)&=-\sgn(\kappa-1/2)\left(\lambda_{\tilde j}+\beta^{-1}\log\frac{\xi_{\tilde j}}{1-\xi_{\tilde j}}\right)+O(e^{-(\lambda_{\tilde j+1}-\lambda_{\tilde j})\beta})\,,\\
\xi_{\tilde j}&=\frac{1}{d_{ \tilde j}} \left(2q \le(|\kappa-
1/2 |- \Theta(-1/2-\kappa)\ri)-  \sum_{q-1/2<j'<  \tilde j} d_{j'}  \right)\,,
} 
where $\Theta(x)=1$ for $x>0$ and zero otherwise.  For this saddle, the quantity $\xi_{\tilde j}$ obeys $0\leq\xi_{\tilde j}(\kappa)\leq1$ and will be given a microscopic interpretation in Section~\ref{QEDmicro} as a filling fraction for the Landau level $\tilde j$.   For reference, $\tilde j$ and $\xi_{\tilde j}$ are plotted in Figure~\ref{saddlesInfo}.   One can check that for $q=0$ or for $k=0$, the physical saddle point is $\alpha=0$, because the sum \eqref{FQED2} is an even function of $\alpha$. 

\begin{figure}[t!]
\begin{center}
    \includegraphics[width=0.49\textwidth]{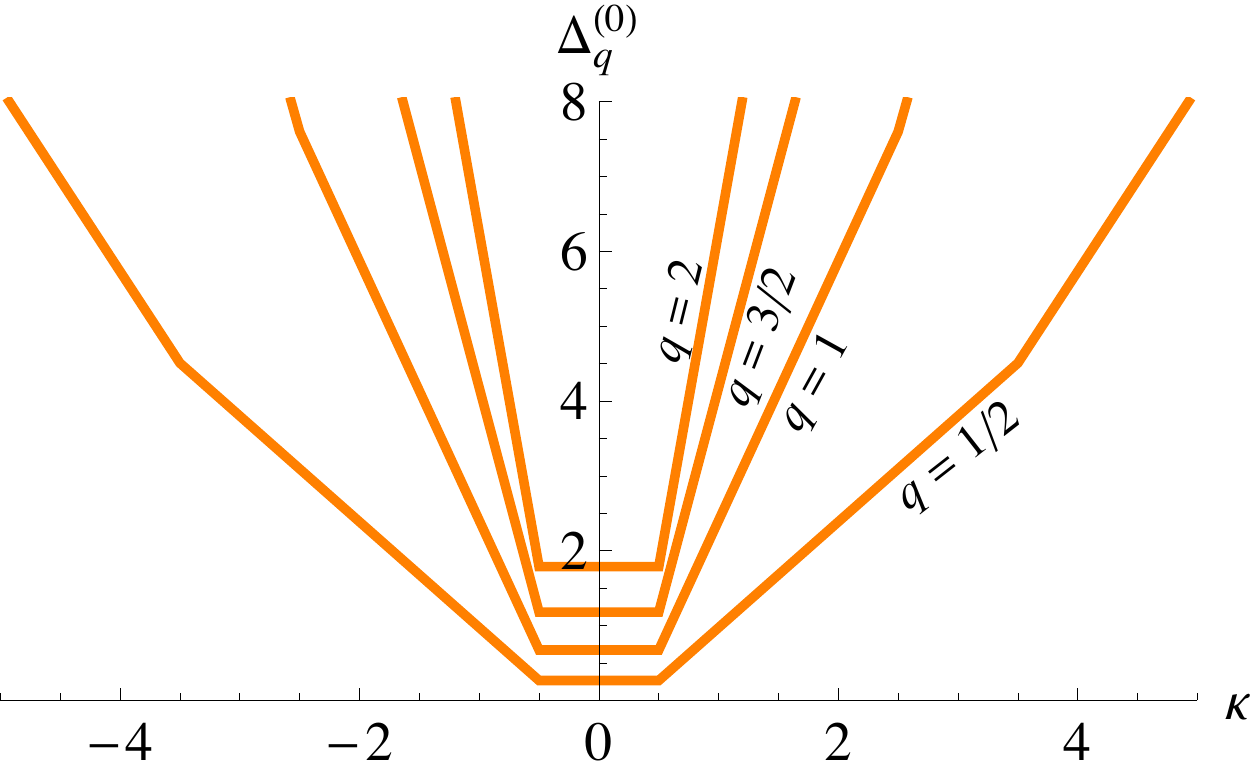} 
    \includegraphics[width=0.5\textwidth]{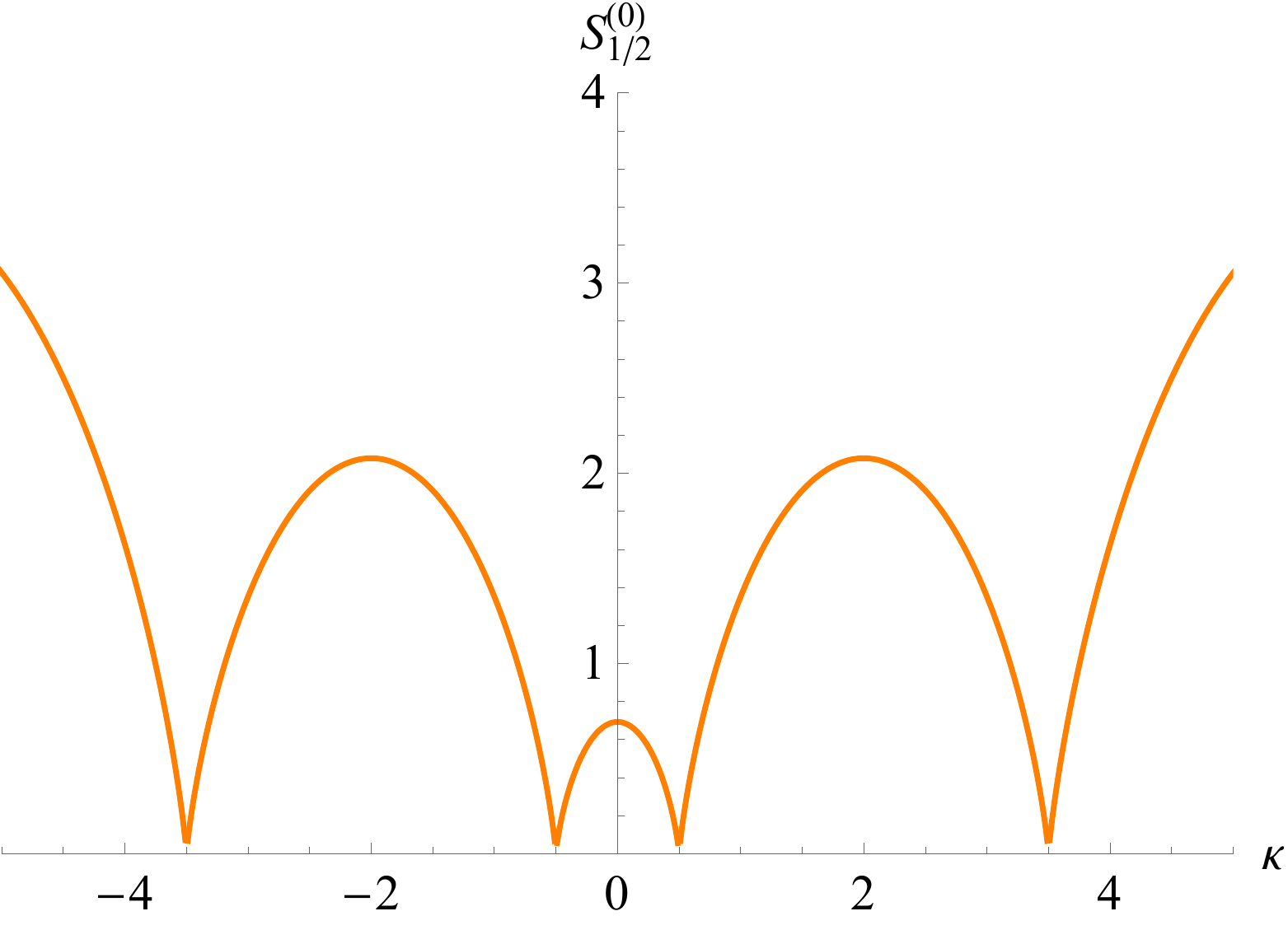}
\caption{\label{saddles} The leading order coefficients $\Delta_q^{(0)}$ and $S_q^{(0)}$ of the scaling dimension and entropy of the lowest-dimension monopole operators in QED$_3$, as a function of $\kappa\equiv k/N$.  The scaling dimension coefficient is plotted for $1/2 \leq q \leq 2$, while the entropy coefficient $S_q^{(0)}$ is plotted only for $q=1/2$ in order to avoid clutter.}
\end{center}
\end{figure}

We can now plug the saddle \eqref{QEDalph} back into \eqref{FQED2} and take the large $\beta$ limit to find the leading order coefficients of the energy and entropy defined in \eqref{FExpansion}--\eqref{FCoeff}: 
\es{QEDFinal}{
\Delta_{q}^{(0)}&=-\sum_{j \geq q-1/2} d_j \lambda_j+\sum_{ q-1/2 \leq j < \tilde j} d_j\lambda_j +\xi_{\tilde j} d_{\tilde j}\lambda_{\tilde j}\,,\\
S_{q}^{(0)}&=-d_{\tilde j}\left(\xi_{\tilde j}\log\xi_{\tilde j}+(1-\xi_{\tilde j})\log[1-\xi_{\tilde j}]\right)\,.
}
Note that the first sum in $\Delta_{q}^{(0)}$ is divergent, but can be regularized using zeta function regularization---see Appendix~\ref{QEDReg} for details.  When $k=0$, $\Delta_{q}^{(0)}$ agrees with the leading order scaling dimension given in \cite{Borokhov:2002ib,Pufu:2013vpa}.  In Figure~\ref{saddles} we plot $\Delta_q^{(0)}$ for $1/2\leq q \leq 2$ as a function of $\kappa$, as well as the leading order coefficient for the entropy, $S_{q}^{(0)}$, in the case $q=1/2$.   

As a consistency check, let us discuss the limit $\kappa \gg 1$.  In this limit, Eq.~\eqref{QEDFinal} reduces to the approximate expression
\es{largeK}{
 \Delta_{q}^{(0)}\approx \frac{2}{3}(2q\kappa)^{3/2}\,, \qquad 
  \text{at large $\kappa$.}
}
This expression actually holds down to $N=1$ (large $N$ is not required), because in the limit $k\to \infty$ the gauge field fluctuations are suppressed for all $N$, and so the saddle point approximation that we used to derive \eqref{largeK} is justified. The large $\kappa$ approximation \eqref{largeK} can be compared to the monopole operator dimension obtained in the 't Hooft limit of $U(N_c)_k$ Chern-Simons-matter theory with $N=1$ discussed in \cite{Radicevic:2015yla}. Indeed, let us take a monopole state with $q$ units of monopole flux in one of the Cartan directions of $U(N_c)$. If we take $k\gg N_c$ the gauge field fluctuations are small, and so we have a fundamental fermion in a fixed monopole background for the $U(N_c)$ gauge field. This fermion can be decomposed into one fermion in an Abelian monopole background of charge $q$, and $N_c-1$ fermions in zero magnetic flux. Thus we claim that the $k\gg N_c$ limit of the result of \cite{Radicevic:2015yla},
\es{Radicevic}{
\Delta_{q}=\frac{2}{3}(2q k)^{3/2}\le[1+O\le({1\ov k},{1\ov N_c}\ri)\ri]\,,
}
 should match the formula \eqref{largeK}. We indeed find agreement, which we take as a consistency check between the results of \cite{Radicevic:2015yla} and ours.

\subsubsection{Subleading free energy}

To find $F_q^{(1)}$, we should consider small fluctuations around the saddle point discussed above.  Let us write $A_\mu = {\cal A}_\mu + a_\mu$ and expand the exponent of \eqref{FQEDdefine} to quadratic order in $a_\mu$.  The linear term in $a_\mu$ vanishes because ${\cal A}_\mu$ is a saddle, so we are left with a Gaussian integral in $a_\mu$: 
\es{subQEDF}{
\exp(-\beta F_q^{(1)})&=\int Da \, \exp\left[-\frac N2 \int d^3xd^3x'\sqrt{g}\sqrt{g'}\, a_\mu(x)K_q^{\mu\nu}(x,x') a_\nu(x')\right]\,,
}
where the kernel $K_q^{\mu\nu}(x,x')$ determining the gauge field fluctuations is
\es{QED3K}{
K_q^{\mu\nu}(x,x')&\equiv- \frac1N \langle J^\mu(x)J^\nu(x')\rangle_{q} -{i\kappa\ov 2\pi}\,\delta(x,x') \,\epsilon^{\mu\nu\rho}\partial'_\rho \,, \qquad 
J^\mu \equiv \psi_i^\dagger\sigma^\mu\psi^i\,.
}
In \eqref{QED3K}, the expectation value is computed in the theory of $N$ free fermions on the gauge field background \eqref{SaddleAnsatz} with $\alpha$ given in \eqref{QEDalph}.  Because this expectation value is proportional to $N$, the quantity $K_q^{\mu\nu}(x,x')$ is independent of $N$. The expression \eqref{subQEDF} is schematic: as written, the integral diverges because it has many flat directions corresponding to pure gauge modes.  To address this problem, one should divide by the volume of the group of gauge transformations.  Equivalently, a computationally simpler procedure that we will employ is to calculate the ratio between $e^{-\beta F_q^{(1)}}$ and $e^{-\beta F_0^{(1)}}$, since in taking this ratio the pure gauge modes cancel.    Moreover, it should be true that $F_0^{(1)} = 0$, because this quantity corresponds to the scaling dimension of the unit operator, so
 \es{F1Ratio}{
  \exp(-\beta F_q^{(1)})&=
   \frac{\int Da \, \exp\left[-\frac N2 \int d^3xd^3x'\sqrt{g}\sqrt{g'}\, a_\mu(x)K_q^{\mu\nu}(x,x') a_\nu(x')\right]}{\int Da \, \exp\left[-\frac N2 \int d^3xd^3x'\sqrt{g}\sqrt{g'}\, a_\mu(x)K_0^{\mu\nu}(x,x') a_\nu(x')\right]} \,.
 }

To perform a Gaussian path integral like the ones in \eqref{F1Ratio}, it is convenient to expand the physical gauge field fluctuations in spherical harmonics / Fourier modes:
 \es{aFourier}{
  a(x) = \mathfrak{a}_{00}^E(0) \frac{d\tau}{\sqrt{4 \pi \beta}}  + 
  \sum_{n=-\infty}^\infty \sum_{\ell=1}^\infty \sum_{m=-\ell}^\ell 
   \left[  \mathfrak{a}_{\ell m}^E(\omega_n) {\cal E}_{n \ell m} (x) 
    + \mathfrak{a}_{\ell m}^B(\omega_n) {\cal B}_{n \ell m} (x) \right]  \frac{e^{-i \omega_n \tau}}{\sqrt{\beta}}+ d \lambda(x)\,,
 }
where $d\lambda$ are pure gauge modes and ${\cal E}_{n \ell m} (\theta, \phi, \tau)$ and ${\cal B}_{n \ell m} (\theta, \phi, \tau)$, together with $d\tau/ (4 \pi \beta)$, form an orthonormal basis of polarizations for the one-form $a(x)$:
 \es{EBDef}{
    {\cal E}_{n \ell m}(x) = \frac{ \ell(\ell+1)  Y_{\ell m}d\tau  -i \omega_n dY_{\ell m}}{\sqrt{\ell(\ell+1)} \sqrt{\omega_n^2 + \ell(\ell+1)}}  \,, \qquad
     {\cal B}_{n \ell m}(x) = \frac{*_2 dY_{\ell m}(\theta, \phi)}{\sqrt{\ell (\ell+1)}}  \,.
 }
Here $\omega_n = 2\pi n / \beta$ are the bosonic Matsubara frequencies, and $*_2$ is the Hodge dual on $S^2$.  From now on we will ignore the pure gauge modes in \eqref{aFourier} because the integral over them cancels between the numerator and denominator of \eqref{F1Ratio}.

After using \eqref{aFourier}, the exponent in the numerator of \eqref{F1Ratio} becomes
 \es{NumExponent}{
  -\frac{N}{2} \abs{\mathfrak{a}_{00}^E(\omega_n)}^2 K^{EE}_{q, 0}(0)  -\frac{N}{2} \sum_{n=-\infty}^\infty \sum_{\ell=1}^\infty \sum_{m=-\ell}^\ell
   \begin{pmatrix} \mathfrak{a}_{\ell m}^E(\omega_n) \\ \mathfrak{a}_{\ell m}^B(\omega_n)
   \end{pmatrix}^\dagger
   {\bf K}_{q, \ell}(\omega_n) \begin{pmatrix} \mathfrak{a}_{\ell m}^E(\omega_n) \\ \mathfrak{a}_{\ell m}^B(\omega_n)
   \end{pmatrix} \,,
 }
where ${\bf K}_{q, \ell}(\omega_n)$ is a $2\times 2$ matrix whose entries we denote with doubled superscripts $E$ or $B$.  Due to rotational invariance, this matrix is independent of the quantum number $m$.  The exponent in the denominator of \eqref{F1Ratio} takes a similar form, but with $q=0$.

The path integrals in both the numerator and denominator of \eqref{F1Ratio} then become infinite products of integrals that are all Gaussian, with an exception to be mentioned shortly.  Evaluating them, one obtains, at large $\beta$,
\es{F1true}{
 \beta F_q^{(1)} \approx \frac{1}{2} \le[\sum_{n=-\infty}^\infty \sum_{\ell=1}^\infty  (2\ell+1)\log \le({\det{\bf K}_{q, \ell}(\omega_n) \ov \det{\bf K}_{0, \ell}(\omega_n) }\ri)
  + \begin{cases}
   \log \le(N\, 2\pi d_{\tilde j} \,\xi_{\tilde j}(1-\xi_{\tilde j})  \ri) & \text{if $\xi_{\tilde j} \neq 0$} \,,\\
   0 & \text{if $\xi_{\tilde j} = 0$}
  \end{cases} \ri]\,.
}
The first term in \eqref{F1true} comes from the integrals over $\mathfrak{a}_{\ell m}^E(\omega_n)$ and $\mathfrak{a}_{\ell m}^B(\omega_n)$, with $\ell \geq 1$---it is the standard formula for a Gaussian integral.  The second term in \eqref{F1true} comes from the integral over $\mathfrak{a}_{00}(0)$, which we now explain.

To understand the last term in \eqref{F1true}, note that the mode $\mathfrak{a}^{EE}_{00}(0)$ is multiplicatively related to the fluctuation $\delta \alpha$ of the holonomy around its saddle point value, $\mathfrak{a}^{EE}_{00}(0)=-i\sqrt{4\pi\beta}\, \delta \alpha$.   Consequently, 
\es{Shortcut}{
K_{q, 0}^{EE}(\omega_n=0)=-{1\ov 4\pi}\, { \partial^2 F_q^{(0)}(\alpha) \ov \partial \alpha^2}
 = \begin{cases}
{\beta\ov 4\pi}\, d_{\tilde j} \,\xi_{\tilde j}(1-\xi_{\tilde j})\,,  \qquad & q\neq 0\,,\\
O(e^{-\beta/2})\,, \qquad & q= 0\,.
\end{cases}\,.
}
The quantity $i \beta \alpha$ being a $U(1)$ holonomy, its integration range is $2 \pi$, and so the integration range of $\mathfrak{a}_{00}^E(0)$ is $\sqrt{4\pi\beta} \, \frac{2 \pi}{\beta}$.  Consequently, at large $\beta$, we have
 \es{IntHol}{
  \int_{- \sqrt{4\pi\beta} \frac{\pi}{\beta}}^{\sqrt{4\pi\beta} \frac{ \pi}{\beta}} d\mathfrak{a}_{00}^E(0)\, 
   e^{-\frac{N}{2} \abs{\mathfrak{a}_{00}^E(\omega_n)}^2 K^{EE}_{q, 0}(0) }
    \approx \begin{cases}
     \sqrt{\frac{16\pi^2}{\beta}} & \text{if $\xi_{\tilde j} = 0$ or $q=0$} \,, \\
     \sqrt{\frac{8\pi^2}{\beta d_{\tilde j} \,\xi_{\tilde j}(1-\xi_{\tilde j})}} & \text{if $q \neq 0$ and $\xi_{\tilde j} \neq 0$}\,.
    \end{cases}
 }
The second term in \eqref{F1true} then follows.

So far, we have reproduced the $\frac 1{2\beta} \log N$ term in the expression for $F_q^{(1)}$ advertised in \eqref{FCoeff}.  The other term advertised in that expression, namely the term proportional to $\frac{1}{\beta} \log \beta$, can be derived as follows.  Just as $K_{q, 0}^{EE}(0)$ in \eqref{Shortcut} was linear in $\beta$ at large $\beta$, some of the entries of the matrices ${\bf K}_{q, \ell}(0)$ with $\ell \geq 1$ will also have linear in $\beta$ entries at large $\beta$.  A tedious computation shows that the coefficient of the linear in $\beta$ contribution to ${\bf K}_{q, \ell}(0)$ takes the form 
 \es{DivInK}{
  \lim_{\beta \to \infty} \frac{{\bf K}_{q, \ell}(0)}{\beta}
   = \overline{\bf K}_{q,\ell} = \xi_{\tilde j}(1-\xi_{\tilde j})  \begin{pmatrix}
    v^E_{q, \ell} \\ v^B_{q, \ell} 
   \end{pmatrix} \begin{pmatrix}
    v^E_{q, \ell} & v^B_{q, \ell} 
   \end{pmatrix} \,,
 } 
for some constants $v^E_{q, \ell}$ and $v^B_{q, \ell}$ that do not vanish only if $\ell \leq d_{\tilde j} - 1$.  Explicit formulas are given in Appendix~\ref{GreenF}.  From those formulas we can convince ourselves that when this linear in $\beta$ contribution is separated out from ${\bf K}_{q, \ell}(\omega_n)$ the remaining part of the kernel can be approximated by a smooth function of $\omega \in \R$ with exponential precision
 \es{SmoothApprox}{
  {{\bf K}}_{q, \ell}(\omega_n) = \beta\overline{\bf K}_{q,\ell} \delta_{n0}+\widetilde{{\bf K}}_{q, \ell}(\omega)\big\vert_{\omega=\omega_n}+ O(e^{-\lambda_{q+1/2} \beta}) \,,
}
where $\lambda_{q+1/2}$ is the lowest nonzero eigenvalue.  Intuitively, the kernel $\widetilde{{\bf K}}_{q, \ell}(\omega)$ should be thought of as the effective kinetic term for the gauge field fluctuations on $S^2\times \R$, hence it is naturally a function of the continuous frequency $\omega$. The sum in \eqref{F1true} then can be rewritten as
\es{SumRewrite}{
 \frac1\beta \sum_{\ell=1}^{d_{\tilde j} - 1}  (2\ell+1) \log\left[ \frac{\det {\bf K}_{q, \ell}(0)}{\det\widetilde{{\bf K}}_{q, \ell}(0)} \right]+  \frac1\beta   \sum_{n=-\infty}^\infty \sum_{\ell=1}^\infty  (2\ell+1)\log \le[{\det{\widetilde{\bf K}}_{q, \ell}(\omega_n) \ov \det{\widetilde{\bf K}}_{0, \ell} (\omega_n) }\ri]+ O(e^{-\lambda_{q+1/2} \beta})\,.
 }
 The second term in this sum can be approximated by an integral, giving  
 \es{Delta1Eq}{
 \Delta_q^{(1)}\equiv\int {d\omega\ov 2\pi }\ \sum_{\ell=1}^\infty  (2\ell+1)\log \le({\det{\widetilde{\bf K}}_{q, \ell}(\omega) \ov \det{\widetilde{\bf K}}_{0, \ell} (\omega) }\ri)
 }
to $O(e^{-\lambda_{q+1/2} \beta})$ precision.   Note that the sub-leading energy term $ \Delta_q^{(1)}$ is manifestly independent of $\beta$, as nothing on the right hand side depends on $\beta$.

Let us now turn our attention to the first term in \eqref{SumRewrite}. Because the matrix $\overline{\bf K}_{q,\ell}$ is written as an outer product of a vector with itself, one can show
\es{detKagain}{
  \sum_{\ell=1}^{d_{\tilde j} - 1}  (2\ell+1) \log\left[ \frac{\det {\bf K}_{q, \ell}(0)}{\det\widetilde{{\bf K}}_{q, \ell}(0)} \right] =   \sum_{\ell=1}^{d_{\tilde j} - 1}  (2\ell+1)\log\left(\beta \xi_{\tilde j}(1-\xi_{\tilde j}) C_{\tilde j, \ell}+1\right)\,,
  }
where we have defined 
 \es{CDef}{
  C_{\tilde j, \ell} \equiv  \begin{pmatrix}
    v^E_{q, \ell} & v^B_{q, \ell} 
   \end{pmatrix}  \widetilde{{\bf K}}^{-1}_{q, \ell}(0) \begin{pmatrix}
    v^E_{q, \ell} \\ v^B_{q, \ell} 
   \end{pmatrix} \,.
 }

With these ingredients, one can see that, because the first sum in \eqref{F1true} equals the sum of \eqref{Delta1Eq} 
 and \eqref{detKagain},  we have
\es{F1smooth}{
 F_q^{(1)}=&\Delta_q^{(1)}  +
  \begin{cases} \frac{\log \le(N\, 2\pi d_{\tilde j} \,\xi_{\tilde j} (1-\xi_{\tilde j}) \ri)  +(d_{\tilde j}^2-1){\log\beta} + \sum_{\ell=1}^{d_{\tilde j}-1} (2\ell+1)\log\left( \xi_{\tilde j}(1-\xi_{\tilde j}) C_{\tilde j, \ell}+\beta^{-1}\right)}{2 \beta} & \text{if $\xi_{\tilde j} \neq 0$}\\
  0 & \text{if $\xi_{\tilde j} = 0$}
  \end{cases} \\
  &{}+ O(e^{-\lambda_{q+1/2}\beta})\,.
}
For $\kappa=0$, $\Delta_q^{(1)}$ was evaluated in \cite{Pufu:2013vpa,Dyer:2013fja}, while for $\kappa\neq0$ we leave its evaluation for a future work.  

In summary, although we have not evaluated $F_q^{(1)}$ fully, we obtained an expression for the temperature dependent part of the free energy.  This correction comes entirely from modes with $\omega_n=0$, hence their contribution on $S^2\times S^1_\beta$ must indeed be suppressed by $1/\beta$ (or $\log \beta/\beta$). Going to higher orders in $1/N$ becomes more challenging and is beyond the scope of our work.  We address the microcanonical interpretation for each term in Eq.~(\ref{F1smooth}) in Section \ref{QEDmicro}.

\subsection{Scalar QED$_3$}
\label{CPNLeading}

\subsubsection{Leading order}

The next theory in which we study monopole operators is scalar QED$_3$, whose action is
 \es{ActionQuartic}{
\mathcal{S}=&\int d^3x \, \sqrt{g} \left[\frac{1}{4e^2} F_{\mu\nu} F^{\mu\nu} +  \sum_{i=1}^N\left[|(\nabla_\mu-i A_\mu)\phi^i|^2+ m^2 \abs{\phi^i}^2 \right]  
  + u \left( \sum_{i=1}^N |\phi^i|^2 \right)^2 \right]  \\
  &{}- \int d^3x \frac{i k}{4 \pi}  \epsilon^{\mu\nu\rho} A_\mu \partial_\nu A_\rho \,,
 }
where $\phi^i$ are complex scalars with unit charge under the $U(1)$ gauge group, $m$ is their mass, $e$ is the gauge coupling constant, and $u$ is the coupling constant for the scalar self-interactions.   This theory is believed to flow to an interacting CFT when $m^2$ is tuned to a critical value.  In studying this theory at large $N$, it is customary to perform a Hubbard-Stratonovich transformation that decouples the quartic term.  This is achieved by introducing a new dynamical field $\mu$ and replacing $ u \left( \sum_{i=1}^N |\phi^i|^2 \right)^2$ by $\mu \sum_{i=1}^N |\phi^i|^2 - \frac{\mu^2}{4u}$, such that the functional integral over $\mu$ reproduces \eqref{ActionQuartic}.  (The integration cycle for $\mu$ is over pure imaginary values.)  This action becomes classically conformally invariant for $m^2 = 0$ and $e,u \to \infty$ and it can be mapped to other conformally flat spaces by covariantizing all derivatives and adding the conformal coupling $\frac{\cal R}{8} \sum_{i=1}^N \abs{\phi^i}^2$.  Thus, a classically conformally-invariant action on $S^2 \times \R$ (which is also the action we will use on $S^2 \times S_\beta^1$) is
 \es{CPNAction}{
\mathcal{S}=&\int d^3x \, \sqrt{g} \sum_{i=1}^N\left[|(\nabla_\mu-i A_\mu)\phi^i|^2+\le(\frac14+\mu\ri) |\phi^i|^2\right]   - \int d^3x \frac{i k}{4 \pi}  \epsilon^{\eta\nu\rho} A_\eta \partial_\nu A_\rho \,,
 }
where the shift of $\mu$ by $1/4$ comes from evaluating the conformal coupling term.  At $k=0$, this theory is in the same universality class as the $\CP^{N-1}$ model.  More generally, gauge invariance requires $k\in\mathbb{Z}$. 

We are interested in studying the theory \eqref{CPNAction} on $S^2 \times S^1_\beta$ with $4\pi q$ magnetic flux through $S^2$ and temperature $T=1/\beta$, so we set $A\equiv\mathcal{A}+a$ and $\mu\equiv\mu_*+i\sigma$,\footnote{The factor of $i$ is consistent with \cite{Dyer:2015zha}.} where $\mathcal{A}$ and $\mu_*$ are the saddle point values of the fields, while $a$ and $\sigma$ are fluctuations. We use the same metric \eqref{metric} and ansatz \eqref{SaddleAnsatz} for $\mathcal{A}_\nu$ as in QED$_3$.  Additionally, the symmetries of $S^2\times S^1_\beta$ require that $\mu_*$ be a constant. In subsequent equation we will drop the asterisk on $\mu_*$ to avoid clutter.

Similar to QED$_3$, after integrating out the matter, the effective action is proportional to $N$ and $k$, so for large $N,k$ we use the saddle point approximation to compute free energy $F_{q}=NF_{q}^{(0)}+O(N^{0})$ with 
\es{FCPN}{
F_{q}^{(0)}(\alpha,\mu)=\beta^{-1}\tr\log\le[-(\nabla_\mu-i\mathcal{A}_\mu)^2+\frac14 +\mu\ri]-2\kappa q\alpha\,,
}
where again $\kappa\equiv k/N$. We will find the saddle point values $\alpha$ and $\mu$ using the saddle point conditions
\es{sadCPN}{
\frac{\partial F_{q}^{(0)}(\alpha,\mu)}{\partial\alpha}\Big\vert_{\alpha,\mu}=\frac{\partial F_{q}^{(0)}(\alpha,\mu)}{\partial\mu}\Big\vert_{\alpha,\mu}=0\,.
}
To evaluate \eqref{FCPN}, we must compute the spectrum of the operator $\left[-(\nabla_\mu-i\mathcal{A}_\mu)^2+\frac14 +\mu\right]$ on $S^2\times S_\beta^1$. This operator has eigenvalues $(\omega_n-i\alpha)^2+\lambda_j^2$, where $ \lambda_j$ are the energies of  modes  of the theory quantized on  $S^2\times \R$  
\es{CPNeigs}{
 \lambda_j=\sqrt{(j+1/2)^2-q^2+\mu} \,, \qquad j\in\{q,\, q+1,\,\dots\}\,,\qquad d_j=2j+1\,,
}
 $d_j$ are the degeneracies of the modes, and we defined the bosonic Matsubara frequencies $\omega_n=\frac{2\pi n}{\beta}\,,n\in\Z$.\footnote{Note in this section that we use the same symbols to denote the eigenvalues, Matsubara frequencies, and degeneracies as in the QED$_3$ case, but they take different values.}
 We now compute
\es{FCPN2}{
   F_{q}^{(0)}(\alpha,\mu) &= -2  \kappa q \alpha +\beta^{-1} \,\sum_{n \in \mathbb{Z}}\sum_{j\geq q} {d_j}  \log \left[ (\omega_n  - i \alpha)^2 +\lambda_j^2 \right]\\
   &= -2  \kappa q \alpha  + \beta^{-1} \sum_{j \geq q} {d_j}\log\le[ 2\left(\cosh(\beta\lambda_j)-\cosh(\beta\alpha)\right)\right] \,.
 }
This sum is divergent, but it can be evaluated in zeta function regularization as we show in Appendix~\ref{scalQEDReg}. Note that the partition function $Z_q^{(0)}(\alpha)=\exp\left[-\beta N{F_q^{(0)}}(\alpha)\right]$ is invariant under the gauge transformation $\alpha\to\alpha+{2\pi i \ov \beta}$ as long as $k\in\Z$.

\begin{figure}[ht!]
\begin{center}
 \includegraphics[width = 0.5\textwidth]{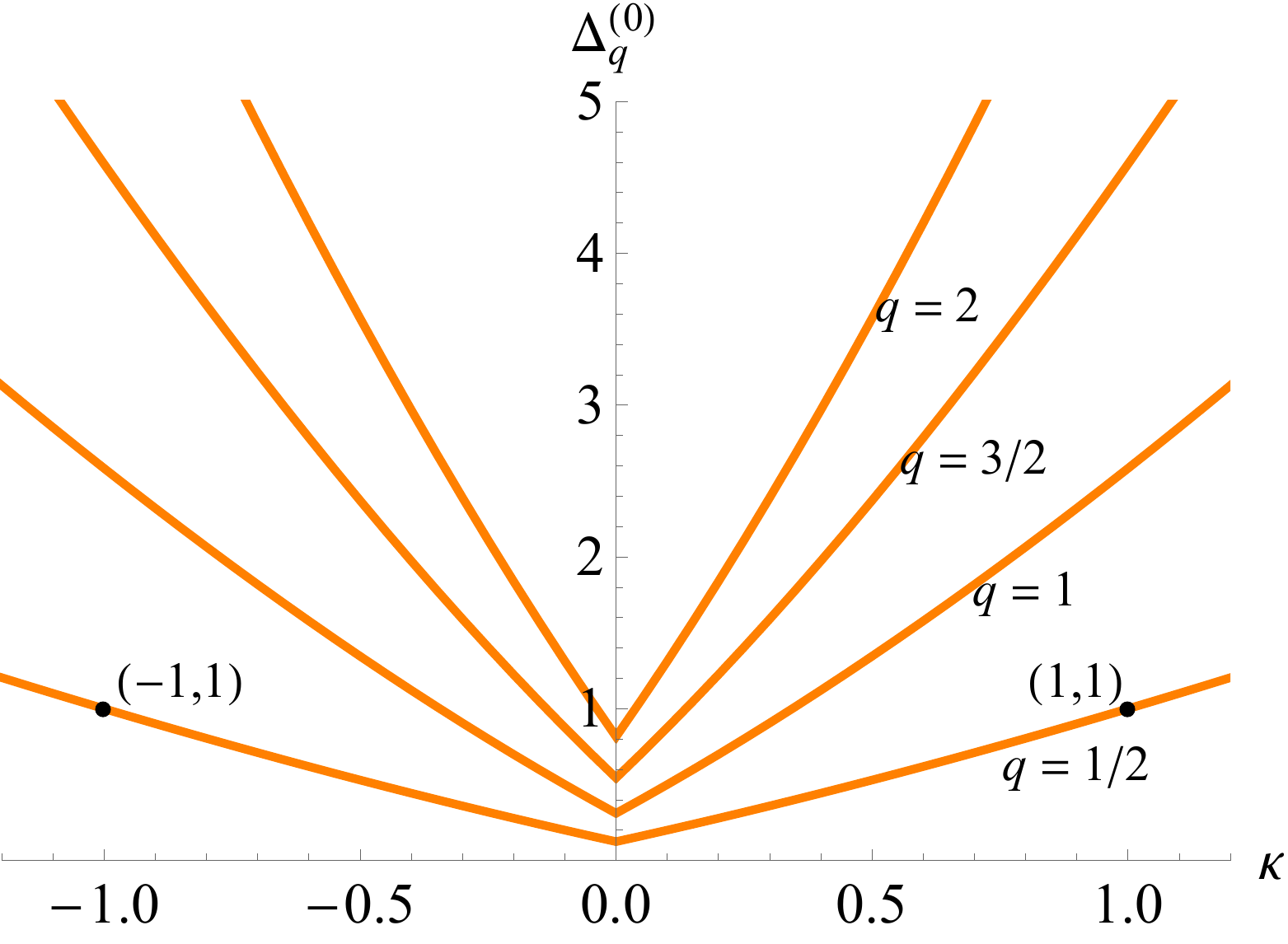}
 \includegraphics[width = 0.49\textwidth]{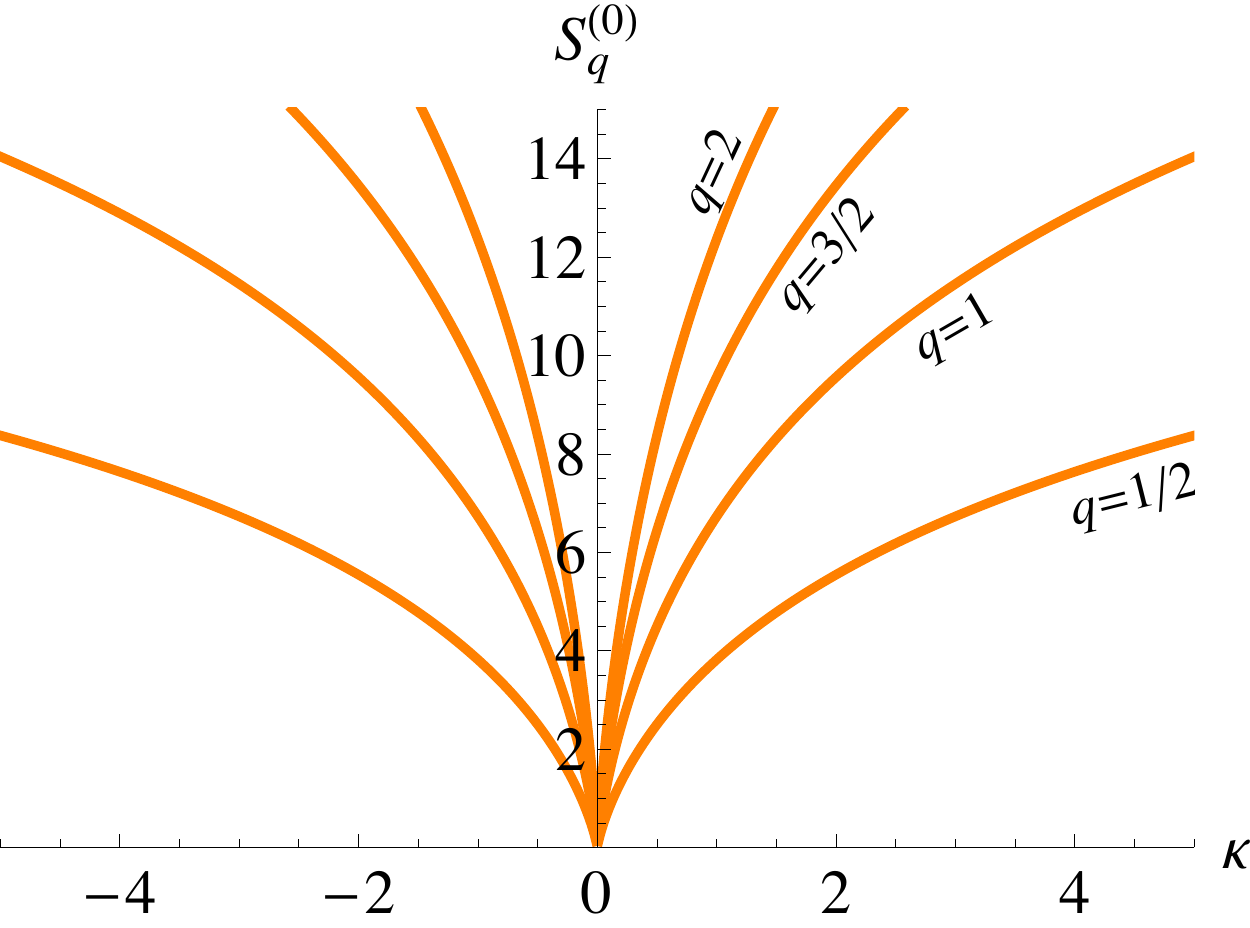} \\
 \includegraphics[width = 0.5\textwidth]{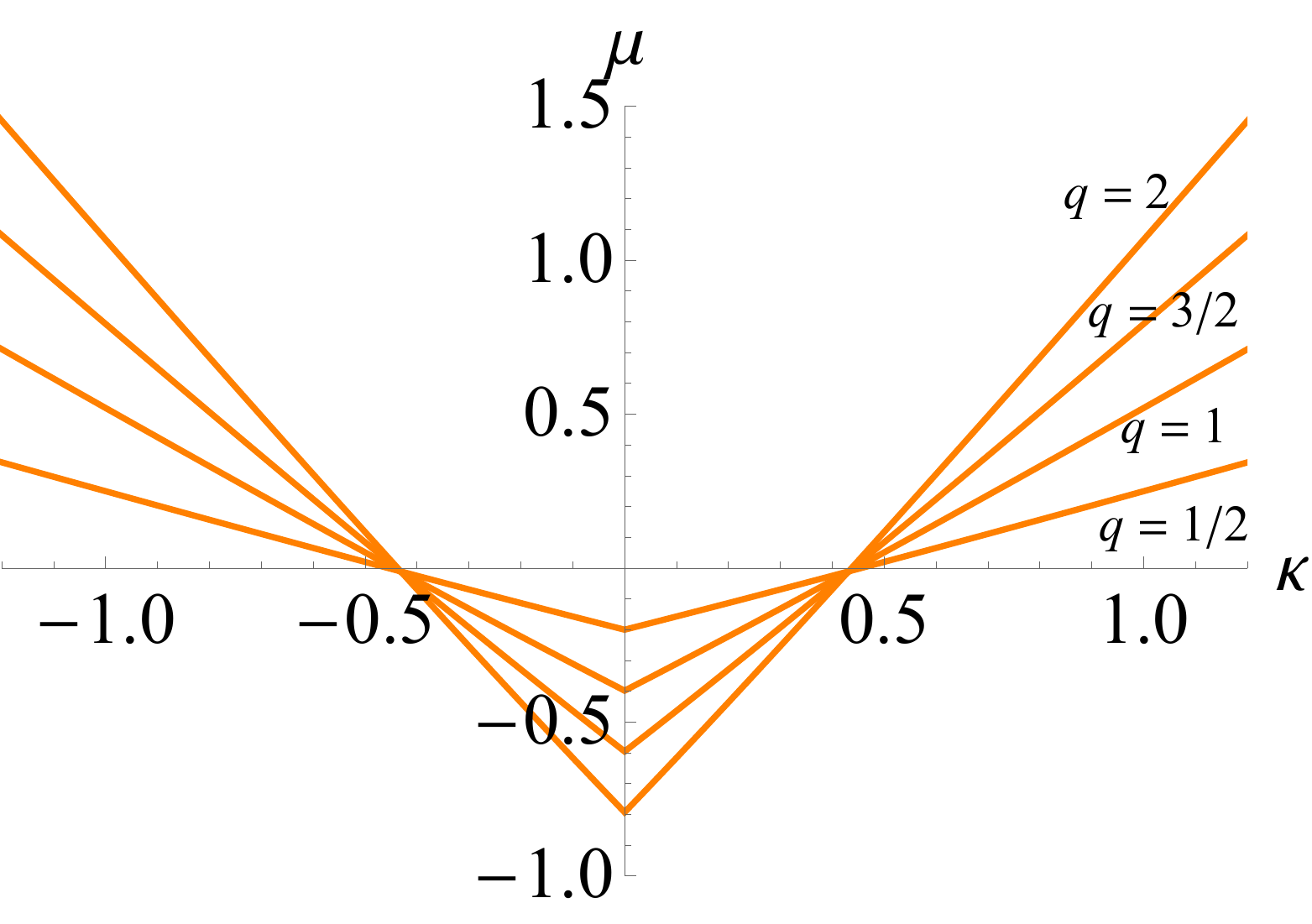}
 \caption{ {\bf Top:}  The leading order coefficients $\Delta_q^{(0)}$ and $S_q^{(0)}$ of the scaling dimension and entropy of the lowest-dimension monopole operators $1/2\leq q \leq 2$ in scalar QED$_3$, as a function of $\kappa\equiv k/N$. The exact value $\Delta_{1/2}^{(0)}=1$ at $|\kappa|=1$ is explained in \eqref{Lucasmagic}. {\bf Bottom:} The saddle point value of the Lagrange multiplier $\mu$ as a function of $\kappa$. Note that $\mu(\kappa)$ is not exactly linear.
 \label{CPN}}
 \end{center}
 \end{figure}

Lastly, we solve \eqref{sadCPN} for large $\beta$ to find a set of possible saddle point values $\alpha$ and $\mu$. For $q=0$, the free energy \eqref{FCPN2} is even in $\alpha$, which implies that $\alpha=0$ is a saddle point. It is easily checked that $\mu=0$ satisfies the saddle point equation.
This result makes sense, as for $\alpha=\mu=0$ we obtain the spectrum of $N$ conformally coupled scalars.  For $q>0$, as in the fermionic case, there are infinitely many candidate saddle points, but unlike in that case, it is now the same saddle that gives the lowest real free energy for all $\kappa$. For $q>0$  this physical saddle point is given by: 
\es{CPNalph}{
\alpha(\kappa)&=-\sgn(\kappa)\left(\lambda_{q}+\beta^{-1}\log\frac{\xi}{1+\xi}\right)+O(e^{-(\lambda_{q+1}-\lambda_q)\beta})\,,\qquad
\xi \equiv \frac{2q |\kappa |}{d_q} \,.
} 
Note that unlike the QED$_3$ term $\log\frac{\xi_{\tilde j}}{1-\xi_{\tilde j}}$, the scalar QED$_3$ term $\log\frac{\xi}{1+\xi}$ is real for all positive $\xi$. Next we take ${\partial F_{q}^{(0)}(\alpha,\mu)\ov \partial \mu}$, and then plug in the value of $\alpha$ \eqref{CPNalph} into this derivative to get the $\mu$ saddle point equation:
\es{muSad}{
\sum_{j\geq q} \frac{d_j}{ \lambda_j(\mu)}+{\xi d_q\ov \lambda_q(\mu)}=0\,.
}
This sum is also divergent, but it can be made finite using zeta function regularization.  For generic $k$ and $q$ we must find $\mu$ numerically, although in the special case $2\abs{\kappa}=d_q$, we find the exact solution $\mu=q^2$.  In Figure~\ref{CPN}, we plot $\mu(\kappa)$ for $1/2\leq q\leq2$, which shows that $\mu(\kappa)$ is almost linear. 

With the saddle point values now fixed, we find the leading order coefficients of the energy and entropy defined in \eqref{FExpansion}--\eqref{FCoeff}:\footnote{Note that the first term in the scalar QED$_3$ scaling dimension does not equal the $k=0$ expression, as it did in QED$_3$, because $\mu$ is a function of $\kappa$.}
\es{CPNFinal}{
\Delta_{q}^{(0)}&=\sum_{j \geq q} d_j\lambda_j+\xi d_q\lambda_{q} \,,\\
S_{q}^{(0)}&=-d_q\left(\xi\log\xi-(1+\xi)\log[1+\xi]\right)\,.
}
In Figure~\ref{CPN} we plot the regularized $\Delta_{q}^{(0)}$ and the corresponding entropies  as a function of $\kappa$ for various small values of $q$.  
 
Interestingly, in the special case $2\abs{\kappa}=d_q$, where we found $\mu = q^2$, we get the simple energy coefficient
\es{Lucasmagic}{
\Delta_{q}^{(0)}&={2\ov 3}q\, (q+1)(2q+1)\,.
}
We observe that this number can be rewritten as a sum of squares:
\es{Lucasmagic2}{
{2\ov 3}q\, (q+1)(2q+1)&=\sum_{0<n\leq2q}n^2\text{   s.t. $n$ is (odd) even for $q$ (half) integer}\,.
}
This rewriting can perhaps help in future explorations of this curious result.

\subsubsection{Subleading order}
The subleading order free energy is computed in analogy with the QED$_3$ case discussed in Section~\ref{QED3All}. The main difference is that we must perform Gaussian integrals over both the gauge field fluctuation $a_\mu$ and the fluctuation $\sigma$ of the Lagrange multiplier field.  The analog of \eqref{subQEDF} thus is
\es{subQEDscal}{
\exp(-\beta F_q^{(1)})&=\int DaD\sigma \, \exp\Biggl[-\frac N2 \int d^3xd^3x'\sqrt{g}\sqrt{g'}\, \biggl( a_{\mu}(x)K_q^{\mu\nu}(x,x') a_{\nu}(x')\\
 &+ \sigma(x) K_q^{\sigma\sigma}(x, x') \sigma(x') + 2\sigma(x) K_q^{\sigma\nu}(x, x') a_\nu(x') \biggr)\Biggr] \,,
}
where
 \es{KernelsScalar}{
K_q^{\mu\nu}(x,x')&\equiv \frac{1}{N} \left[ -\langle J^\mu(x)J^\nu(x')\rangle_{q}+2g^{\mu\nu}\delta(x-x')\langle J(x) \rangle_{q} \right] -{i\kappa\ov 2\pi}\,\delta(x,x') \,\epsilon^{\mu\nu\rho}\partial'_\rho \,,\\
K_q^{\sigma\nu}(x,x')&\equiv-\frac{i}{N} \langle J(x) J^\nu(x')\rangle_{q}  \,,\qquad K_q^{\sigma\sigma}(x,x')\equiv \frac{1}{N} \langle J(x) J(x')\rangle_{q}  \,, 
}
with 
\es{QED3J}{
J^\mu\equiv i\left[\phi^*_i\left(\nabla_\mu-i\mathcal{A}_\mu^q\right)\phi^i-\phi^i\left(\nabla_\mu+i\mathcal{A}_\mu^q\right)\phi^*_i   \right] \,, \qquad J \equiv \phi^*_i \phi^i \,.
}
As in \eqref{F1Ratio}, in order to remove the divergences associated with flat directions, we will compute the ratio between $e^{-\beta F_q^{(1)}}$ and $e^{-\beta F_q^{(0)}}$.  To compute this ratio, we should expand all the fluctuations in spherical harmonics / Fourier modes.  Thus, in addition to expanding $a_\mu$ in harmonics as in \eqref{aFourier}, we should also expand $\sigma$:
 \es{sigmaFourier}{
  \sigma(x) =  
  \sum_{n=-\infty}^\infty \sum_{\ell=0}^\infty \sum_{m=-\ell}^\ell 
    \mathfrak{b}_{\ell m}(\omega_n) Y_{\ell m} (\theta, \phi) \frac{e^{-i \omega_n \tau}}{\sqrt{\beta}} \,.
 }
The result of plugging \eqref{aFourier} and \eqref{sigmaFourier} into the exponent of \eqref{subQEDscal} yields an expression similar to \eqref{NumExponent}:
 \es{NumExponentScal}{
  &{}-\frac{N}{2} \begin{pmatrix} b_{00}(0) \\ \mathfrak{a}_{00}^E(0)
   \end{pmatrix}^\dagger
   {\bf K}_{q, 0}(0) \begin{pmatrix} b_{00}(0) \\ \mathfrak{a}_{00}^E(0)
   \end{pmatrix}
    - \frac{N}{2} \sum_{n \neq 0} \abs{b_{00}(\omega_n)}^2 K_{q, 0}(\omega_n) \\
  &{}-\frac{N}{2} \sum_{n=-\infty}^\infty \sum_{\ell=1}^\infty \sum_{m=-\ell}^\ell
   \begin{pmatrix} b_{\ell m}(\omega_n) \\ \mathfrak{a}_{\ell m}^E(\omega_n) \\ \mathfrak{a}_{\ell m}^B(\omega_n)
   \end{pmatrix}^\dagger
   {\bf K}_{q, \ell}(\omega_n) \begin{pmatrix} b_{\ell m}(\omega_n) \\ \mathfrak{a}_{\ell m}^E(\omega_n) \\ \mathfrak{a}_{\ell m}^B(\omega_n)
   \end{pmatrix} \,,
 }
where now ${\bf K}_{q, \ell}(\omega_n)$ is a $3\times 3$ matrix for $\ell \geq 1$, ${\bf K}_{q, 0}(0)$ is a $2\times 2$ matrix, and $K_{q, 0}(\omega_n)$ for $n \neq 0$ is just a number.

Just as in the fermionic case, we should perform the Gaussian integrals with exponents \eqref{NumExponent} and reproduce the $\frac{1}{\beta} \log N$ and $\frac{1}{\beta} \log \beta$ terms advertised in \eqref{FCoeff}.  The computation is similar, with the only exception that the Lagrange multiplier fluctuations $\sigma$ mix with those of the gauge fields.  The final answer takes the form
\es{F1smoothCPN}{
 F_q^{(1)}=&\Delta_q^{(1)}  +
  \begin{cases} \frac{\log \le(N\, 2\pi d_q \,\xi (1+\xi) \ri)  +(d_q^2-1){\log\beta} + \sum_{\ell=1}^{d_{q}-1} (2\ell+1)\log\left( \xi(1+\xi) C_{q, \ell}+\beta^{-1}\right)}{2 \beta} & \text{if $\xi \neq 0$}\\
  0 & \text{if $\xi_{\tilde j} = 0$}
  \end{cases} \\
  &{}+ O(e^{-\left(\lambda_{q+1}-\lambda_q\right)\beta})\,
}
for some constants $C_{q, \ell}$ and $\Delta_q^{(1)}$. See Appendix~\ref{GreenB} for an expression for $C_{q, \ell}$. When $\kappa = 0$, $\Delta_q^{(1)}$ was evaluated in \cite{Dyer:2015zha}, while for $\kappa\neq0$ we leave the evaluation for a future work.
The expression \eqref{F1smoothCPN} is very similar to the expression \eqref{F1smooth} we obtained in the fermionic QED$_3$ case, the only differences being that $d_{\tilde j}$ is replaced by $d_q$ in \eqref{F1smoothCPN}, and $\xi_{\tilde j} (1 - \xi_{\tilde j})$ is replaced by $\xi(1+\xi)$.  As we will see in Section \ref{CPNmodes}, these differences are precisely what one would expect between fermions and bosons.

\subsection{$\mathcal{N}=1$ SQED$_3$}\label{sec:sqed1}

We can repeat the analysis of the previous two sections in a theory with charged bosons and fermions and minimal ${\cal N} = 1$ supersymmetry.  The ${\cal N} = 1$ vector multiplet contains the vector field $A_\mu$ as well as a gaugino $\lambda$, which is a Majorana spinor.  The minimal matter multiplet is a real multiplet $(\phi, \psi, F)$ containing a real scalar $\phi$, a Majorana fermion $\psi$ and a real auxiliary field $F$.  In order to have matter charged under a $U(1)$ gauge group, we start with $2N$ real multiplets which we then group pairwise into $N$ complex multiplets $(\phi^i, \psi^i, F^i)$, $i = 1, \ldots, N$.  We assign the complex multiplets $(\phi^i, \psi^i, F^i)$ gauge charge $+1$.

On $S^2 \times S^1_\beta$, the Euclidean $\mathcal{N}=1$ SQED$_3$ action  is thus 
 \es{SQEDN1Action}{
\mathcal{S}=&\,\int d^3x\sqrt{g}\sum_{i=1}^N\left[|(\nabla_\mu-i A_\mu)\phi^i|^2+\frac14|\phi^i|^2-\bar{\psi}_i (i \slashed \nabla + \slashed A) \psi^i +\sqrt{2}i\le(\bar\phi_i\lambda\psi^{i}+\bar{\psi}_i\bar\lambda\phi^{i}\ri)- F^i\bar F_i\right]\\
&   - \int d^3x \frac{ \hat k}{4 \pi}  \left[i\epsilon^{\mu\nu\rho} A_\mu \partial_\nu A_\rho +2\sqrt{g}\, i\bar\lambda\lambda\right] \,,
 }
where $\hat k$ is the bare CS level.  Since the auxiliary fields $F^i$ and $\bar F_i$ only appear quadratically, they can be easily integrated out.   The theory \eqref{SQEDN1Action} preserves an $SU(N)$ flavor symmetry under which the matter multiplets transform in the fundamental representation.  Note that there is no quartic scalar interaction term, because such an ${\cal N} = 1$-preserving interaction would come from a cubic superpotential, but there is no such gauge invariant cubic superpotential that preserves $SU(N)$.
 
Gauge invariance requires $\hat k\in\Z$.   As with QED$_3$, a gauge invariant regularization of the fermions induces a CS term of level $-N/2$, which we combine with the bare CS level to define $k\equiv \hat k-N/2$.\footnote{We note that in the supersymmetry literature $k$ is usually defined with the opposite sign.} It is this effective $k$ that contributes magnetic flux to Gauss's law, and so we label $\mathcal{N}=1$ SQED$_3$ using the effective $k$, not $\hat k$. 

Let us now evaluate the $S^2 \times S^1_\beta$ free energy of this theory in the saddle point approximation.    The ansatz for the saddle point configuration of the gauge field is $A = {\cal A}$, with ${\cal A}$ given in \eqref{SaddleAnsatz}.  There cannot be a saddle point value for a fermion field, so we set $\lambda$ to zero, and then we can follow the same steps as the previous sections to compute the free energy $F_{q}=NF_{q}^{(0)}+O(N^{0})$:
\es{FSQEDN12}{
   F_{q}^{(0)}(\alpha) =&\,-2\kappa q \alpha+ \beta^{-1} \sum_{j \geq q} {d_j}\log\left[ 2\left(\cosh(\beta\lambda_j)-\cosh(\beta\alpha)\right)\right]\\
   &- \beta^{-1} \left[\widehat{\sum}_{j \geq q+1/2} {d_j}\log\left[ 2\left(\cosh(\beta\hat{\lambda}_j)+\cosh(\beta\alpha)\right)\right]+d_{q-1/2}\log\left[2\cosh\left(\beta\alpha/2\right)\right]\right]\,,\\
 }
where  $\lambda_j$ are the eigenvalues of the Klein-Gordon operator \eqref{CPNeigs} with $\mu=0$, $\hat{\lambda}_j$ are the eigenvalues of the Dirac operator given in \eqref{QEDeigs}, and $d_j$ is the degeneracy of eigenvalues for both operators. Here and in the rest of this paper wherever both bosonic and fermionic quantities are used, fermionic quantities will be distinguished from the bosonic ones with a hat.  Note that the two sums in \eqref{FSQEDN12} run over different values of $j$:  in the first sum, $j-q$ runs over non-negative integers, and in the second sum $j - q - 1/2$ runs over non-negative integers.  The notation $\widehat\sum$ will serve as a reminder of this fact. 

As before, we fix the saddle point value $\alpha$ using the usual saddle point equation 
\es{sadSQED}{
\frac{\partial F_{q}^{(0)}(\alpha) }{\partial\alpha}\Big\vert_{\alpha}=0\,
}
and find the real $\alpha$ that gives a real free energy to be:
\es{SQEDN1alph}{
\alpha(\kappa)&=
\begin{cases}
   0+\beta^{-1}\log\frac{\hat{\xi}}{1-\hat{\xi}} +O(e^{-(\hat\lambda_{q+1/2})\beta})\,, \qquad  &\hat{\xi}(\kappa)=1/2-\kappa \,, \quad\hspace{0.2cm} |\kappa| \leq 1/2\,,   \\
-\sgn(\kappa)\left(\lambda_q+\beta^{-1}\log\frac{\xi}{1+\xi}\right) +O(e^{-(\lambda_{q+1}-\lambda_q)\beta}) \,,\qquad   &\xi(\kappa)=\frac{2q(|\kappa|-1/2)}{d_q}\,,\quad |\kappa|>1/2\,,
  \end{cases}
 }
where for $|\kappa|\leq1/2$ we have the same QED$_3$ saddle as in \eqref{QEDalph} and $0\leq\hat\xi(\kappa)\leq1$, while for $|\kappa|>1/2$ we have the scalar QED$_3$ saddle given in \eqref{CPNalph} with $\mu=0$, $|\kappa|$ shifted by $-1/2$, and $\xi(\kappa)$ any positive number. 

\begin{figure}[t!]
\begin{center}
\includegraphics[width=0.49\textwidth]{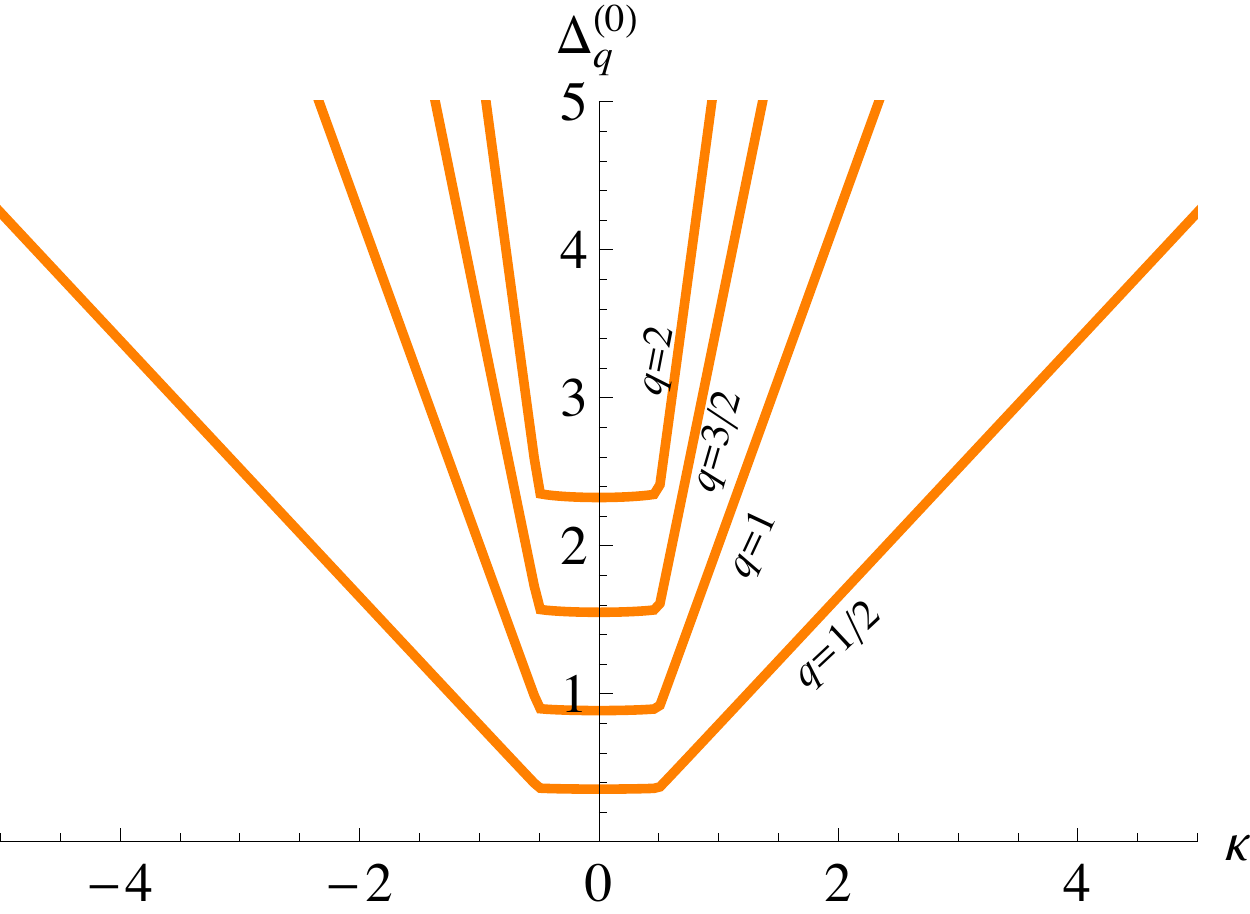}
\caption{
The leading order in $1/N $ monopole scaling dimension $\Delta_{1/2}=N\Delta_{1/2}^{(0)}+O(N^0)$ for $1/2 \leq q \leq 2$ as a function of $\kappa\equiv k/N$. }
\label{SQEDN1}
\end{center}
\end{figure}

With the saddle point values now fixed, we find the leading coefficients in the energy and entropy (compare to \eqref{FExpansion}--\eqref{FCoeff}):
\es{SQEDN1Final}{
\Delta_{q}^{(0)}&=\begin{cases}
\sum_{j \geq q-1/2} d_j\lambda_j -\widehat{\sum}_{j \geq q-1/2} d_j\hat{\lambda}_j\,,\qquad &|\kappa|\leq1/2\\
\sum_{j \geq q} d_j\lambda_j -\widehat{\sum}_{j \geq q-1/2} d_j\hat{\lambda}_j+\xi d_{q}\lambda_{q} \,,\qquad &|\kappa|>1/2\\
\end{cases}\\
S_{q}^{(0)}&=\begin{cases}
-d_{q-1/2}\left(\hat{\xi}\log\hat{\xi}+(1-\hat{\xi})\log(1-\hat{\xi})\right)\,,\qquad &|\kappa|\leq1/2\\
-d_q\left(\xi\log\xi-(1+\xi)\log[1+\xi]\right)\,,\qquad &|\kappa|>1/2\\
\end{cases}
}
The sum in $\Delta_{q}^{(0)}$ is divergent, but can be regularized using zeta functions just as in QED$_3$ and scalar QED$_3$. In Figure~\ref{SQEDN1} we plot the regularized $\Delta_{1/2}^{(0)}$ as a function of $\kappa$. The computation of the subleading free energy  is more complicated than the previous cases due to the gaugino $\lambda$, and we do not carry it out here.
\pagebreak

\subsection{$\mathcal{N}=2$ SQED$_3$}\label{sec:Neq2}

Let us now repeat the same analysis for ${\cal N} = 2$ supersymmetric QED$_3$.  ${\cal N} = 2$ SUSY requires the vector multiplet $(A_\mu, \lambda, \sigma, D)$ to contain a gauge field $A_\mu$, a Majorana fermion $\lambda$, as well as real scalar fields $\sigma$ and $D$.  We will consider the theory of a $U(1)$ vector multiplet with an ${\cal N} = 2$-preserving Chern-Simons term, coupled to $N^+$ chiral multiplets $(\phi^+,\psi^+,F^+)$ of gauge charge $+1$ and $N^-$ chiral multiplets $(\phi^-,\psi^-,F^-)$ of gauge charge $-1$.

The Euclidean action on $S^2 \times S^1_\beta$ that we work with is
\es{SQEDN2Action}{
\mathcal{S}=&\int d^3x\sqrt{g}\sum_\pm\sum_{i=1}^{N^\pm}\left[|(\nabla_\mu\mp i A_\mu)\phi^{\pm,i}|^2+\le(\frac14+\sigma^2\pm D\ri)|\phi^{\pm,i}|^2\right.\\
&\left.-\bar{\psi}_{i}^{\pm}(i \slashed \nabla \pm (\slashed A + i\sigma)) \psi^{\pm,i}+\sqrt{2}i\le(\bar\phi_i^{\pm}\lambda\psi^{\pm,i}+\bar{\psi}_i^{\pm}\bar\lambda\phi^{\pm,i}\ri)-{F^{\pm,i}}\bar F^{\pm}_i\right]\\
& -   \int d^3x \frac{\hat k}{4 \pi}  \left[i\epsilon^{\mu\nu\rho} A_\mu \partial_\nu A_\rho+2\sqrt{g}\le(-D\sigma +i \bar\lambda\lambda\ri)\right]\,,
}
where $\hat k$ is the bare CS level, which is required by gauge invariance to obey $\hat k\in\Z$.\footnote{We note that in the supersymmetry literature $k$ is usually defined to be the opposite sign. }   The action~\eqref{SQEDN2Action} preserves ${\cal N} =2$ superconformal symmetry in the limit $\beta \to \infty$.  At $\beta < \infty$, SUSY is broken by the anti-periodic boundary conditions on the fermions, although if one imposes periodic boundary conditions on the fermions, then one can preserve half the number of supersymmetries for all $\beta$.  The latter construction corresponds to the path integral representation of the superconformal index and is discussed in detail in Appendix~\ref{SQEDN2Index}.

Up to quotients or multiplications by discrete groups, which will be addressed in footnote \ref{N=2SQEDFullSymmetry}, the action \eqref{SQEDN2Action} is invariant under the global symmetry
\es{manySymmetries}{
U(1)_R \times U(1)_{\text{top}}\times U(1)_A \times SU(N^+)\times SU(N^-) \,,
}
where $SU(N^\pm)$ are the flavor symmetries under which the chiral multiplets of gauge charge $\pm1$ transform as a fundamental, $U(1)_A$ is an axial symmetry that only exists if both $N^\pm\neq0$, $U(1)_\text{top}$ is the usual topological symmetry, and $U(1)_R$ is a symmetry under which $(\phi^\pm, \psi^\pm, F^\pm)$ and $(A_\mu, \lambda, \sigma, D)$ have charges $(\frac 12, -\frac 12, -\frac 32)$ and $(0, 1, 0, 0)$, respectively.  (When $\beta \to \infty$, the theory is supersymmetric and the $U(1)_R$ symmetry is an R-symmetry because it does not commute with supersymmetry.)  See Table~\ref{chargeList} for a summary.  Note that requiring these symmetries as well as SUSY in the limit $\beta \to \infty$ uniquely determines the action \eqref{SQEDN2Action}, because no gauge-invariant superpotential is possible.

\begin{table}[!h]
\begin{center}
\begin{tabular}{c||c|c|c|c|c}
  & $SU(N^{\pm})$ &  $U(1)_A$ &$U(1)_R$ &$U(1)_T$&$U(1)_\text{gauge}$ \\
 \hline \hline
 $\phi^{\pm,i}$   & $\bold{N}^\pm $&  $1$&$\frac 12$&$0$&$\pm1$ \\
   \hline
   $\psi^{\pm,i}$   & $\bold{N}^\pm $&  $1$&$-\frac 12$&$0$&$\pm1$ \\
   \hline
  \end{tabular}
\caption{Representations of matter fields for the $\pm$ charged chiral field for the global and gauge symmetries.  
\label{chargeList}}
\end{center}
\end{table}

In defining the non-supersymmetric QED$_3$ theory, we had to specify the prescription \eqref{FreeFerm} for computing the fermion functional determinant in the presence of a background gauge connection.  Likewise, here we also should specify a prescription, which we take to be 
\es{FreeFermN2}{
Z[A,\sigma,D]_\text{free chiral}=&\,{\abs{\Det(i \slashed \nabla \pm (\slashed A-i\sigma))}\ov \Det\le( -\le(\nabla_\mu \mp iA_\mu \ri)^2+\le(\frac14+\sigma^2\mp D\ri)\ri)}\\&\times\exp\le[{-{\pm 1\ov 2}\, \le(i\pi \eta(A)-{1\ov 2\pi}\int d^3x\sqrt{g}\, (D\sigma - i \bar \lambda \lambda) \ri) }\ri]\,.
} 
We chose this prescription because in the limit $\beta \to \infty$ it preserves supersymmetry.  The exponential in \eqref{FreeFermN2} then induces a supersymmetric CS term of level $-\frac{N^{+}-N^-}{2}$, which we combine with the bare CS level to define $k\equiv \hat k-\frac{N^{+}-N^-}{2}$. It is this effective $k$ that contributes magnetic flux to Gauss's law, and so we label $\mathcal{N}=2$ SQED$_3$ using the effective $k$, not $\hat k$. To simplify the subsequent equations, let us further  define
\es{ndef}{
N=N^++N^-\,,\qquad n^\pm=N^\pm/N\,,\qquad \delta n=n^+-n^-\,,\qquad \kappa=k/N\,.
}

We are interested in studying this theory on $S^2\times S^1_\beta$ with metric \eqref{metric} and with $4\pi q$ magnetic flux through $S^2$, so we set $A\equiv\mathcal{A}+a$ where $\mathcal{A}$ is the monopole background defined in \eqref{SaddleAnsatz}, which depends on $q$ and contains a parameter $\alpha$ to be determined at large $N$ by the saddle point condition.  In performing the saddle point approximation, we should also expand the other bosonic fields in the vector multiplet around their saddle point values:  
 \es{OtherSaddle}{
  \sigma = \sigma_* + \delta \sigma \,, \qquad D = D_* + \delta D \,, \qquad
  }
where the $S^2$ rotation invariance together with Euclidean time-translation invariance require $\sigma_*$ and $D_*$ to be constants.  So the large $N$ saddles are characterized by $q$ as well as the values of $\alpha$, $\sigma_*$, $D_*$, which should be determined in terms of $q$ and $\kappa$.

To put in perspective what we find, we note that $\mathcal{N}=2$ SQED$_3$ (in flat space) contains protected BPS monopole operators.  On $S^2 \times \R$, they can be associated with a background on which half of the supersymmetry variations $\delta\lambda=\delta\bar\lambda=0$ vanish.  These equations lead to the unique rotationally-invariant and time-translation invariant solution of a monopole given in \eqref{SaddleAnsatz}, with the vector multiplet scalars taking the value
\es{BPS}{
D=0\,,\qquad \sigma=\pm q\,,
}
where the (negative) positive sign gives the (anti-)BPS monopole background.  
If our large $N$ computations give that $D_*$ and $\sigma_*$ differ from \eqref{BPS} only by order $1/\beta$ terms, then we expect that the lowest energy state with $U(1)_\text{top}$ charge $q$ is indeed (anti-)BPS\@.  If these saddle point values differ from \eqref{BPS} even in the $\beta \to \infty$ limit, then we expect this lowest energy state to not be BPS.

\begin{figure}[H]
\begin{minipage}{0.85\textwidth}
\includegraphics[width=0.85\textwidth]{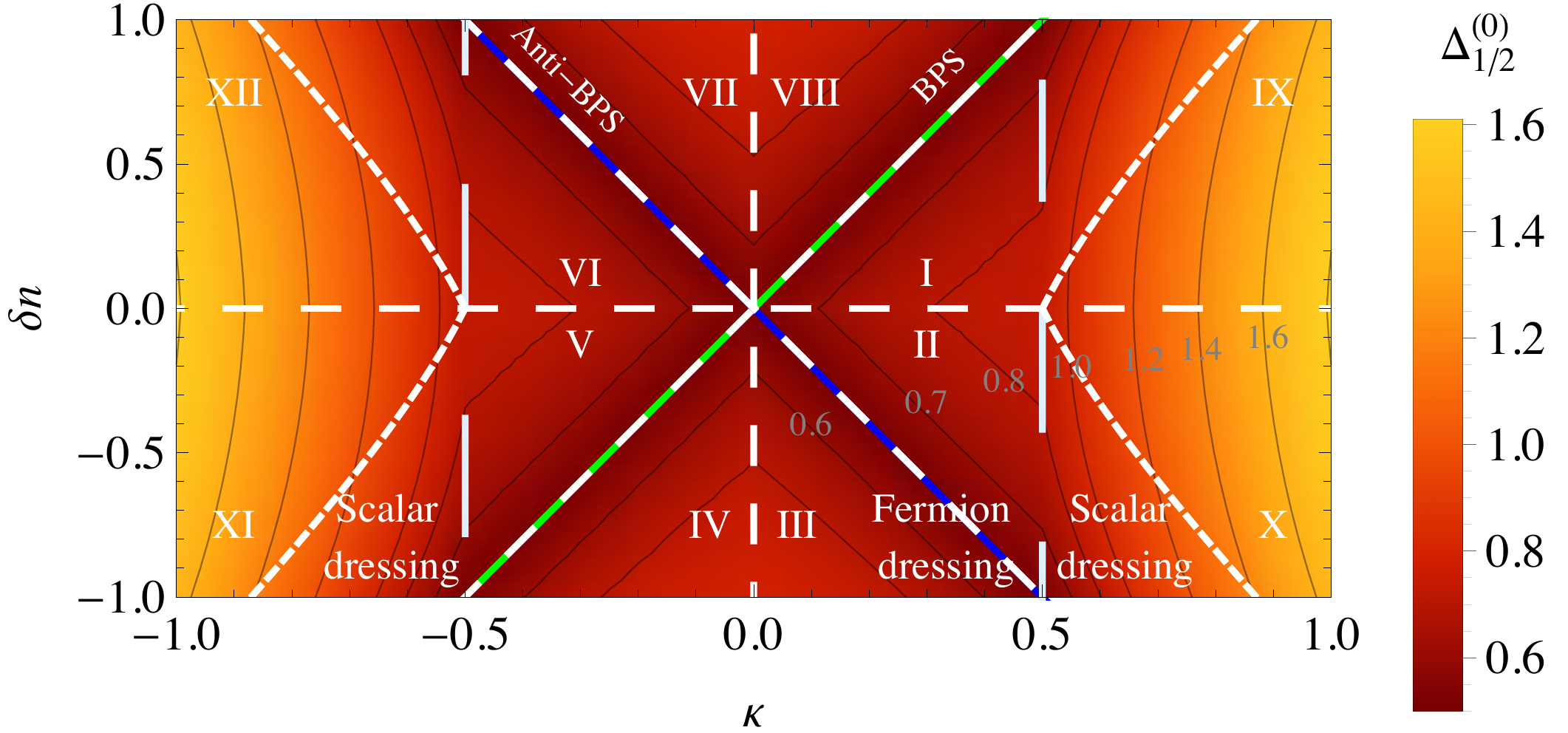}% 
\end{minipage}%
\begin{minipage}{0.15\textwidth}
\hspace{-2cm}
\begingroup
\renewcommand{\arraystretch}{1}
\renewcommand{\tabcolsep}{1.5pt}
\tiny\begin{tabular}{c||c |c| c}
 & $\sgn(\alpha)$ & $\sgn(\sigma)$ & $\sgn(D)$ \\
\hline\hline
$I$ &  $-$&$+$&$+$\\
$II$ & $+$&$-$&$-$\\
$III$ &$- $&$-$&$+$\\
$IV$  & $-$&$+$&$+$\\
$V$ &$+$&$+$&$-$\\
$VI$ & $-$&$-$&$+$\\
$VII$ &$+$&$-$&$-$\\
$VIII$ & $+$&$+$&$-$\\
$IX$ &  $-$&$+$&$-$\\
$X$ & $+$&$-$&$+$\\
$XI$ &$+$&$+$&$+$\\
$XII$ & $-$&$-$&$-$\\
\end{tabular}
\endgroup
\end{minipage}
\includegraphics[width=0.47\textwidth]{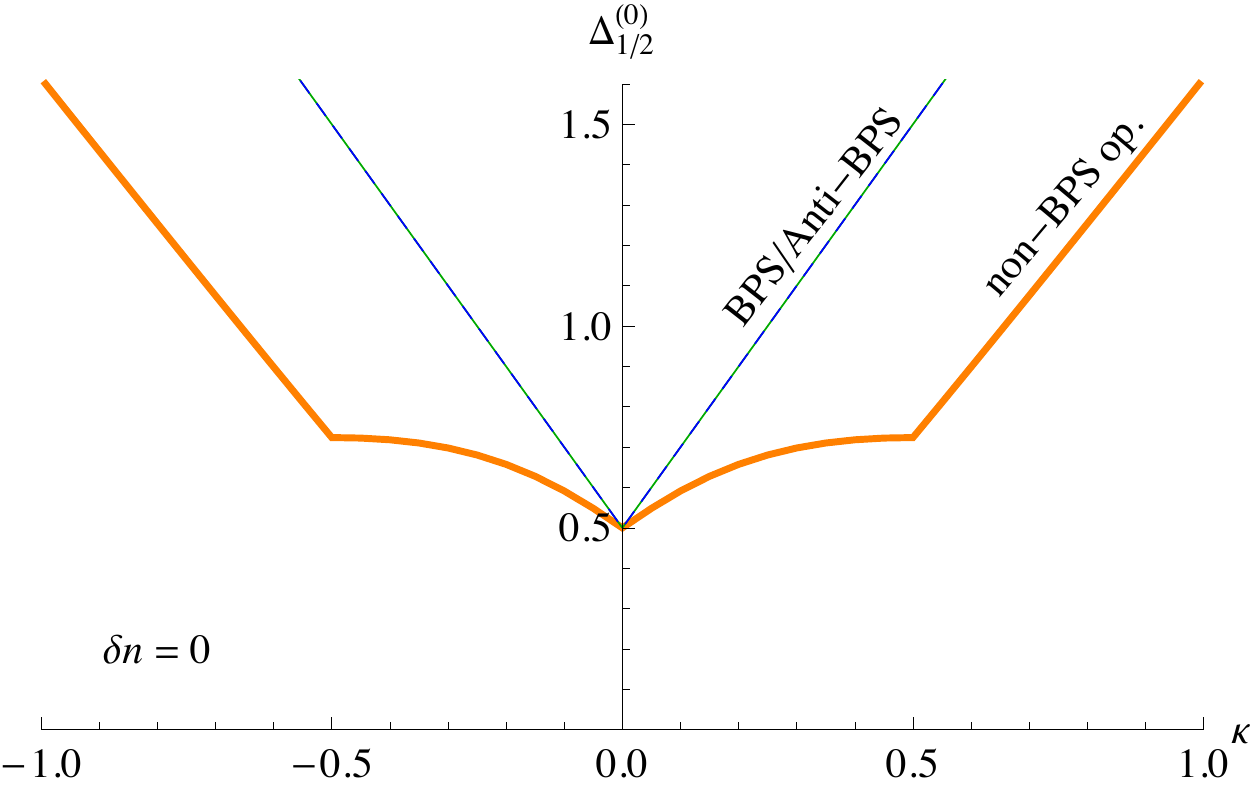}\includegraphics[width=0.47\textwidth]{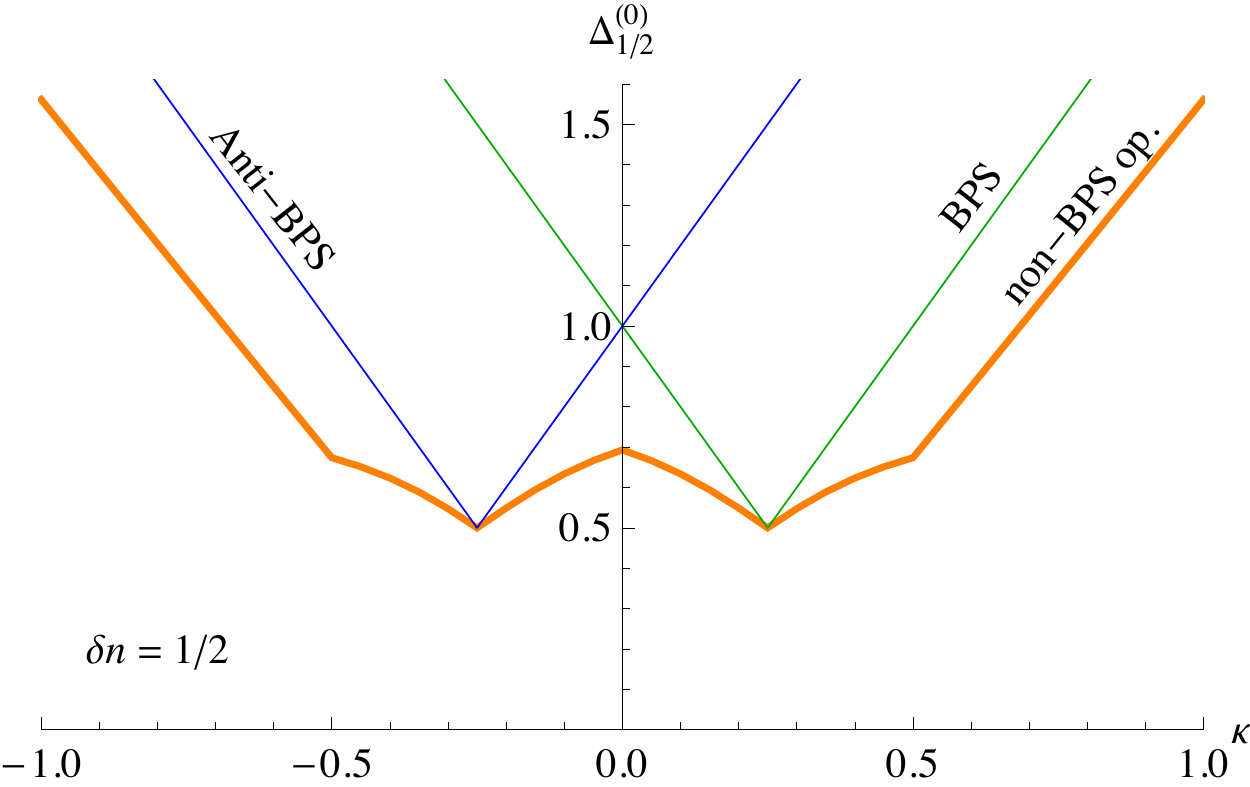}\\
\includegraphics[width=0.47\textwidth]{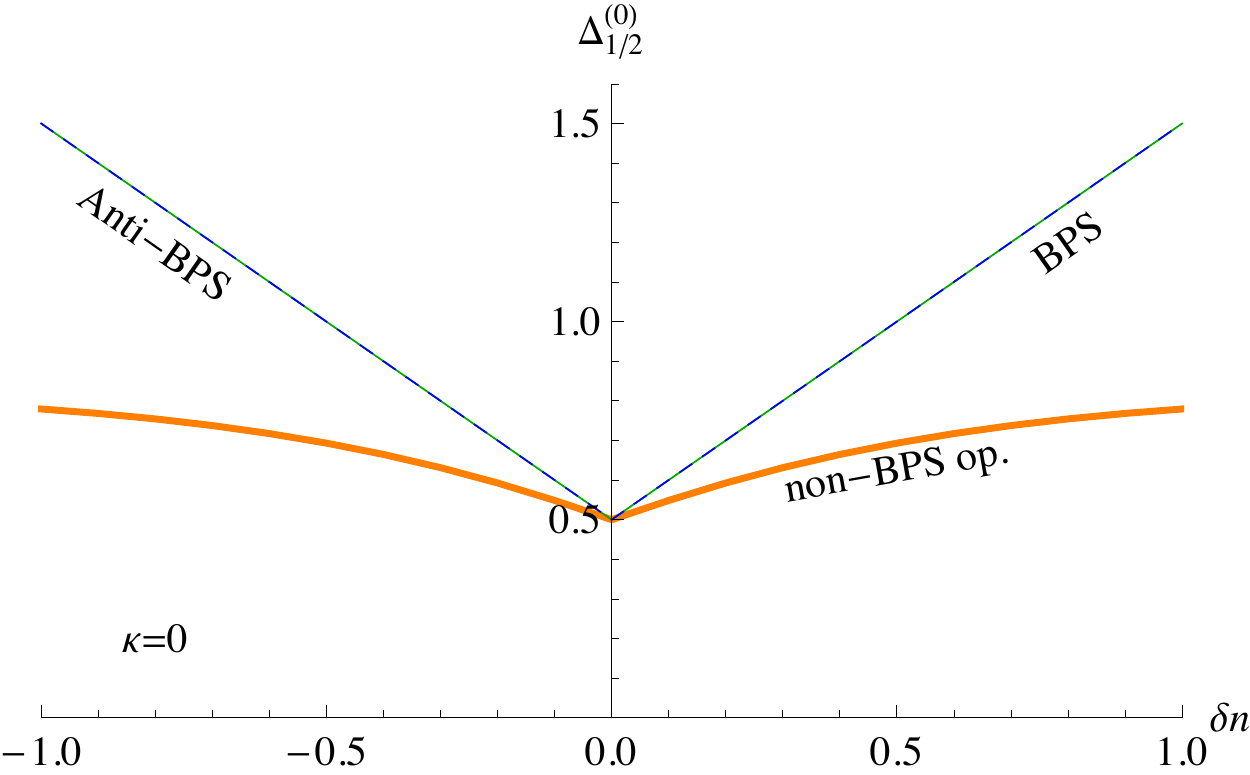}\includegraphics[width=0.47\textwidth]{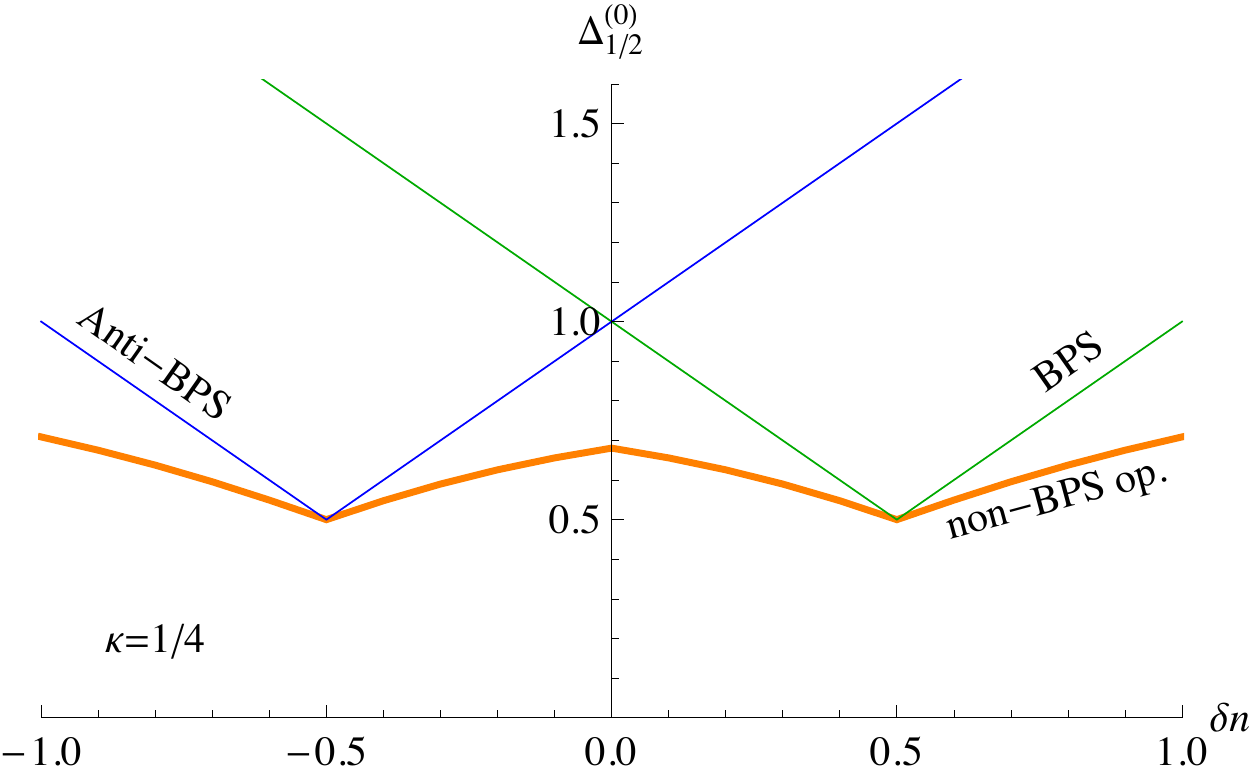}
    \caption{\label{fig:univ-bounds-sigmaT-zoomIn} {\bf Top: }Leading order scaling dimension $\Delta_{1/2}^{(0)}$ of lowest lying $q=1/2$ monopole operator as a function of $\kappa\equiv k/N$ and $\delta n\equiv\frac{N^+-N^-}{N^++N^-}$. Thick dashed lines separate different phases described in the table on the right: the transitions I to IX, II to X, V to XI and VI to XII are smooth in the value of $\sigma$ and $D$, with $D$ changing sign, while all other transitions are discontinuous. For convenience, $\sigma$ and $D$  are shown  in Appendix \ref{SQEDN2SigmaAndD}, in Figure \ref{fig:univ-bounds-sigmaAndD}, for several values of $\delta n$ and $\kappa$. Dashed lines with large dashing separate the regions in which the bare monopole  is dressed with fermionic ($-1/2<\kappa<1/2$) or scalar modes. Along the (anti-)diagonal line the lowest lying monopole is (anti-)BPS and is shown in (blue) green.  {\bf Center:} Horizontal bisection of the density plot  along the line $\delta n=0$ (left) and $\delta n =1/2$ (right) showing the scaling dimension $\Delta_{1/2}^{(0)}$ as a function of $\kappa$. The (blue) green line shows the (anti-)BPS monopole operator dimension \eqref{localBPSb} and the orange  curves show the dimension of the lowest monopole operator. {\bf Bottom: } Vertical bisection of the density plot for $\kappa = 0$ (left) and $1/4$ (right) showing the scaling dimension $\Delta_{1/2}^{(0)}$ as a function of $\delta n$. The color coding is the same as in the center.
    }\label{SQEDN2}
  \end{figure}

As in the previous sections, the functional determinant of the matter fields gives a free energy as a function of $(\alpha, D, \sigma)$ (to avoid clutter, we drop the star subscript on $D$ and $\sigma$):
\es{SQEDN2F2}{
 F_{q}^{(0)}(\alpha, \sigma, D)&= \beta^{-1}\sum_\pm n^\pm \bigg[\sum_{j\geq q} d_j \log(2 \cosh(\beta \lambda_j^{\pm}) - 2 \cosh(\beta\alpha) ) - d_{q-1/2} \log\left(2\cosh\left(\beta(\alpha\mp \sigma )/2\right)\right) \\
 &- \widehat{\sum}_{j\geq  q+1/2} d_j \log(2 \cosh(\beta \hat{\lambda}_j) +2 \cosh(\beta \alpha)) \bigg]  -2\kappa(q\alpha-D\sigma)\,,\\
} 
where 
\es{eigsSQEDN2}{
&\hat\lambda_j=\sqrt{(j+1/2)^2-q^2+\sigma^2} \,,\\
&\lambda^{\pm}_j=\sqrt{(j+1/2)^2-q^2+\sigma^2\pm D} \,,
}
and $d_j=2j+1$.  The saddle point equations are
 \es{SQEDN2SaddleEqs}{
\frac{\partial F_{q}^{(0)}}{\partial \alpha}\Big\vert_{\alpha,D,\sigma}=\frac{\partial F_{q}^{(0)}}{\partial D}\Big\vert_{\alpha,D,\sigma}=\frac{\partial F_{q}^{(0)}}{\partial \sigma}\Big\vert_{\alpha,D,\sigma}=0\,.
}
Similarly to $\mathcal{N}=1$ SQED$_3$, we find the value of $\alpha$ obeying \eqref{SQEDN2SaddleEqs} that gives the real free energy to be 
\es{SQEDN2alph}{
\alpha(\kappa)&=
\begin{cases}
   \mp_\text{tot}\left(\abs{\sigma}+\beta^{-1}\log\frac{\hat\xi}{1-\hat\xi}\right) +O(e^{-(\hat\lambda_{q+1/2}-\hat\lambda_{q-1/2})\beta})\,, \quad  &\hat{\xi}(\kappa)=\frac{ |2\kappa-\delta n|}{2n^{\mp_\text{tot}}}\,, \quad\qquad |\kappa| \leq 1/2 \,,  \\
\mp_\text{tot}\left(\lambda^{{\mp_D}}_q+\beta^{-1}\log\frac{\xi}{1+\xi}\right)+O(e^{-(\lambda_{q+1}-\lambda_q)\beta})  \,,\quad &\xi(\kappa)=\frac{  2q\le(|\kappa|-1/2 \ri)}{d_q n^{\mp_D}} \,, \hspace{.75cm} |\kappa|>1/2\,,\\
  \end{cases}  
 }
where we defined the symbols $\pm_\text{tot}$ and $\pm_D$ as
 \es{pmDef}{
  \pm_\text{tot}\equiv \sgn(G)\sgn(\sigma)\,, \qquad  \pm_D\equiv\sgn(D) \,, 
   \qquad G \equiv 2qN\le(\kappa- \sgn \sigma {\delta n\ov 2}\ri) \,.
 }
While from the thermodynamic point of view, $G$ is just a convenient quantity that simplifies formulas, we will see that from the canonical quantization point of view, $G$ is the gauge charge of the bare monopole.  
For $|\kappa| \leq 1/2$ we have the lowest fermionic saddle and $0\leq\hat{\xi}(\kappa)\leq1$, while for $|\kappa| > 1/2$ we have the lowest bosonic saddle and $\xi(\kappa)$ can be any positive number. There is no simple closed form expression for the solution of the other two saddle point equations in \eqref{SQEDN2SaddleEqs}, but they can be solved numerically: we plug the saddle point value of $\alpha$ back to \eqref{SQEDN2F2} and find the saddle point value of $\sigma,\, D$ numerically. We deal with the appearance of  $\sgn(\sigma),\,\pm_D$ in \eqref{SQEDN2alph} simply by working with all four options at once, finding a saddle point with the assumed sign, and picking the one with the lowest free energy.\footnote{In some regions of parameter space there are multiple physically acceptable saddles giving real free energy, and it would be interesting to understand their significance. Here we always pick the saddle point with the lowest free energy.} In Figure~\ref{SQEDN2}, for $q=1/2$, we split the $(\kappa, \delta n)$ space in 12 regions labeled $I$ through $XII$ which determine the signs of $\sigma,\,D$ (shown in the table next to the density plot in Figure~\ref{SQEDN2}) with which one can determine the value of $\hat{\xi}(\kappa),\, \xi(\kappa)$. For convenience, the precise values for the Lagrange multipliers $\sigma$ and $D$ are shown in Appendix \ref{SQEDN2SigmaAndD}, in Figure \ref{fig:univ-bounds-sigmaAndD}, for several values of $\delta n$ and $\kappa$. 

In terms  of $D$ and $\sigma$, we find the leading order coefficients in the large $N$ expansion of the energy and entropy (compare to \eqref{FExpansion}--\eqref{FCoeff}):  
\es{SQEDN2Final}{
\Delta_{q}^{(0)}&=\begin{cases}
\Delta_\text{bare}+n^{\mp_\text{tot}}\hat{\xi} d_{q-1/2}\abs{\sigma} \,,\qquad &|\kappa| \leq 1/2\,,\\
\Delta_\text{bare}+n^{\mp_\text{tot}} d_{q-1/2}\abs{\sigma}+n^{\mp_D}\xi d_{q}\lambda^{{\mp_D}}_{q} \,,\qquad &|\kappa| \geq 1/2\,,\\
\end{cases}\\
S_{q}^{(0)}&=\begin{cases}
-n^{\mp_\text{tot}}d_{q-1/2}\left(\hat{\xi}\log\hat{\xi}+(1-\hat{\xi})\log[1-\hat{\xi}]\right)\,,\qquad &|\kappa| \leq 1/2\,, \\
-n^{\mp_D}d_{q}\left(\xi\log\xi-(1+\xi)\log[1+\xi]\right)\,,\qquad &|\kappa| > 1/2\,,
\end{cases}
}
with 
 \es{DeltaBareDef}{
  \Delta_\text{bare}&\equiv\sum_\pm n^\pm\left(\sum_{j\geq q}d_j\lambda^{\pm}_j -\widehat{\sum}_{j\geq  q+1/2}d_j\hat{\lambda}_j -q\abs{\sigma}\right)\,.
 }
In Figure~\ref{SQEDN2} we plot the regularized $\Delta_{1/2}^{(0)}$ as a function of $\kappa$ and $\delta n$.  From this figure we learn that in general the BPS (or anti-BPS) operator is not the lowest dimension operator in a sector with monopole charge $q$. (For the properties of the (anti-)BPS operators see Appendix~\ref{SQEDN2Index}.) The exception is the case $\kappa= \delta n/2$ (or $\kappa= -\delta n/2$), where with the signs summarized in the table in Figure~\ref{SQEDN2}, $G=0$, and the bare monopole is already gauge invariant and BPS (or anti-BPS). In Appendix~\ref{SQEDN2Index}, the interested reader can find a detailed computation of the superconformal index for which only BPS states contribute. 

As with $\mathcal{N}=1$ SQED$_3$, the computation of the subleading entropy correction is more complicated than the previous cases due to the gaugino $\lambda$ and other auxiliary fields, and we leave its complete evaluation to future work.
The answer will contain a $-\frac12 \log N$ term as in all our examples, as it  comes from the holonomy fluctuations common to all theories.

The discussion above refers to the thermal free energy on $S^2$, which is computed by the $S^2 \times S^1_\beta$ partition function with anti-periodic boundary conditions for the fermions and periodic boundary conditions for the bosons along the $S^1$.  As mentioned below \eqref{SQEDN2Action}, one can also consider a supersymmetric $S^2 \times S^1_\beta$ theory where both fermions and bosons have periodic boundary conditions.  Such a partition function calculates the superconformal index \cite{Bhattacharya:2008zy}, and it can be evaluated exactly using supersymmetric localization \cite{Imamura:2011su,Kim:2009wb}.  It can also be evaluated in the large $N$ expansion in a similar way as the thermal $S^2 \times S^1_\beta$ computation.  In Appendix~\ref{SQEDN2Index} we show explicitly that the two methods agree in the large $N$ limit.  We view this agreement as a check of our method.

\section{Microstate construction}
\label{MICROSTATES}

In Section~\ref{PARTITION} we determined the $S^2$ thermal free energy at small temperature $T = 1/\beta$ for four different gauge theories with large numbers of flavors $N$.  In this section, we provide a partial interpretation of our results based on an oscillator construction.  Our interpretation is not complete because, as we will see, this oscillator construction is accurate only in the limit of small gauge coupling (UV limit), $e^2 N \to 0$, which is different from the $e^2 N \to \infty$ limit (IR limit) we took in calculating the thermal free energy.  At large $e^2 N$, one has to resum quantum corrections with arbitrarily many loops.  Indeed, it is well-known that in theories with a large number of flavors, flat space calculations of scaling dimensions in the $1/N$ expansion involve resummations of subsets of loop diagrams at every order in $1/N$.  The results are usually quite simple, as most operators built from a finite number of fundamental fields acquire only very small anomalous dimensions.  We expect that such a picture would also apply to how the energy levels on $S^2$ change between the $e^2 N \to 0$ and $e^2 N \to \infty$ limits, but we leave a more thorough investigation to future work.

Here, we would like to take a more pragmatic approach.  Starting from the thermal results derived in Section~\ref{PARTITION}, we will work backwards and deduce the form of the $1/N$ corrections to the energy of the monopole states in the limit $e^2 N \to \infty$.  We will find an interesting picture of many flavor representations that are degenerate to leading order in $N$, and we will provide evidence for how the degeneracy is lifted at subleading orders.  It would be very interesting to derive these results more directly from Feynman diagram computations on $S^2 \times \R$.

\subsection{QED$_3$}
\label{QEDmicro}

\subsubsection{Mode construction}

Let us start with the QED$_3$ case.  As we will now explain, at leading order in $N$, the expressions for $\Delta_q^{(0)}$ and $S_q^{(0)}$ in \eqref{QEDFinal} can be associated with the lowest-energy gauge-invariant state in the theory of free massless fermions on $S^2$ with $4 \pi q$ magnetic flux.  

Briefly, this state can be constructed as follows.  On Lorentzian $S^2 \times \R$, with background magnetic flux $4 \pi q$ uniformly distributed through the $S^2$, the solution to the classical equations of motion for the fermionic fields and their conjugates can be expanded in modes: 
\es{psiDecomp}{
\psi^i(t,x)&=\sum_{m=1/2-q}^{q-1/2}c^{i,\dagger}_{q-1/2, m}\, C_{q-1/2,m}(x)+\sum_{j>q-1/2,m} \le(a^{i,\dagger}_{j m}\, A_{j m}(x)e^{i\lambda_j t}+b_{j m}^i\, B_{j m}(x)e^{-i\lambda_j t}\ri)\,, \\
\psi_i^\dagger(t,x)&=\sum_{m=1/2-q}^{q-1/2}c_{q-1/2, m, i}\, C^\dagger_{q-1/2,m}(x)+\sum_{j>q-1/2,m} \le(a_{j m, i}\, A_{j m}^\dagger(x)e^{i\lambda_j t}+b^\dagger_{j m, i}\, B_{j m, i}^\dagger(x)e^{-i\lambda_j t}\ri)\,.
}
Here, $A_{qjm}(x)$, $B_{qjm}(x)$, and $C_{q,q-1/2,m}(x)$ are the spinor monopole spherical harmonics, the coefficients $c^{i,\dagger}_{q-1/2, m}$, $a^{i,\dagger}_{j m}$, $b^\dagger_{j m, i}$ can be interpreted as fermionic creation operators, and $c_{q-1/2, m, i}$, $a_{j m, i}$, $b_{j m}^i$ are the corresponding annihilation operators.\footnote{In terms of the $S_{q,j=\ell-1/2,m}$ and $T_{q,j=\ell+1/2,m}$ spinor harmonics defined in \cite{Pufu:2013vpa}, we have $A_{qjm}(x)=\frac{(qT_{qjm}(x)+(\lambda_j+j+1/2)S_{qjm}(x))}{\sqrt{(2j+1)(j+1/2+\lambda_j)}}\,,\quad B_{qjm}(x)=\frac{(qT_{qjm}(x)+(\lambda_j-j-1/2)S_{qjm}(x))}{\sqrt{(2j+1)(j+1/2-\lambda_j)}}\,, \quad C_{q,q-1/2,m}(x)=S_{q,q-1/2,m}(x)\,.$}    

We can construct the states by first defining the Fock vacuum state (``bare monopole'') $\vert M_\text{bare}\rangle$ annihilated by all the annihilation operators,
\es{Mbare}{
a_{j m,i} \vert M_\text{bare}\rangle =b_{j m}^i \vert M_\text{bare}\rangle=c_{q-1/2,m,i} \vert M_\text{bare}\rangle=0\,.
} 
This Fock vacuum is non-physical because, as we explain shortly, it has a non-zero gauge charge $G$.  But it can be used to construct the other states in this theory obtained by acting with any number of creation operators.  Since the $(c, c^\dagger)$ fermionic oscillators have zero energy, the vacuum of this theory is $2^{2 qN}$-fold degenerate:  the degenerate states are obtained by acting with any number of the $2q N$ $c^\dagger_{q-1/2,m,i}$ operators on the Fock vacuum.  

The gauge charge $G$ of the Fock vacuum can be determined as follows.  Because the $c^\dagger_{q-1/2,m,i}$ have gauge charge $+1$ (as they appear in the expansion of $\psi_i$), the $2^{2qN}$ vacua have gauge charges that range between $G$ and $G + 2 qN$, distributed symmetrically about the average $G + qN$.  This average gauge charge of the vacua can be identified with the large $\beta$ limit of minus the derivative of the (Gibbs) free energy with respect to the chemical potential $\alpha$, evaluated at $\alpha = 0$:
 \es{dFdmu}{
  - \frac{d (NF_q^{(0)})}{ d\alpha} \bigg|_{\alpha = 0} =  2 N \kappa q \,.
 }
Setting \eqref{dFdmu} equal to $G +qN$, we deduce that the gauge charge of the bare monopole is
 \es{GaugeBare}{
  G =  -qN - \frac{d (NF_q^{(0)})}{ d\alpha} \bigg|_{\alpha = 0}  = {2q}\left( k- \frac N2 \right) \,.
 }
The energy of the bare monopole is obtained by summing up the zero point energies of all the modes
 \es{E0}{
  \Delta_\text{bare} = -N \sum_{j=q-1/2}^\infty d_j \lambda_j \,.
 } 
See Table~\ref{QEDmodes} for a summary of the properties of the bare monopole state and of the creation operators.
 
 \begin{table}[!h]
\begin{center}
\begin{tabular}{c||c|c|c|c|c}
  & energy & spin & gauge charge &$SU(N)$ irrep&degeneracy \\
 \hline \hline
  $a^{i,\dagger}_{j m}$ & $\lambda_{j}$ & $j$ & $+1$ & $\bold{N}$&$Nd_j$ \\
  \hline
 $b^{i,\dagger}_{j m,i}$ & $\lambda_{j}$ & $j$ & $-1$& $\overline{\bold{N}}$ & $Nd_j$ \\
  \hline
 $c^{i,\dagger}_{q-1/2, m}$ & $0$ & $q-1/2$ & $+1$ & $\bold{N}$& $Nd_{q-1/2}$ \\
  \hline
 $M_\text{bare}$ & $-N\sum_j d_j\lambda_j$ & $0$ & $2qN(\kappa-1/2)$ & $\bold{1}$& 1 \\
\end{tabular}
\caption{The UV properties of matter modes and of the bare monopole in QED$_3$. \label{QEDmodes}}
\end{center}
\end{table}
 
The Fock space description above was for fermions in a background gauge field.  With a dynamical gauge field, all physical states must obey Gauss's law:
\es{monoGauge}{
  \le(Q_\text{osc}+G\ri)\vert \Psi\rangle_\text{phys} = 0 \,,
}
where $Q_\text{osc}$ is the contribution of the fermionic oscillators to the electric charge, 
 \es{Qosc}{
  Q_\text{osc}&\equiv \sum_{m=1/2-q}^{q-1/2}c^{i,\dagger}_{q-1/2,m} c_{q-1/2,m,i}+\sum_{j\geq q+1/2} \sum_{m=-j}^j\le(a^{i,\dagger}_{j m} a_{j m,i}-b^\dagger_{j m,i}b_{j m}^i\ri) \,.
 }
 It is clear from \eqref{monoGauge} that $G$ receives contributions from normal ordering  the oscillators in \eqref{Qosc}, and its expression \eqref{GaugeBare} can indeed also be determined this way.
 
\begin{figure}[!t]
\begin{center}
\includegraphics[width = 0.65\textwidth]{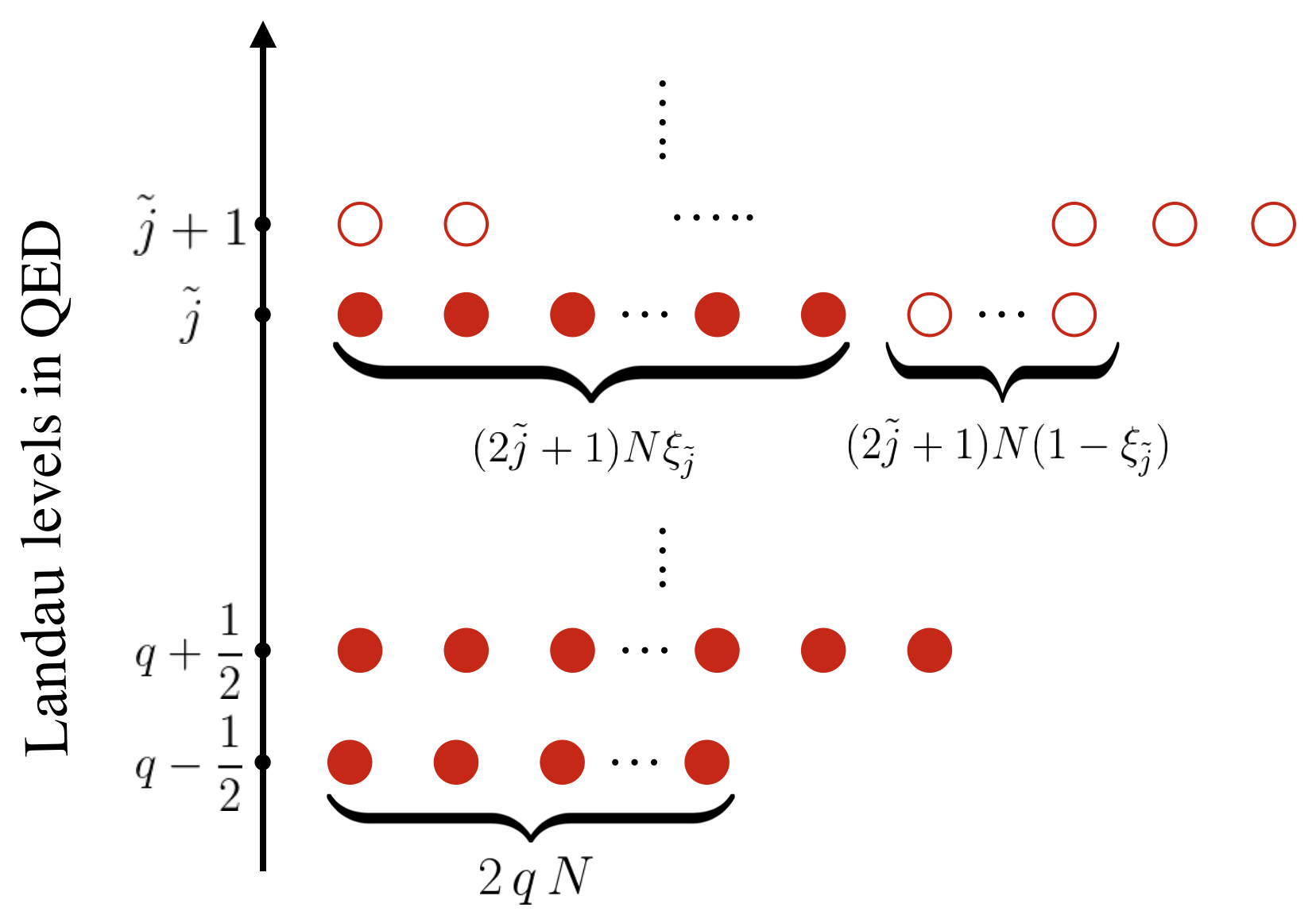}
\caption{ Cartoon showing Landau level filling through the oscillator construction of the microstates. As suggested by the filled red circles, for a given Chern-Simons level, we fill all the Landau levels up to $\tilde j$ and $(2 \tilde j+1)  N \xi_{\tilde j} $ modes at level $\tilde j$. 
 }\label{fig:fermiModeFilling}
 \end{center}
 \end{figure}

The lowest dimension physical states (i.e.~the ones captured by Eq.~\eqref{QEDFinal} which we are trying to interpret), are obtained by dressing the bare monopole with $\abs{G}$ modes of the lowest possible energy (acting with $b_{j m, i}^\dagger$ if $G>0$ and with $a^{i,\dagger}_{j m}$ or $c^{i,\dagger}_{q-1/2, m}$ if $G<0$).  Since there are only finitely many modes for any given $j$, this dressing results in the Landau level picture in Figure~\ref{fig:fermiModeFilling}.  Quantitatively, it is not hard to check that the lowest dimension state constructed as we just described is obtained by filling all Landau levels up to $\tilde j$ and acting with an additional $ d_{\tilde j} N \xi_{\tilde j}$ modes from the level $\tilde j$, with $\tilde j$ and $\xi_{\tilde j}$ given in \eqref{QEDexplicit}--\eqref{QEDalph}.  The energy and entropy in \eqref{QEDFinal} also agree precisely with this Landau level interpretation.  For instance, the first term in $\Delta_q^{(0)}$ in \eqref{QEDFinal} represents the contribution of the bare monopole, the second term is the contribution of the completely filled Landau levels, and the third term is the contribution of the partially filled Landau level.  The entropy only depends on the Landau level that is partially filled.

\subsubsection{Leading order degeneracy}
 
The above picture gives $e^{N S_q^{(0)}}$ degenerate physical states.  Of course, this picture is only valid at large $N$ and also in the UV limit $e^2 N \to 0$, and it is a priori not clear whether it survives in the IR limit $e^2 N \to \infty$.  But the thermal calculation in Section~\ref{QED3All} suggests that most of these states do survive in this limit, and they are still degenerate to leading order in $1/N$.  The immediate questions are:  1)  Is this degeneracy protected by a symmetry?  2)  If not, how is the degeneracy broken?

The symmetries that could protect the degeneracy are conformal symmetry, $U(1)_\text{top}$, and the $SU(N)$ flavor symmetry under which the $\psi^i$ transform as a fundamental.  We know the leading order energy and $U(1)_\text{top}$ charge $q$ of the degenerate monopole states, so what is left to be determined are the $SU(2)_\text{rot}$ spin and $SU(N)$ representations.  If the physical states transform in an irreducible representation of $SU(N) \times SU(2)_\text{rot}$, then the degeneracy is protected;  if they transform in a reducible representation, then one should expect that the degeneracy is lifted at subleading orders in $1/N$.

For simplicity, let us describe the case $\kappa < -1/2$, where the different states differ by which $a^{i,\dagger}_{\tilde j m}$ we act with, but they are all built by acting with $d_{\tilde j} N \xi_{\tilde j}$ such creation operators.  To determine the $SU(N) \times SU(2)_\text{rot}$ irreps, we can use a trick:  we can regard the $m$ indices as labeling the basis states of the fundamental of an  $SU(d_{\tilde j})$ group, which is an enlargement of the $SU(2)_\text{rot}$ rotation group.  $SU(d_{\tilde j})$ is not a symmetry of QED$_3$, but is nevertheless a convenient bookkeeping device \cite{Dyer:2013fja}. 
We then formally combine the $(i, m)$ indices into a fundamental index of the larger group $SU(Nd_{\tilde j})$, which is not a symmetry of the theory either. After building monopoles that transform in the representation of $SU(Nd_{\tilde j})$ we can decompose these representations into representations of the true global symmetry group $SU(N)\times SU(2)_\text{rot}$.

 The fermionic modes are all anti-commuting, so the monopole states transform under $SU(Nd_{\tilde j})$ as the rank-$N\xi_{\tilde j}d_{\tilde j}$ totally antisymmetric irrep, which decomposes under $SU(N)\times SU(d_{\tilde j})$ as the sum
\es{antisum}{
\bigoplus_\nu(\Upsilon_\nu,\tilde\Upsilon_\nu)
}
over all possible irreps $\nu$ with Young diagrams\footnote{\label{MaxDepthFootnote}In \eqref{antisum}, the $SU(N)$ Young diagrams $\Upsilon_\nu$ may have any number of columns of $N$ boxes, and similarly the Young diagrams $\tilde \Upsilon_\nu$ may have any number of columns of $d_{\tilde j}$ boxes.  When reading off the $SU(N)$ and $SU(d_{\tilde j})$ irreps given by the Young diagrams, these columns are redundant and may be deleted, but their boxes are included in the total box count.} $\Upsilon_\nu$ with a total of $d_{\tilde j}N\xi_{\tilde j}$ boxes (whose conjugates are denoted by $\tilde\Upsilon_\nu$) such that $\Upsilon_\nu$ has maximum width $d_{\tilde j}$ and height $N$. Each ordered pair $(\Upsilon_\nu,\tilde\Upsilon_\nu)$ appears once in this decomposition.\footnote{There is a further consistency check for the mode construction. The global symmetry of the theory including discrete groups and quotients is
\be
\label{trueGlobalSymmetry} 
   \frac{U(1)_{\text{top}} \times SU(N)}{\mathbb Z_N} \rtimes \mathbb Z_2^C\,,
\ee
where $U(1)_\text{top}$ is the topological symmetry under which the monopoles are charged,  $SU(N)$ is a flavor symmetry under which the $\psi^i$ transform as a fundamental and the charge conjugation symmetry, $\mathbb Z^C_2$, exchanges the fundamental with the antifundamental representation of the fermionic fields and flips the sign of the monopole charge. The action of $\mathbb Z_N$ is generated by $ ( e^{4\pi i (k-N/2)/N}, e^{2\pi i/N} \mathbf 1_{N}) \in U(1)_\text{top} \times SU(N)$.   

The action of $\mathbb Z_{N}$ on $U(1)_{\text{top}}$ is determined by considering  the properties of the bare monopole under gauge transformations and $U(1)_{\text{top}}$ transformations. Under the $\Z_{N}$ group element $( e^{4\pi i (k-N/2)/N}, e^{2\pi i/N} \mathbf 1_{N})$ the fermion field transforms as $\psi\to \psi \,  e^{2\pi i/N}$, while the bare monopole transforms as $M_\text{bare}\to M_\text{bare}\,e^{4\pi i q(k-N/2)/N}=M_\text{bare}\,\exp\le[ {2\pi i\ov N}G\ri]$. The transformation of the fermion and the bare monopole can be undone by a $U(1)$ gauge transformation of angle $-2\pi/N$, hence we correctly identified the $\Z_{N}$ quotient.

Monopole operators need to transform faithfully under \eqref{trueGlobalSymmetry}, hence they are forced to have $SU(N)$ representations of $N$-ality $-2q(k- N/2)$ mod $N$ due to the $\mathbb Z_N$ quotient.  All $SU(N)$ representations $(\Upsilon_\nu,\tilde\Upsilon_\nu)$, listed in (\ref{antisum}) are indeed of this precise $N$-ality.} The next step is to decompose the $SU(d_{\tilde j})$ representations into $SU(2)_\text{rot}$ representations. This can be done case by case using the fact that the (anti)fundamental of $SU(d_{\tilde j})$, $({\bf \bar{d}_{\tilde j}})$ ${\bf d_{\tilde j}}$ maps to the spin-${\tilde j}$ representation of $SU(2)_\text{rot}$. 

As one can certainly see in explicit examples, this construction generically gives monopole states transforming in a reducible representation of $SU(N) \times SU(2)_\text{rot}$.  (They are, generically, also transforming in a reducible representation of the larger group $SU(N) \times SU(d_{\tilde j})$.)  The $SU(N) \times SU(2)_\text{rot}$ representation is irreducible only when $d_{\tilde j} = 1$, which happens for instance when $\kappa = 0$ and $q = 1/2$, or when $\xi_{\tilde j} = 0$.  An example of a case where we have a reducible representation is $\kappa = 0$ and $q=1$, where $SU(d_{\tilde j}) = SU(2)_\text{rot}$, so we have one $SU(N)$ irrep for every spin $j$.  When $\kappa = 0$ and $q>1$, we generically have several degenerate $SU(N)$ irreps for a given spin.

Thus, the decomposition \eqref{antisum} suggests that, in the UV, there is a large degeneracy among different $SU(N) \times SU(2)_\text{rot}$ representations at leading order in $1/N$.  As we will now show, in the IR this degeneracy is broken at higher orders in $1/N$, with a pattern that we make explicit shortly.

\subsubsection{The lifting of the degeneracy}

Our evidence for the breaking of the degeneracy between the various $SU(N)$ irreps is given by the $\frac{\log \beta}{\beta}$ terms in the large $\beta$ expansion of the order $N^0$ free energy \eqref{F1smooth}.  As explained in Section~\ref{STRATEGY}, such a $\frac{\log \beta}{\beta}$ term appears because the many irreps get split by different amounts, and in the large $N$ limit the distribution of energy levels effectively becomes continuous.  (Taking $N \to \infty$ first before $\beta \to \infty$ is very important here.)  Note that in \eqref{F1smooth} there are no such terms precisely when we expect a single $SU(N)$ irrep in \eqref{antisum}, so there is no accidental degeneracy to be broken.

Let us start by studying the splitting in the simplest case where there is a degeneracy, namely where $SU(d_{\tilde j}) = SU(2)_\text{rot}$, or $\tilde j = 1/2$.  According to \eqref{QEDexplicit}, this corresponds to $q = 1$ and $-1/2 < \kappa < 1/2$.     In this case, the index $\nu$ in \eqref{antisum} can be taken to simply be the $SU(2)_\text{rot}$ spin $\ell$, and Eq.~\eqref{antisum} gives the $SU(N) \times SU(2)_\text{rot}$ representations to be
 \es{SumSpins}{
  \bigoplus_{\ell = 0}^{\ell_\text{max}} ({\bf R}_\ell, {\bf 2 \ell + 1}) \,, \qquad
   \ell_\text{max} = \min \left(N \xi, N(1-\xi) \right) \,,
 }
where the $SU(N)$ representation ${\bf R}_\ell$ is
 \es{Rell}{
  {\bf R}_{\ell} \equiv N \xi + \ell \left\{ \rule{0pt}{1.65cm} \right. \begin{ytableau}
 {}& {}  \\
{} & {}  \\
\none[\vdots] & \none[\vdots] \\
{} & {}  \\
{}  \\
\none[\vdots]\\
{}
\end{ytableau}\raisebox{.62cm}{$\left. \rule{0pt}{.95cm} \right\}N \xi -\ell$}\,,  \qquad
 \dim {\bf R}_\ell = \frac{(2 \ell + 1)}{N + 1} \begin{pmatrix} N + 1\\ N \xi - \ell \end{pmatrix}
  \begin{pmatrix} N + 1\\ N (1-\xi) - \ell \end{pmatrix}  \,.
 } 
Thus, the lifting of the degeneracy can only depend on the spin $\ell$ in this case.

We would like to propose an $\ell$-dependent energy formula that reproduces \eqref{F1smooth}.  In the case $q=1$, $\tilde j = 1/2$ we are studying, Eq.~\eqref{F1smooth} added to the leading free energy gives
 \es{F1smoothq1}{
  F_1=& \left[ N \Delta_1^{(0)} + \Delta_1^{(1)} \right] +
 \frac{ - N S_1^{(0)} + \log \le(N\, 4\pi \,\xi (1-\xi) \ri)  +3{\log\beta} + 3 \log\left( \xi(1-\xi) C_{1, 1}+\beta^{-1}\right)}{2 \beta} \\
 &{}+ O(e^{-\lambda_{q+1/2}\beta})\,.
 }
From the discussion in Section~\ref{STRATEGY}, if the energy levels in \eqref{SumSpins} become dense  and the states are approximated by a continuum, then the density of states should take the form 
 \es{DE}{
  {\cal D}(E) \approx
   \frac{e^{N S_1^{(0)}}}{N^{1/2} \pi \xi^2 (1 - \xi)^2  C_{1, 1}^{3/2}} (E - N \Delta^{(0)}_1 - \Delta^{(1)}_1 )^{1/2} 
 }
in order to reproduce the $\frac{3}{2\beta} \log \beta$ term in the free energy in \eqref{F1smoothq1}.  On the other hand, the explicit construction of the states gives
 \es{dos}{
  {\cal D}(E) \approx \frac{(2 \ell+1) \dim {\bf R}_\ell}{\Delta E_\ell} 
 }  
where $\Delta E_{\ell}$ is the so-far unknown energy difference between the states with spin $\ell+1$ and those with spin $\ell$.  In order to reproduce the $1/\sqrt{N}$ behavior in \eqref{DE}, we should rescale the spins by introducing $y = \ell / \sqrt{N}$, and then take the limit as $N \to \infty$.  In this limit, we obtain
 \es{dosApprox}{
  {\cal D}(E) \approx  e^{N S^{(0)}_1} \frac{2}{N^{1/2} \pi \xi^2 (1-\xi)^2}  y^2 e^{-\frac{1}{\xi(1-\xi)} y^2} 
   \frac{dy}{dE} \,.
 }
Assuming that states with small $y$ have the lowest energy, we should equate the small $y$ limit of \eqref{dosApprox} with \eqref{DE}.  This yields
 \es{dEdy}{
   \frac{dE}{dy}  (E - N \Delta^{(0)}_1 - \Delta^{(1)}_1 )^{1/2} 
    =   2 C_{1, 1}^{3/2}  y^2 \,,
 }
which can be integrated to give
 \es{Ey}{
  E \approx   N \Delta^{(0)}_1 + \Delta^{(1)}_1  + C_{1, 1}  y^2 \,.
 }
This is our main result for the energy splitting in QED$_3$.  It implies that for  monopoles with spins $\ell \sim O(\sqrt{N})$ the energy is split to constant order while for higher spins we cannot determine the exact splitting without going to the next order in $1/N$.  Note that while \eqref{Ey} was derived only in the small $y$ limit, it actually holds for all $y$.  Indeed, one can check that the integral $\int dE\, {\cal D}(E) e^{-\beta E}$, with ${\cal D}(E)$ being the full density of states in \eqref{dosApprox} (not just its $y \to 0$ limit, as was used above), reproduces \eqref{F1smoothq1} precisely. Also note that for $\ell\sim O(N^0)$ the splitting is only at $O(1/N)$.

This discussion can be generalized to $d_{\tilde j} > 2$.  The main difference between $d_{\tilde j} = 2$ and $d_{\tilde j} > 2$ is that when $d_{\tilde j} > 2$, while in \eqref{antisum} a given $SU(N)$ irrep is paired up with a single $SU(d_{\tilde j})$ irrep, upon decomposing these irreps under $SU(N) \times SU(2)_\text{rot}$, there are several $SU(N)$ representations for a given $SU(2)_\text{rot}$ representation.  Thus, the energy splitting would not only depend on the spin $\ell$, like in the $d_{\tilde j} = 2$ case, but also on the extra labels which specify the $SU(d_{\tilde j})$ irrep from which the $SU(2)_\text{rot}$ irrep of spin $\ell$ comes from.  It would be interesting to derive the precise energy splitting formula.  One thing worth noting is that the coefficient of the $\frac{1}{\beta} \log \beta$ is proportional to $d_{\tilde j}^2 - 1$, which is the number of generators of the auxiliary group $SU(d_{\tilde j})$.

\subsubsection{Comments on the Gauss law}

Before moving on to the scalar QED$_3$ case, let us comment on an issue that may be confusing.  In the IR limit of the QED$_3$ theory \eqref{QEDAction}, we take $e^2 N \to \infty$ thus ignoring the Maxwell term.  The equation of motion for the gauge fields sets
 \es{eomGauge}{
  j^\mu  + \frac{k}{4 \pi} \epsilon^{\mu\nu\rho} F_{\nu\rho} = 0 \,,
 }
where $j^\mu = \bar \psi \gamma^\mu \psi$ is the matter gauge current.  So why, then, should we not require that the physical states are only those in which \eqref{eomGauge} is obeyed?  Instead, we are requiring that the physical states obey the much weaker constraint \eqref{monoGauge}, which is the integrated version of the $\tau$-component of \eqref{eomGauge}.

The resolution is that \eqref{eomGauge} does hold, when appropriately interpreted.   The right interpretation of \eqref{eomGauge} is as the Heisenberg equation of motion.  In a perturbative expansion, both $j^\mu$ and $F_{\nu\rho}$ must be expanded in modes;  these expressions can be given as power series in $e^2$, with the leading term of $j^\mu$ obtained from \eqref{psiDecomp} and $F_{\mu\nu}$ obtained from an oscillator decomposition of the gauge field.  As is the case in gauge theories, when the oscillator decompositions for both the matter fields and gauge field are appropriately performed, the only condition needed to enforce \eqref{eomGauge} is the integrated Gauss law \eqref{monoGauge}.   When interpreting \eqref{eomGauge}, it is not correct to plug in the saddle point value for $F_{\mu\nu}$ and keep only the leading terms in $j^\mu$ coming from the mode decomposition \eqref{psiDecomp}, because the former is derived in the $e^2 N \to \infty$ limit, while the latter in the $e^2N \to 0$ limit.

\subsection{Scalar QED$_3$}
\label{scalarQED3modes}

We now use a similar mode interpretation in scalar QED\@.   This interpretation suffers from similar shortcomings as in the fermionic QED case, namely that it is accurate only as $e^2 N, u N \to 0$, and that at this point we can only work backwards and deduce the structure of the $1/N$ corrections to the energies of the various states as $e^2 N, u N \to \infty$ from our free energy computations in Section \ref{CPNLeading}.  

In the scalar case, the Fock space construction of the monopole states is based on bosonic creation and annihilation operators, $a_{j m}^{i, \dagger},\, b_{j m, i}^\dagger$ and $a_{j m,i},\, b_{j m}^i$, respectively, which appear in the mode expansion of the scalar fields on Lorentzian $S^2 \times \R$:
\es{ModeExpansionScal}{
\phi^{i}(x)&=\sum_{j=q}^\infty \sum_{m=-j}^j \frac{1}{\sqrt{ 2 \lambda_j}} \left[ a^{i,\dagger}_{j m}\, Y^*_{q j m}(\theta, \phi)e^{i\lambda_j t}+b^{i}_{j m}\,Y_{q j m}(\theta, \phi)e^{-i\lambda_j t} \right] \,,\\
\phi_i^\dagger(x)&=\sum_{j=q}^\infty \sum_{m=-j}^j  \frac{1}{\sqrt{ 2 \lambda_j}} \left[ a_{j m, i}\, Y^*_{q j m}(\theta, \phi)e^{i\lambda_j t}+b^\dagger_{j m, i}\,Y_{q j m}(\theta, \phi)e^{-i\lambda_j t} \right] \,,
}
where $Y_{q j m}(x)$ are scalar monopole spherical harmonics given in \cite{Wu:1976ge,Wu:1977qk}, and $\lambda_j$ are the bosonic eigenvalues in \eqref{CPNeigs}, computed after plugging in the saddle point value of the Lagrange multiplier $\mu$ (see Figure~\ref{CPN} for instance).  We then have a Fock space of states whose Fock vacuum (bare monopole) state $|M_\text{bare} \rangle$ is annihilated by all the annihilation operators.  Unlike in the fermionic QED case, this vacuum state is unique.

When $k\neq 0$, the bare monopole is unphysical because it carries gauge charge 
\es{monoGaugeCPN}{
G= -\frac{d (NF_q^{(0)})}{ d\alpha} \bigg|_{\alpha = 0}  = 2qN\kappa\,,
}
where $F_q^{(0)}$ is given in \eqref{FCPN2}.  As in the fermionic case, the bare monopole state has a non-zero energy obtained by summing up the ground state energies of all the bosonic oscillators.  The properties of the modes and the bare monopole are summarized in Table~\ref{CPNmodes}. 
 \begin{table}[!h]
\begin{center}
\begin{tabular}{c||c|c|c|c|c}
  & energy & spin & gauge charge &$SU(N)$ irrep&degeneracy \\
 \hline \hline
  $a^{i,\dagger}_{j m}$ & $\lambda_{j}$ & $j$ & $+1$ & $\bold{N}$&$Nd_j$ \\
  \hline
 $b^{\dagger}_{j m,i}$ & $\lambda_{j}$ & $j$ & $-1$& $\overline{\bold{N}}$ & $Nd_j$ \\
  \hline
  $M_\text{bare}$ & $N\sum_j d_j\lambda_j$ & $0$ & $2qN\kappa$ & $\bold{1}$& 1 \\
\end{tabular}
\caption{UV properties of modes and the bare monopole in scalar QED$_3$. \label{CPNmodes}}
\end{center}
\end{table}

Similarly to the reasoning followed in QED$_3$, Gauss's law requires that physical states have zero gauge charge.  Since the modes are bosonic, we can minimize the energy by always dressing with the lowest $j=q$ mode. This picture is, once again, exact in the UV and it matches the expression for the leading order energy \eqref{CPNFinal}. Indeed, in this expression, the first term is the energy of the bare monopole, given by summing the zero point energies of all modes. The second term is the energy of the $d_q N\xi =|G|$ number of $j=q$ modes we dress with, where $\xi$ can in general be any positive number, unlike in QED$_3$. However, the energy $\lambda_j$ of each individual quantum is affected by the presence of the others through the dependence on $\mu$: instead of a proper free boson in a monopole background, we have a mean field-like description.

For the special case $q=1/2$ and $\kappa=1$, the simplified energy \eqref{Lucasmagic} has a particularly simple interpretation, as the Casimir term is zero and $\lambda_q=1$, so that the energy of the state is equal to  the energy of the $N$ free fermions.\footnote{The bosons transmute into fermions in the presence of a monopole background of half integer $q$, as can be read from Table~\ref{CPNmodes}.} It would be very interesting to understand if this is a coincidence, or a hint towards new possible dualities as those suggested \cite{Komargodski:2017keh}. 

The determination of the possible $SU(N)$ irreps of the dressed monopole state is similar to the QED$_3$ case described in Section~\ref{QEDmicro}, except we now have commuting creation operators.  When $\kappa<0$, we dress with positively charged modes $a^{i, \dagger}_{jm}$ such that the physical states transform under the auxiliary group $SU(Nd_q)$ as the rank-$N\xi d_q$ totally symmetric irrep.  
This irrep decomposes under $SU(N)\times SU(d_q)$ as the sum
\es{symsum}{
\bigoplus_\nu(\Upsilon_\nu,\Upsilon_\nu)
}
over all possible irreps $\nu$ with Young diagrams\footnote{The same comment as in Footnote~\ref{MaxDepthFootnote} applies.} $\Upsilon_\nu$ with a total of $d_qN\xi $ boxes such that $\Upsilon_\nu$ has maximum height $\min(d_q,N)$.\footnote{Similarly to the fermionic case discussed in Section \ref{QEDmicro}, the global symmetry of the theory is
\be 
\label{trueGlobalSymmetry-scalarQED} 
 \frac{U(1)_{\text{top}} \times SU(N)}{\mathbb Z_N} \rtimes  \mathbb Z_2^C\,,
\ee
where the action of the $\mathbb Z_N$ quotient is generated by $g = ( e^{4\pi i k/N}, e^{2\pi i/N} \mathbf 1_{N})$ and $\Z_2^C$ is the charge conjugation symmetry. Consequently, the representations of monopole operators under $SU(N)$ should be of $N$-ality $-2kq$ mod $N$. Indeed, the representations in (\ref{symsum}) satisfy this condition. 
}   Each pair $(\Upsilon_\nu,\Upsilon_\nu)$ appears once in this decomposition.  When $\kappa>0$ and we dress with negatively charged modes then we should take the conjugate of these representations.   

As in the fermionic QED case, the $SU(N) \times SU(d_q)$ irreps determined as above should be further decomposed under $SU(N) \times SU(2)_\text{rot}$, where $SU(2)_\text{rot}$ is the rotation symmetry of $S^2$.   It is not hard to check that, unless $\kappa = 0$, this decomposition results in many degenerate $SU(N) \times SU(2)_\text{rot}$ irreps.

We expect that this degeneracy is lifted in a way similar to the fermionic QED$_3$ case.  As we did there, let us explain how it is lifted in the simplest case in which there is a degeneracy, namely for $d_q = 2$.  In this case, $SU(d_q) = SU(2)_\text{rot}$ and the $SU(N) \times SU(2)_\text{rot}$ irreps are uniquely labeled by their $SU(2)_\text{rot}$ spin $\ell$.  This is what happens for $q=1/2$ and any value of $\kappa \neq 0$.  The representations \eqref{symsum} become in this case
 \es{SumSpinsScalar}{
  \bigoplus_{\ell = 0}^{N \xi} ({\bf R}_\ell, {\bf 2 \ell + 1})  \,,
 }
where the $SU(N)$ representation ${\bf R}_\ell$ is
 \es{RellScalar}{
  {\bf R}_{\ell} \equiv    \raisebox{-0.15in}{\makebox[0in][l]{$\underbrace{\rule{0.68in}{0in}}_{N \xi - \ell}$}}
   \raisebox{0.25in}{\makebox[0in][l]{$\overbrace{\rule{1.2in}{0in}}^{N \xi + \ell}$}}
   \begin{ytableau}
 {}& {} & \none[\cdots] & {} & {} & \none[\cdots] & {} \\
 {}& {} & \none[\cdots] & {} 
\end{ytableau}\,,  \quad
 \dim {\bf R}_\ell =  \frac{(2 \ell + 1)}{N - 1} \begin{pmatrix} N (1 + \xi) + \ell - 1\\ N \xi + \ell + 1 \end{pmatrix}
  \begin{pmatrix} N(1 + \xi) - \ell - 2 \\ N \xi - \ell \end{pmatrix} \,.
 } 

We would again like to provide an energy-splitting formula that explains the thermal result.  The sum of the leading free energy and the subleading correction \eqref{F1smoothCPN} is in this case
 \es{F1smoothq12}{
  F_{1/2}=& \left[ N \Delta_{1/2}^{(0)} + \Delta_{1/2}^{(1)} \right] +
 \frac{ - N S_{1/2}^{(0)} + \log \le(N\, 4\pi \,\xi (1+\xi) \ri)  +3{\log\beta} + 3 \log\left( \xi(1+\xi) C_{1/2, 1}+\beta^{-1}\right)}{2 \beta} \\
 &{}+ O(e^{-(\lambda_{q+1}-\lambda_q)\beta})\,.
 }
The derivation of the $\ell$-dependent energy splitting is very similar to that in the fermionic QED case.  After defining $y = \ell / \sqrt{N}$ and taking the large $N$ limit, it gives 
\es{EyCPN}{
  E \approx   N \Delta^{(0)}_{1/2} + \Delta^{(1)}_{1/2}  + C_{1/2, 1}  y^2 \,.
 }
Thus, again, for spins $\ell \sim O(\sqrt{N})$ the energy is split to constant order, but for spins of $O(N^0)$ the splitting is only at $O(1/N)$.  It would be interesting to provide a first-principles derivation of this result, and also to generalize it to $q > 1/2$.

\subsection{${\cal N} = 1$ SQED$_3$}

We can perform a similar analysis in ${\cal N} = 1$ SQED\@.  Indeed, we can interpret the leading order results in \eqref{SQEDN1Final} as coming from a Fock space picture in this case too.   To build this Fock space, we consider bosonic annihilation operators $a_{j m,i},\, b_{j m}^i$ and  fermionic annihilation operators $\hat{a}_{j m,i},\, \hat{b}_{j m}^i$, $\hat{c}_{q-1/2, m,i}$, as well as the corresponding creation operators. The bare monopole $|M_\text{bare} \rangle$ is defined as the vacuum state of $S^2\times \mathbb{R}$ of the free complex multiplets in a monopole background, annihilated by all bosonic and fermionic annihilation operators. As in fermionic QED$_3$, at leading order in $1/N$ there are $2^{\abs{q} N}$ degenerate vacua obtained by acting with any number of $\hat{c}_{q-1/2, m,i}^\dagger$ on $|M_\text{bare} \rangle$.  The same argument that led to \eqref{GaugeBare} shows that the gauge charge $G$ of $|M_\text{bare} \rangle$ is given by
\es{monoGaugeSQEDN1}{
G=- qN  - \frac{d (NF_q^{(0)})}{ d\alpha} \bigg|_{\alpha = 0} = 2qN(\kappa-1/2)\,,
}
where $F_q^{(0)}$ is given in \eqref{FSQEDN12}. The properties of the modes and the bare monopole are summarized in Table~\ref{SQEDN1modes}. 
 \begin{table}[!h]
\begin{center}
\begin{tabular}{c||c|c|c|c|c}
  & energy & spin & gauge charge &$SU(N)$ irrep&degeneracy \\
 \hline \hline
  $a^{i,\dagger}_{j m}$ & $\lambda_{j}$ & $j$ & $+1$ & $\bold{N}$&$Nd_j$ \\
  \hline
 $b^{\dagger}_{j m,i}$ & $\lambda_{j}$ & $j$ & $-1$& $\overline{\bold{N}}$ & $Nd_j$ \\
  \hline
    $\hat{a}^{i,\dagger}_{j m}$ & $\hat{\lambda}_{j}$ & $j$ & $+1$ & $\bold{N}$&$Nd_j$ \\
  \hline
 $\hat{b}^{\dagger}_{j m,i}$ & $\hat{\lambda}_{j}$ & $j$ & $-1$& $\overline{\bold{N}}$ & $Nd_j$ \\
  \hline
 $\hat{c}^{i,\dagger}_{q-1/2, m}$ & $0$ & $q-1/2$ & $+1$ & $\bold{N}$& $Nd_{q-1/2}$ \\
  \hline
  $M_\text{bare}$ & $N(\sum d_j\lambda_j-\widehat{\sum}d_j\hat{\lambda}_j)$ & $0$ & $2qN(\kappa-1/2)$ & $\bold{1}$& 1 \\
\end{tabular}
\caption{Properties of modes and the bare monopole in $\mathcal{N}=1$ SQED$_3$.\label{SQEDN1modes}}
\end{center}
\end{table}

As in the previous cases, the physical states obey Gauss's law and thus have zero total gauge charge.  This can be achieved by acting on the bare monopole with creation operators that carry total gauge charge $-G$.  In order to construct the lowest-energy states, we should act with the lowest available modes.  For $|\kappa|\leq1/2$ we thus dress $|M_\text{bare} \rangle$ with fermionic zero modes, which as described in Section~\ref{QEDmicro} leads to monopoles with degenerate energy \eqref{SQEDN1Final} whose $SU(N)$ irreps are tableaux built from $d_{q-1/2} N \hat{\xi}$ boxes of maximum width $d_{q-1/2}$.  For $|\kappa|>1/2$ we run out of zero modes and are forced to dress with the next lowest mode, which is the $j=q$ bosonic mode. In this case, the calculation of the possible $SU(N)$ irreps follows the analysis presented in Section~\ref{scalarQED3modes} for scalar QED$_3$.  In particular, when $\kappa<-1/2$, we dress with positively charged bosonic modes, and the resulting monopoles transform in $SU(N)$ irreps with tableaux built from $d_q N\xi $ boxes of maximum height $\min(d_q,N)$.  When $\kappa>1/2$, we dress with negatively charged modes, and the monopole operators transform in the conjugates of these representations.\footnote{These representations are consistent with the precise global symmetry of the theory, which is identical to that in QED$_3$, with the only difference that the $\mathcal{N}=1$ theory has a $\Z_2$ R-symmetry that acts the same way as $(-1)^F$.}  This construction precisely matches \eqref{SQEDN1Final}.  Generically, as in the previous two cases, we find reducible flavor symmetry representations at leading order in $N$.

While in Section~\ref{sec:sqed1} we have not completed the free energy computation at subleading order in $1/N$, we expect that, as for QED$_3$ and scalar QED$_3$, the degeneracy between the various irreducible components of the flavor symmetry representation is lifted by the $1/N$ corrections.  In particular, we expect that the energy splitting is quadratic in the spin (and possibly other labels), and that it will be $O(N^0)$  when the spin is $O(\sqrt{N})$.

\subsection{${\cal N} = 2$ SQED$_3$}

We now move on to an analysis of the mode construction in ${\cal N} = 2$ SQED$_3$. Let us first assume that at the saddle point $\sigma>0$, and study the case $\sigma< 0$ later.  If we treat the vector multiplet fields as background fields with a charge $q$ monopole profile for $A$ and constant $\sigma$ and $D$, as is the case for the saddles found in Section~\ref{sec:Neq2}, then we can decompose the matter fields in modes as in  \eqref{psiDecomp} and \eqref{ModeExpansionScal}:
\es{ModeExpansion}{
\phi^{\pm,i}(t,x)&=\sum_{j \geq q} \sum_{m=-j}^j  \frac{1}{\sqrt{2 \lambda^{\pm}_j}} \left[a^{(\pm)i,\dagger}_{j m}\, Y^*_{qj m}(x)e^{i\lambda^\pm_j t}+b^{(\pm)i}_{j m}\,Y_{qj m}(x)e^{-i\lambda^\pm_j t} \right] \,,\\
\psi^{\pm,i}(t,x)&=\widehat{\sum}_{m=1/2-q}^{q-1/2}\hat c^{(\pm)i,\dagger}_{q-1/2,m}\, C_{q,q-1/2,m}(x)e^{i \sigma t}\\
&{}+\widehat{\sum}_{j\geq q+1/2}\sum_{m=-j}^j \le(\hat{a}^{(\pm)i,\dagger}_{j m}\, A^\pm_{qj m}(x)e^{i \hat\lambda_j t}+\hat{b}^{(\pm)i}_{j m}\, B^\pm_{qj m}(x)e^{-i \hat\lambda_j t}\ri)\,,\\
}
where $Y_{qj m}(x)$ are scalar monopole harmonics, and $A^{\pm},\, B^{\pm},\, C$ are spinor monopole harmonics.\footnote{ $A^{\pm},\, B^{\pm}$ are $\sigma$-dependent linear combinations of standard spinor monopole harmonics, and hence they are not the same as the spinor harmonics in  \eqref{psiDecomp}. $C$ is the same as in \eqref{psiDecomp}.  } A similar decomposition holds for the complex conjugate fields.  The properties of the modes and the bare monopole state $\vert M_\text{bare}\rangle$ defined as the state which is annihilated by all the annihilation operators, 
\es{MbareSUSY}{
\{a^{(\pm)}_{j m,i} ,\, b^{(\pm)i}_{j m},\,\hat{a}^{(\pm)}_{j m,i} ,\, \hat{b}^{(\pm)i}_{j m},\,\hat{c}^{(\pm)}_{q-1/2,m,i}\} \vert M_\text{bare}\rangle=0\,,
} 
are summarized in Table~\ref{SQEDN2modes}.  The bare monopole has gauge charge $G$ that can be determined by noticing that, at leading order in $1/N$, there are $2^{q (N^+ + N^-)}$ degenerate vacua with gauge charges ranging between $G - 2q N^-$ and $G + 2q N^+$.  The gauge charges are symmetrically distributed about the average gauge charge $G +q(N^+ - N^-)$, which can be identified with $-\frac{d (NF_q^{(0)})}{ d\alpha} \big|_{\alpha = 0}$, with $F_q^{(0)}$ as in \eqref{SQEDN2F2}.  Solving for $G$, one finds
\es{SQEDN2gauge2}{
G=2q\le(k-{N^+-N^-\ov2}\ri) \,, \qquad \sigma>0\,.
}
(This quantity already appeared in \eqref{pmDef}.) The bare monopole also has $R$ and $A$ charges given in Table~\ref{SQEDN2modes}, which can be found by introducing chemical potentials for the $U(1)_R$ and $U(1)_A$ symmetries, and taking the derivative of the partition function with respect to these chemical potentials. (Or by a careful analysis of normal ordering constants in the oscillator expressions of the corresponding charges.) 
 \begin{table}[!h]
\begin{center}
\scriptsize
\begin{tabular}{c||c|c|c|c|c|c|c}
  & energy & spin & gauge charge & $R$-charge &$SU(N^\pm)$ irrep&  $A$-charge & degeneracy \\
 \hline \hline
  $a^{(\pm)i,\dagger}_{j m}$ & $\lambda^{{\pm}}_{j}$ & $j$ & $\pm1$ & $1/2$ &$\bold{N}^\pm$& $1$ &$N^{\pm}d_j$ \\
  \hline
 $b^{(\pm)\dagger}_{j m,i}$ &  $\lambda^{{\pm}}_{j}$ & $j$ & $\mp1$& $-1/2$ &$\overline{\bold{N}^\pm}$ & $-1$ &$N^{\pm}d_j$ \\
  \hline
    $\hat{a}^{(\pm)i,\dagger}_{j m}$ & $\hat{\lambda}_{j}$ & $j$ & $\pm1$ & $-1/2$ & $\bold{N}^\pm$& $1$ &$N^{\pm}d_j$ \\
  \hline
 $\hat{b}^{(\pm)\dagger}_{j m,i}$ & $\hat{\lambda}_{j}$ & $j$ & $\mp1$&$1/2$ & $\overline{\bold{N}^\pm}$ & $-1$ & $N^{\pm}d_j$ \\
  \hline
 $\hat{c}^{(\pm)i,\dagger}_{q-1/2, m}$ & $\sigma$& $q-1/2$ & $\pm1$ & $-1/2$ & $\bold{N}^\pm$& $1$ &$N^{\pm}d_{q-1/2}$ \\
  \hline
   $M_\text{bare}$ & $N\Delta_\text{bare}$ (see \eqref{DeltaBareDef}) & $0$ &  $2qN(\kappa-\delta n/2)$ &${Nq\ov 2} $& $\bold{1}$&$-{Nq}$& 1 \\
\end{tabular}
\caption{Properties of modes for the $\pm$ charged chiral field and the bare monopole for $\mathcal{N}=2$ SQED$_3$ for $\sigma>0$. For $\sigma<0$ see the discussion below. \label{SQEDN2modes}}
\end{center}
\end{table}

Let us now discuss $\sigma<0$ case. Note that for  $\sigma<0$ the mode expansion \eqref{ModeExpansion} is still valid, but the coefficients of the $C_{q,q-1/2, m}$ harmonics should be interpreted as annihilation operators in order for creation operators to create positive energy states:
\es{ModeExpansion2}{
\psi^{\pm,i}(t,x)&=\widehat{\sum}_{m=1/2-q}^{q-1/2}c^{(\pm)i}_{q-1/2,m}\, C_{q,q-1/2,m}(x)e^{i \sigma t}+\dots\,, \qquad \sigma<0\,.
} 
With this renaming the bare monopole (defined as the Fock vacuum \eqref{MbareSUSY}) remains the lowest energy state of the free chiral multiplets on $S^2\times \R$.\footnote{Had we not done the renaming, we would be referring to an excited state as the bare monopole.}  A similar argument to the one that gave \eqref{SQEDN2gauge2} shows that the gauge charge of the bare monopole is now given by
\es{SQEDN2gauge3}{
G=2q\le(k+{N^+-N^-\ov2}\ri) \,, \qquad \sigma<0\,.
}
Its energy for any value of $\sigma$ is given by $N\Delta_\text{bare}$ defined in \eqref{SQEDN2Final}. In order to be consistent with our assignment of $R$ and $A$ charges, the bare monopole for $\sigma<0$ has to carry charges $R_\text{bare}=-\frac{Nq}{2}$ and $A_\text{bare}=Nq$.

When the vector multiplet is dynamical, Gauss's law requires that gauge invariant states have zero total gauge charge, so all physical states are obtained by acting on $|M_\text{bare} \rangle$ with creation operators carrying total gauge charge $-G$. To minimize the energy, we dress the bare monopole with the lowest available modes. Because the energies of the modes depend on the values of $\sigma,\, D$, we have to refer to the table in Figure~\ref{SQEDN2} for the region of interest to decide what sign $\sigma,\, D$ take. Intuitively, this decides the hierarchy of mode energies.\footnote{The magnitude of $ D$ is never so big on the saddle point to disrupt the hierarchy, so only the signs of $\sigma$ and $D$ are important.} That the quanta interact and set the values of $\sigma,\, D$ is as in the scalar QED$_3$ case.

  For $|\kappa| \leq 1/2$ we can dress with the lowest fermionic modes as described in Section~\ref{QEDmicro}. Because the charge of the lowest fermionic mode depends on the sign of $\sigma$, the notation $\pm_\text{tot}\equiv \sgn(G)\sgn(\sigma)$ introduced in \eqref{SQEDN2alph} is quite natural: we dress with $\abs{G}$ modes of the field $\psi^{\mp_\text{tot},i}$. The resulting monopole states transform in $SU(N^{\mp_\text{tot}})$ irreps whose tableaux are built from $\abs{G}=d_{q-1/2} N^{{\mp_\text{tot}}}\hat{\xi} $ boxes of maximum width $d_{q-1/2}$, and are singlets under $SU(N^{\pm_\text{tot}})$. 
  
For $|\kappa| > 1/2$ we run out of lowest fermionic modes and are forced to dress with the next lowest mode, which is the $j=q$ bosonic mode. In this case, the calculation of the possible $SU(N)$ irreps follows the case presented in Section~\ref{scalarQED3modes}: when $\kappa<-1/2$ and we dress with positively charged bosonic modes, and the resulting monopoles have $SU(N^{\mp_D})$ irreps with tableaux built from $d_{q} N^{\mp_D}\xi $ boxes of maximum height $\min(d_{q},N^{\mp_D})$ and are singlets under  $SU(N^{\pm_D})$;  while when $\kappa>1/2$, we dress with negatively charged bosonic modes and they transform in the conjugates of these representations.\footnote{\label{N=2SQEDFullSymmetry}Similarly to the previous cases the global symmetry of the theory contains discrete factors and quotients
\be 
\label{trueGlobalSymmetry-scalarQED2} 
\le(U(1)_R \times \frac{U(1)_{\text{top}}\times U(1)_A \times SU(N^+)\times SU(N^-)}{\mathbb Z_{N^+} \times \mathbb Z_{N^-}}\ri) \rtimes \mathbb Z_2^C \,,
\ee
where the action of the $\mathbb Z_{N^+}$ and $\mathbb Z_{N^-}$ quotients are generated by $g^+=\left( e^{2\pi i \left(k- N^+\right)/N^+},  e^{-\pi i/N^+}, e^{2\pi i/N^+} \mathbf 1_{N^+}, \mathbf 1_{N^-} \right)$ and $g^-=\left( e^{-2\pi i\left(k+N^-\right)/N^-},  e^{-\pi i/N^-},  \mathbf 1_{N^+}, e^{2\pi i/N^-} \mathbf 1_{N^-} \right)$, and $\Z_2^C$ is the charge conjugation symmetry. For even $N$ a $4\pi$ $R$-symmetry and a $2\pi$ axial rotation are equal to the identity, but for odd $N$ these periodicities increase to $8\pi$ and $4\pi$ respectively due to monopoles having fractional charges, see Table~\ref{SQEDN2modes}. At the same time the action of a $4\pi$ $R$-symmetry, $\pi$ $U(1)_{\text{top}}$, and a $2\pi$ axial rotation is to multiply monopole operators by $(-1)^q$, and hence these group elements have to be identified. Generically, this global symmetry is anomalous, so the representations of monopole operators under $SU(N^\pm)$ should have $N^\pm$-ality $q\left(\mp k+N^\pm\right) + {A\ov 2} \text{ mod } N^\pm$, which they do. Once again the action of $\mathbb Z_{N^\pm}$ on $U(1)_{\text{top}}$ is determined by considering  the properties of the bare monopole under gauge transformations, $U(1)_{\text{top}}$, and $U(1)_A$ transformations. Under the $\Z_{N^+}$ group element $g^+$ an elementary field transforms as $\phi^{\pm}\to \phi^{\pm} e^{\pm {\pi i\ov N^+}}$, while the bare monopole transforms as $M_\text{bare}\to M_\text{bare}\,\exp\le[ {2\pi i\ov N^+}\le( q \left(k- N^+\right)-{A_\text{bare}\ov 2}\ri) \ri]=M_\text{bare}\,\exp\le[ {\pi i\ov N^+}G\ri]$. The transformation of the elementary fields and the bare monopole can be undone by a $U(1)$ gauge transformation of angle $-\pi/N^+$, hence we correctly identified the $\Z_{N^+}$ quotient. The $\Z_{N^-}$ case can be worked out similarly.} For all values of $\kappa$, we again expect an energy splitting similar to that found for QED and scalar QED in Sections~\ref{QEDmicro} and~\ref{scalarQED3modes}. 

A microscopic interpretation similar to the one provided above for the thermal $S^2 \times S^1_\beta$ partition function can also be provided for the supersymmetric $S^2 \times S^1$ partition function that computes the superconformal index.  We explain this construction in Appendix~\ref{SQEDN2Index}.  In short, the superconformal index receives contributions only from states whose energy equals the sum of their R-charge and the eigenvalue of the $j_3$ component of angular momentum.  At large $\beta$, the state that dominates can be constructed by acting on $|M_\text{bare} \rangle$ with $a^{(\mp_\text{tot})i,\dagger}_{qq}$, because this is the lowest-energy creation operator that obeys the condition $\Delta = R + j_3$.  Thus, generically, the BPS state constructed this way will have larger energy  than the energy of the lowest non-BPS state, with the only exception being when $|M_\text{bare} \rangle$ itself is gauge invariant.  For more details, see Appendix~\ref{SQEDN2Index}.

\section{Conformal bootstrap for $k=0$ QED$_3$}
\label{QED_Conformal_bootstrap}

The previous sections demonstrated that low spin monopoles are degenerate to leading order in $1/N$. We now show evidence from the non-perturbative conformal bootstrap that this feature persists even for small $N$ in some cases. 

The numerical bootstrap places rigorous bounds on the scaling dimensions of the lowest lying operators in CFT spectra. Curiously, known theories can appear as kinks on the boundary of the allowed region. For instance, when placing bounds on the spectrum of scalar operators for 3d theories with $\mathbb{Z}_2$ symmetry,  one finds a kink corresponding to the critical exponents of the 3d Ising model \cite{El-Showk:2014dwa, Kos:2014bka, Simmons-Duffin:2015qma}. A similar phenomenon was observed in a previous numerical bootstrap study for theories with $SU(N)\times U(1)_{\text{top}}$ global symmetry \cite{Chester:2016wrc}, although this feature depended on further assumptions about the scaling dimensions of operators in the topologically neutral sector.  Ref.~\cite{Chester:2016wrc} found that for $N=2,\, 4\,$ and $ 6$, the large $N$ prediction for both monopole operator scaling dimensions and some OPE coefficients of conserved currents in QED$_3$ with $k=0$ were close to kinks or almost saturated the numerical bounds found by the bootstrap. This previous study only focused on spin zero monopole operators; in this work we investigate monopoles in the $N=4$ theory\footnote{In \cite{Chester:2016wrc} this value was found to yield the most stable numerical results, because the number of crossing equations grows with $N$.} that have spin and transform in flavor symmetry representations that we expect to have the lowest dimension for their respective spin based on the results of the previous sections. For completeness, we also include representations in our analysis that are not expected to have the lowest dimension for their spin, and indeed we find very lax bounds for them.

To ease notation, we will identify the flavor $SU(4)$ symmetry as $SU(4) \cong SO(6)$ and denote a monopole operator $M_{q, R, \ell}$ by its $U(1)_\text{top}$ charge $q$, spin $\ell$, and $SO(6)$ representation $R$. The monopoles of interest will be in either the vector ($V$), singlet ($S$), symmetric traceless ($T$), or anti-symmetric ($A$) representations of $SO(6)$. 

 For $N=4$, the spins and $SU(4)$ irreps for the lowest dimension monopoles given by the decomposition (\ref{antisum}) are, 
\es{exRep}{
&q=1/2:\qquad \le(\,\,\begin{ytableau}
 {} \\
 {} \\
\end{ytableau}\,,0\ri)\\
&q=1:\hspace{1.3cm} \le(\,\,\begin{ytableau}
 {} &{} \\
{} &{}  \\
\end{ytableau}\,,0\ri)\,,\qquad \le(\,\,\begin{ytableau}
 {} &{} \\
{}  \\
{}\\
\end{ytableau}\,,1\ri)\,,\qquad \le(\,\,\begin{ytableau}
 {} \\
{}   \\
{}   \\
{}\\
\end{ytableau}\,,2\ri)\,.\\
}  
For $\abs{q}=1/2$, there is only one monopole operator, which is a Lorentz scalar in the vector representation of $SO(6)$, $M_{\pm 1/2,V,0}$. For $\abs{q}=1$, there are three monopoles: $M_{\pm 1,S,2}$, $M_{\pm1,A,1}$, and $M_{\pm1,T,0}$. The operators appearing in the OPEs of two $\abs{q_{1,2}}=1/2$ monopole operators have the following properties
 \es{eq:monopole-OPE}{
M_{+1/2,V,0} \times M_{-1/2,V,0} &\sim 
 \text{operators with  $q=0$; $S$, $A$, or $T$ with any spin}\,, \\
M_{+1/2,V,0}\times M_{+1/2,V,0} &\sim \text{operators with $q=1$; $S$ or $T$ with even spin, $A$ with odd spin} \,, \\
M_{-1/2,V,0}\times M_{-1/2,V,0} &\sim \text{operators with  $q=-1$; $S$ or $T$ with even spin, $A$ with odd spin} \,.
}
 The neutral operators ($q = 0$) in the theory are built from gauge-invariant combinations of $\psi_i$ and $A_\mu$:  for example, the lowest dimension scalar operators in the $S$, $A$, and $T$ sectors are ${\cal O}_S = \bar \psi_i \psi^i$,  ${\cal O}_A = \bar \psi_i \psi^j - \frac{1}{4} \delta_i^j \bar \psi_k \psi^k$, and ${\cal O}_T = \bar \psi_{[i_1} \psi^{[j_1} \bar \psi_{i_2]} \psi^{j_2]} - \text{(traces)}$, respectively.  The sectors $q=1$ and $q=-1$ contain monopole operators including $M_{\pm 1,S,2}$, $M_{\pm1,A,1}$, and $M_{\pm1,T,0}$, which are expected to have the lowest dimension for their respective irrep, as well as monopoles in other irreps that are expected to have higher dimensions.

 Following \cite{Chester:2016wrc}, we can bound the scaling dimensions of the internal operators appearing in the OPE \eqref{eq:monopole-OPE} using semi-definite programming.  As mentioned above, if we impose no assumptions about the spectrum besides the existence of the global symmetry $SU(N) \times U(1)_{\text{top}}$, then the bootstrap bounds do not seem to make contact with the scaling dimensions of QED$_3$. However, if we impose a gap on the scaling dimension $\Delta_T$ of $\mathcal O_T$, which is the lowest lying scalar operator in the $T$ representation, then we find kinks on the boundary of the allowed region that are close to the large $N$ values of the scaling dimensions of the two monopole operators. This $\cO_T$ is a four fermion operator with engineering dimension $4$ and negative anomalous dimension in the large $N$ expansion, which motivates the range $\Delta_T\leq4$ that we consider.

\begin{figure}[H]
\begin{center}
   \includegraphics[width=0.8\textwidth]{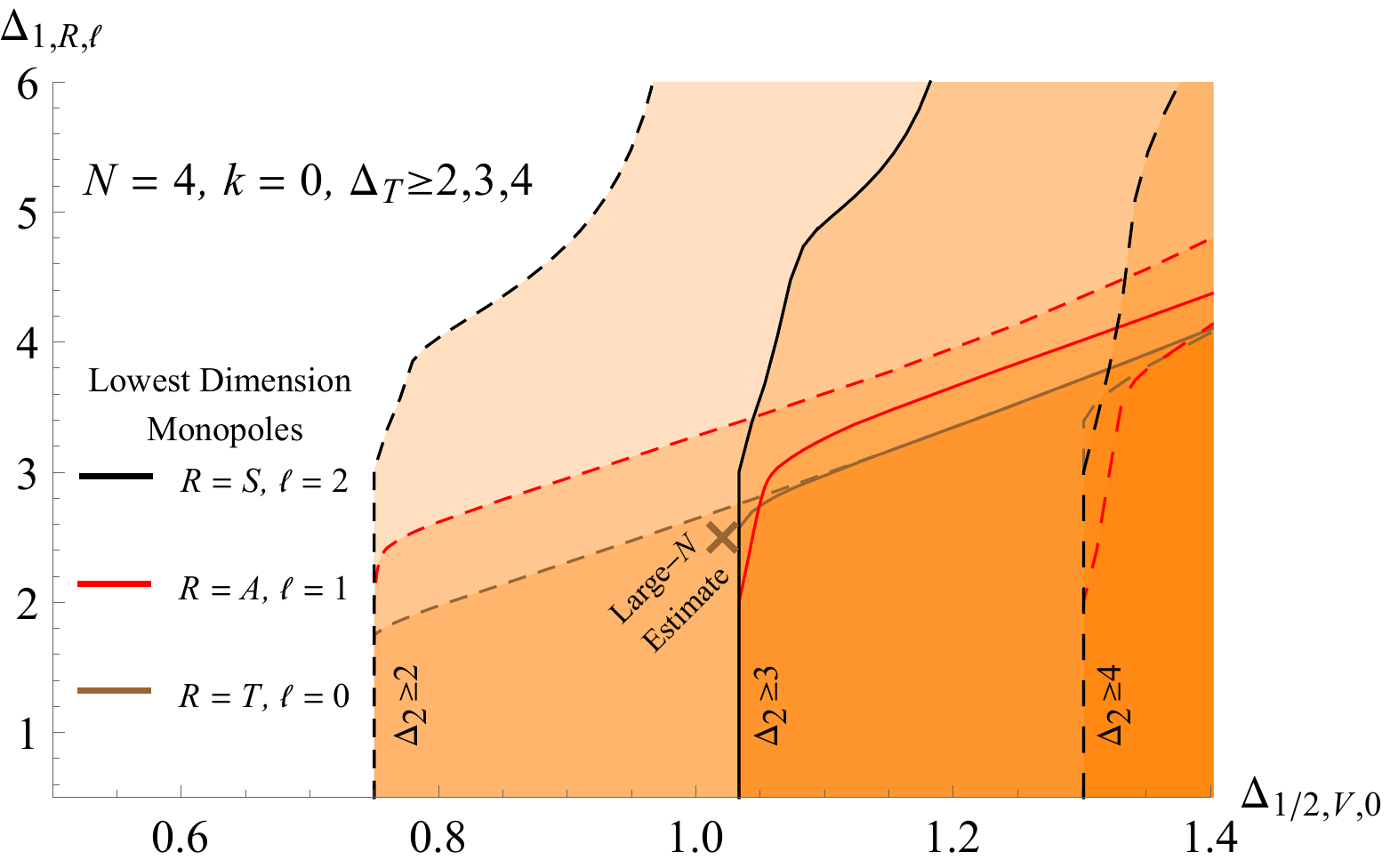}
    \includegraphics[width=0.8\textwidth]{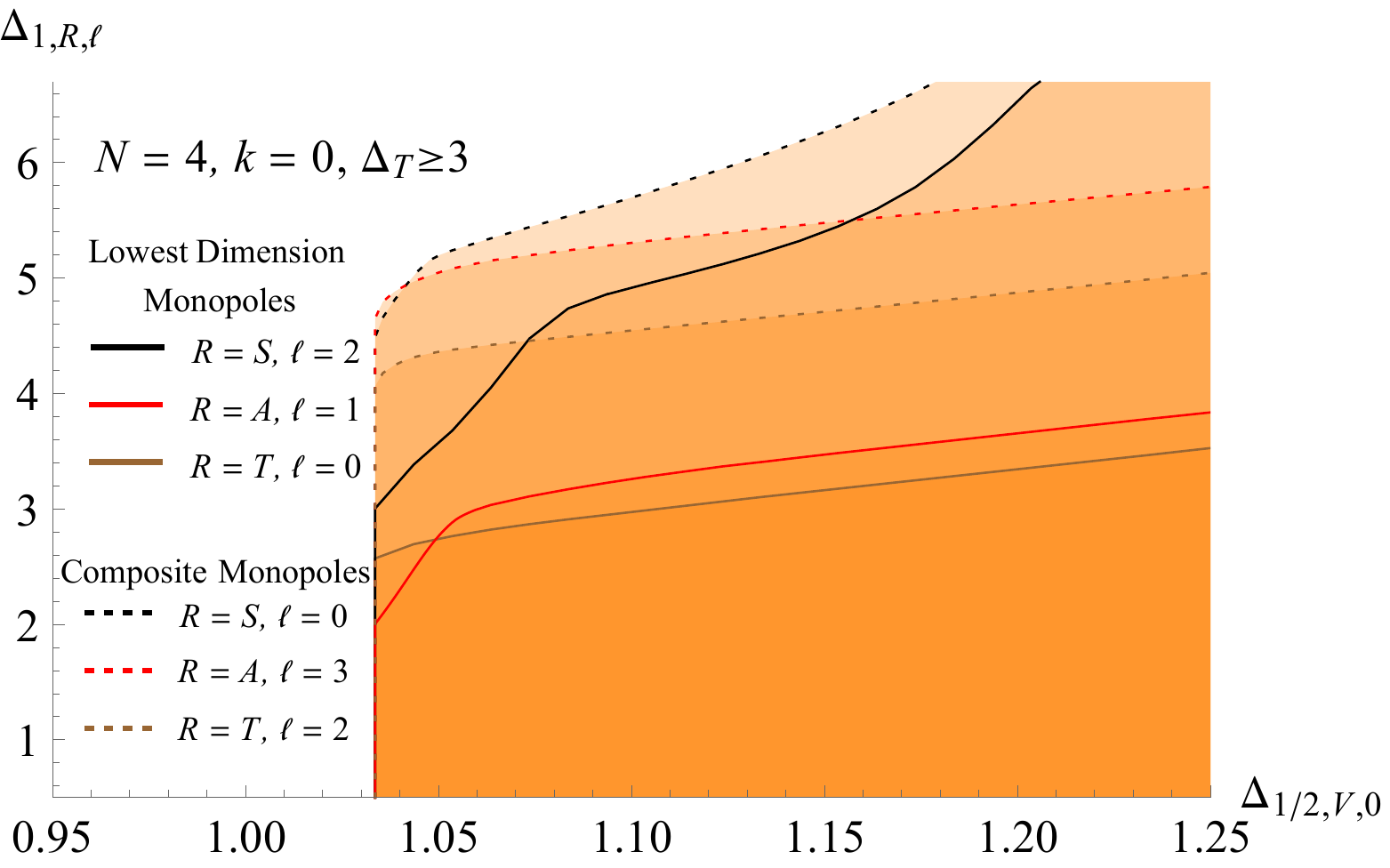}
\caption{
Upper bounds on scaling dimensions $\Delta_{{1,R,\ell}}$ of the lowest dimension $q=1$ monopole operators in the singlet $S$ (brown line), antisymmetric $A$ (red line), and traceless symmetric $T$ (black line) irreps of $SO(6)\cong SU(4)$ in terms of the scaling dimension $\Delta_{{1/2,V,0}}$ of the lowest dimension $q=1/2$ monopole operator in the vector $V$ irrep with a gap imposed on the lowest $q=0$ irrep $T$ scaling dimension $\Delta_T$. {\bf Top:} The lowest dimension monopoles, with spins $0\,,1\,,2$ for $T,\,A,\,S$ respectively, for gaps $\Delta_T\geq2\,,3\,,4$, where the brown cross denotes the large $N$ expansion values of $(N\Delta^{(0)}_{1/2}+\Delta^{(1)}_{1/2},N\Delta^{(0)}_{1}+\Delta^{(1)}_{1})$ determined in \cite{Pufu:2013vpa,Dyer:2013fja}.  {\bf Bottom:} For $\Delta_T\geq3$, the next to lowest dimension monopoles in each sector, with spins $2\,,3\,,0$ for $T,\,A,\,S$ respectively. These bounds were computed with the numerical bootstrap parameters $j_\text{max}=25$ and $\Lambda=19$, for details on their meaning see \cite{Chester:2016wrc}.  }
\label{bootPlots}
\end{center}
\end{figure}

In the top plot of Figure~\ref{bootPlots}, we show upper bounds on the scaling dimensions $\Delta_{{1,S,2}}$, $\Delta_{{1,A,1}}$, and $\Delta_{{1,T,0}}$ as a function of $\Delta_{{1/2,V,0}}$, with gaps $ \Delta_T \geq 2,3,4$ in the $q=0\,,T$ sector. Note that the two lowest spin monopole operator scaling dimensions, $\Delta_{{1,A,1}}$ and $\Delta_{{1,T,0}}$, have bounds that are very close, whereas the largest spin monopole operator scaling dimension $\Delta_{{1,S,2}}$ seems to have a much higher upper bound. This is consistent with our expectation that only the lower spin monopoles have similar scaling dimension. We can even estimate the splitting between these lowest spin monopole using the large $N$ formula \eqref{Ey}, which combined with \eqref{QEDKvals} and the $k=0$ kernel from \cite{Dyer:2013fja} gives the spin-dependent energy splitting
\es{splittingq1}{
\delta E_\ell\sim 0.145\, \ell^2\,.
}
For the case $\ell=1$, this is the same magnitude that we observe in Figure~\ref{bootPlots}.

In the bottom plot, we focus on the gap $\Delta_T\geq3$, which from the top plot we see has a bound on $\Delta_{{1,T,0}}$ that is reasonably close to the large $N$ prediction. In this plot, we also show bounds on the scaling dimensions of $M_{1,S,0}$, $M_{1,A,3}$, and $M_{1,T,2}$, which we think of as composite operators built from the product of the lowest dimension monopoles $M_{1,S,2}$, $M_{1,A,1}$, and $M_{1,T,0}$ and fermion operators (and their derivatives). We expect that these composites will have higher scaling dimensions than the lowest dimension monopole operators. Our numerical bounds match this expectation. 

\section{Conclusion and future directions}
\label{conclusion}

In this paper, we determined the scaling dimensions and degeneracies of the lowest energy monopole operators in QED$_3$, scalar QED$_3$, ${\cal N} =1$ SQED$_3$, and ${\cal N} = 2$ SQED$_3$ with Chern-Simons level $k$, in the regime of large $k$ and $N$ with fixed $\kappa\equiv k/N$.   Generically, at leading order in $1/N$, in each case we found many degenerate monopole operators that transform as a reducible representation under the symmetry group of the theory.  Because this representation is reducible, one expects that the degeneracy between the irreducible components is lifted by the $1/N$ corrections.  For QED$_3$ and scalar QED$_3$, we found evidence that this degeneracy is broken at sub-leading order in $1/N$, and we computed the energy splitting in the simplest case in which there is a degeneracy in the two theories (namely for $q=1$ in QED$_3$ with $-1/2 < \kappa < 1/2$ and $q=1/2$ in scalar QED$_3$ with any $\kappa$).  In the case of QED$_3$ at Chern-Simons level $k=0$, we performed a bootstrap study for $N=4$ and found indications that the large $N$ picture we provided survives down to small values of $N$.  

It is worth noting that in ${\cal N} = 2$ SQED the lowest monopole operators for a given $q$ are generically non-BPS\@.  Each $q$ sector also contains BPS operators of higher dimension that transform in an irreducible representation of the flavor symmetry, and thus for them the energy-splitting picture mentioned above does not apply.  At fixed $q$, it is possible to have the scaling dimension of a BPS operator be larger than that of a non-BPS operator because these operators also have different R-charges, in agreement with the unitarity bound.

Looking ahead, there are several questions we left unanswered and tasks that we left for future work.  In the future, it would be desirable to have a more complete picture of the subleading corrections to the free energy in all the cases we studied.\footnote{For scalar QED$_3$, a sub-leading analysis of the special case $k=N, q=1/2$ will be reported in \cite{shai}.}  While we only explored the energy splitting in detail in the simplest cases, we left a generalization of these results for the future.  In the supersymmetric cases, such a generalization would be much more complicated, because it would require an analysis of the fluctuations of the gaugino and of the other auxiliary fields.

Our results so far mostly come from a path integral approach.  Indeed, we extracted information about the $S^2$ Hilbert space from the thermal partition function on $S^2 \times S^1_\beta$, which we evaluated starting from its path integral representation.  It would be very interesting to perform the same computations starting from a canonical quantization perspective on $S^2 \times \R$.   Such a computation would allow us to compute separately the scaling dimension of each irreducible component of the flavor symmetry representation, which we could not access in our current setup.   A canonical quantization approach would also allow us to potentially compute the energies of the excited states on $S^2$.

Another future direction is a generalization of the conformal bootstrap analysis we performed in Section~\ref{QED_Conformal_bootstrap}. Now that we have determined the reducible representations of the lowest dimension monopoles, this information can be used to bootstrap a wider class of Abelian gauge theories. In particular, it would be nice to generalize our bootstrap results to the case of $k\neq0$, and to gauge theories with scalars.

Our analysis in QED$_3$ and scalar QED$_3$ is also relevant for the recently proposed web of dualities \cite{Son:2015xqa,Aharony:2015mjs,Karch:2016sxi,Murugan:2016zal,Seiberg:2016gmd,Hsin:2016blu,Radicevic:2016wqn,Kachru:2016rui,Kachru:2016aon,Karch:2016aux,Metlitski:2016dht,Aharony:2016jvv,Benini:2017dus,Komargodski:2017keh}.  For instance, when $k \geq N/2$, monopole operators in fermionic QED$_3$ are dual to baryons in a $SU(k+N/2)_{-1}$ gauge theory coupled to $N$ fundamental scalars, the monopole charge being mapped to baryon number \cite{Benini:2017dus}. Thus, the leading order scaling dimension presented in Section~\ref{QED3All} as well as the degeneracy and splitting discussed in Section~\ref{QEDmicro} should be reproduced in the non-Abelian dual.   Similarly, it would be interesting to see if the analytical results obtained in scalar QED$_3$ in \eqref{SpecialCase} can also be interpreted as the dimensions of a baryonic operator in some non-Abelian dual description.

We hope to come back to these issues in the future.

\subsection*{Acknowledgments}

We thank E.~Dyer for collaboration during the initial stages of the project and many helpful discussions. We also thank T.~Dumitrescu, S.~Giombi, P.-S.~Hsin, Z.~Komargodski, B.~Le Floch, and G.~Tarnopolsky for  useful discussions. The research of MM was supported in part by the U.S. Department of Energy under grant No.~DE-SC0016244.  MM's work was performed in part at the Aspen Center for Physics, which is supported by National Science Foundation grant PHY-1607611. SSP was supported in part by the US NSF under Grant No.~PHY-1418069 and by the Simons Foundation Grant No.~488653.

\appendix

\section{Zeta function regularization}
\label{regularization}

\subsection{QED$_3$}
\label{QEDReg}

In the main text we obtained the following leading order energy expression in QED$_3$ in \eqref{QEDFinal}:
\es{QEDFinalAgain}{
\Delta_{q}^{(0)}&=-{\sum}_{j\geq q-1/2} d_j \hat \lambda_j+{\sum}_{q+1/2\leq  j<\tilde j}d_j\hat\lambda_j +\hat\xi_{\tilde j} d_{\tilde j}\hat\lambda_{\tilde j}\,,\\
}
where $\hat\xi_{\tilde j}$ is the fermionic filling faction given in \eqref{QEDalph}. The first term is the divergent Casimir energy, which is the  leading order scaling dimension at $k=0$. Following \cite{Pufu:2013vpa}, we regularize it using zeta function regularization:
\es{QEDCas}{
\Delta_\text{Cas}^{(0)}=-\lim_{s\to0}{\sum}_{j\geq q-1/2} (2j+1) ((j+1/2)^2-q^2)^{1/2-s}\,,
}
where $s$ is a regularization parameter that we let be large enough so that the sum is absolutely convergent. We then add and subtract quantities that are divergent at $s=0$ to get
\es{QEDCas2}{
\Delta_\text{Cas}^{(0)}=&-\lim_{s\to0}{\sum}_{j\geq q-1/2}\left[  (2j+1) ((j+1/2)^2-q^2)^{1/2-s}-2(j+1/2)^{2-2s}+q^2(1-2s)(j+1/2)^{-2s}\right]\\
&+\lim_{s\to0}{\sum}_{j\geq q-1/2}\left[ -2(j+1/2)^{2-2s}+q^2(1-2s)(j+1/2)^{-2s}\right] \,,
}
where the first line is now absolutely convergent, so we can take $s\to0$ and evaluate it numerically. The second line is divergent, but can be regularized using zeta functions. The result is
\es{QEDCas3}{
\Delta_\text{Cas}^{(0)}=&-{\sum}_{j\geq q-1/2}\left[ (2j+1)  \sqrt{(j+1/2)^2-q^2}-2(j+1/2)^2+q^2\right]-\frac{q}{6}(q+2)(2q-1)\,.
}

\subsection{Scalar QED$_3$}
\label{scalQEDReg}

In the main text we obtained the following leading order energy expression in scalar QED$_3$ in \eqref{CPNFinal}:
\es{CPNFinalAgain}{
\Delta_{q}^{(0)}&=\sum_{j\geq q} d_j\lambda_j+\xi d_{q}\lambda_{q} \,,\\
}
where $\xi_{\tilde j}$ is given in \eqref{CPNalph}. The first term is the divergent Casimir energy, which we will regularize using zeta function regularization. Note that unlike QED$_3$, this Casimir energy is not equal to the leading order scaling dimension at $k=0$, because $\mu$ is a function of $k$. The calculation is very similar to the QED$_3$ case described above, and yields 
\es{CPNCas3}{
\Delta_\text{Cas}^{(0)}=&\sum_{j\geq q}\left[ (2j+1) \sqrt{(j+1/2)^2+\mu-q^2}-2(j+1/2)^2+(q^2-\mu)\right]-q\mu+\frac{q(1+2q^2)}{6}\,,
}
where the saddle point value for $\mu$ is found from the zeta function regularized version of its saddle point equation \eqref{muSad}:
\es{muSadReg}{
\sum_{j\geq q}\left(\frac{d_j}{ \lambda_j(\mu)}-1\right)-q+{\xi d_{q}\ov \lambda_q(\mu)}=0\,.
}

\section{Checking the method of steepest-descent for QED$_3$}
\label{steep-desc}

In this appendix we check that for QED$_3$ in the $\kappa =k/N \rightarrow \infty$ limit the real saddle chosen in Section~\ref{QED3All}   indeed dominates the path integral over gauge field configurations. To confirm that a saddle dominates, one needs to prove the possibility of deforming the integration contour through the real saddle point configuration in such a way that the saddle is the global maximum of the exponent, $-S[A]$, over the chosen path. While it is difficult to prove that such a contour deformation is possible in the infinite dimensional space of all gauge field configurations, when $\kappa \rightarrow \infty$ there is no gauge dynamics besides the holonomy \cite{Sundborg:1999ue, Aharony:2003sx, Schnitzer:2004qt, Shenker:2011zf}. Thus, in this limit, the paritition function becomes 
\be
Z_q/Z_0 = {\int_{-i \pi/\beta}^{i \pi /\beta} d \alpha\,  e^{-\beta N F_q^{(0)}(\alpha)}}/{\int_{-i\pi/\beta}^{i \pi/\beta}d\alpha}\,.
\ee 
This allows for an accurate check of the validity of the method of steepest-descent for the real saddle given by  \eqref{QEDalphGen}. As mentioned in Section~\ref{QED3All}, among the saddles listed in equation \eqref{QEDalphGen} only one is real while the others are complex conjugate pairs with large $\beta$ solutions close to $+\lambda_j$ and $-\lambda_j$ up to $O(\beta^{-1})$. For convenience we rewrite the factor in the exponent as
\be \label{eq:periodic}
-\beta N F_{q}^{(0)} = -2 \hat k q \beta  \alpha + N \left[d_{q-1/2} \log(1+e^{\beta \alpha})+ \sum_{j\geq q+1/2} d_j \log(2\cosh(\beta \lambda_j)+ 2\cosh(\beta \alpha))\right]
\ee
where gauge invariance requires that the CS level $\hat k \in \mathbb Z$. 

\begin{figure}[H]
\includegraphics[width=0.55\textwidth]{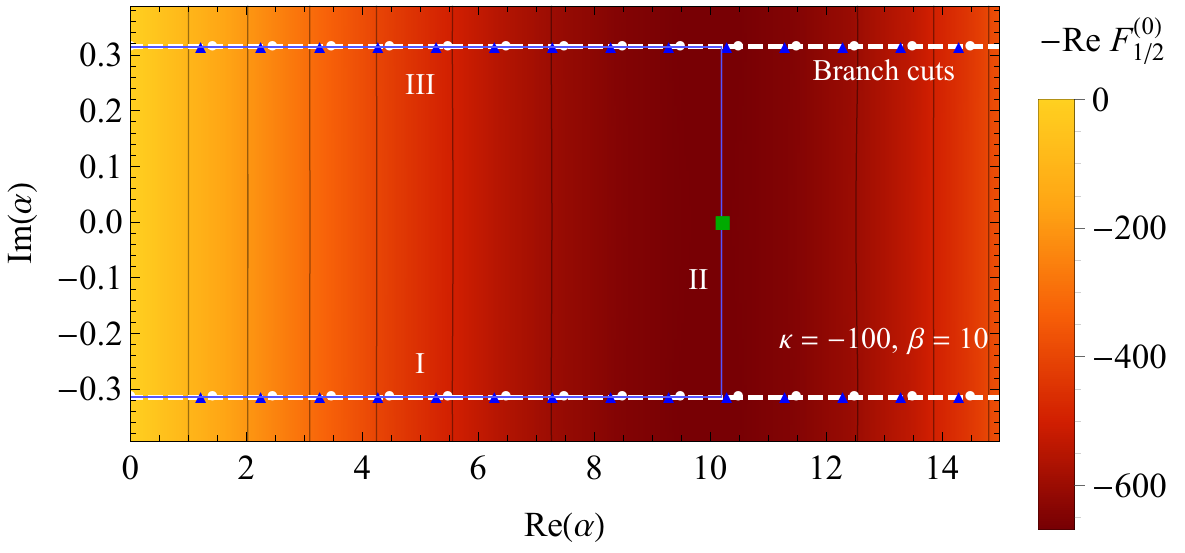} 
\includegraphics[width=0.4\textwidth]{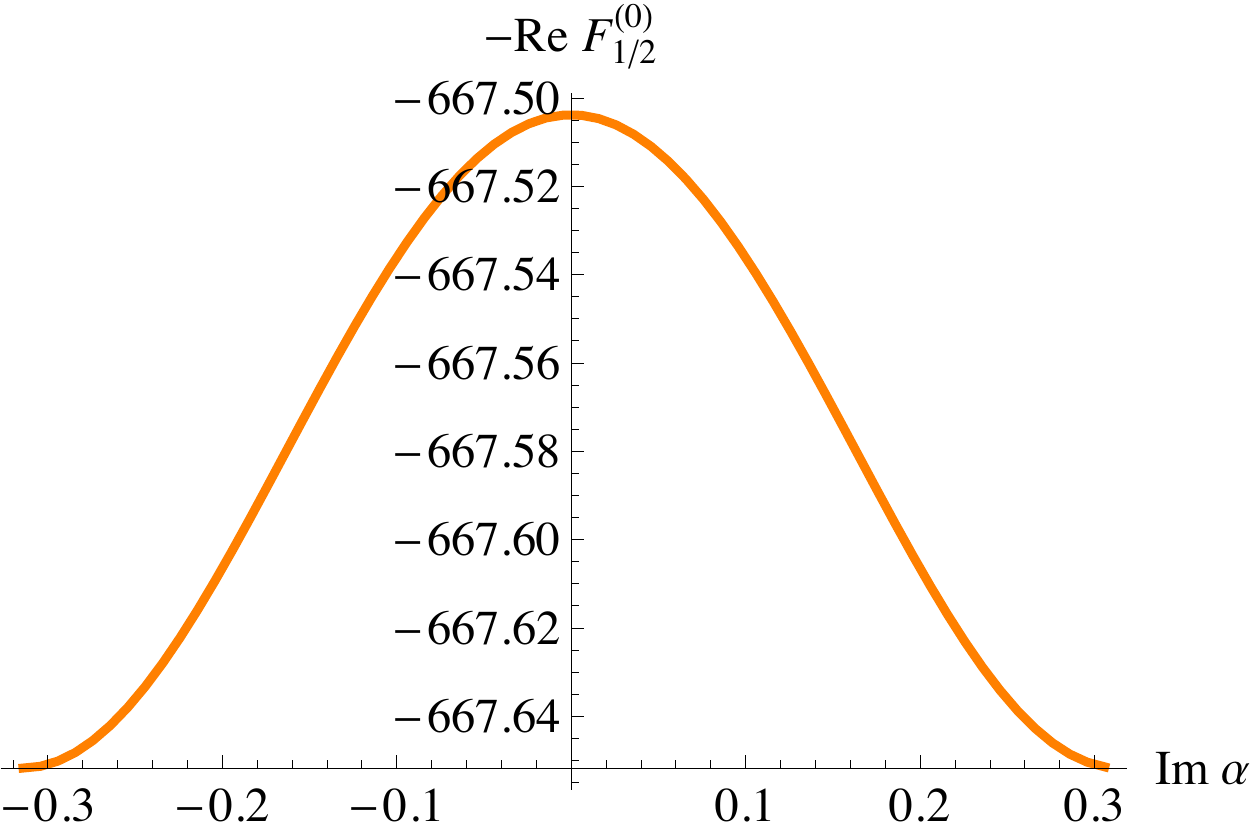} 
\caption{
\label{fig:complex-alpha-plane}{\bf Left:} We show the value of $-\Re(\beta F_q^{(0)})$ for complex values of $\alpha$ for $\beta=10$ and $\kappa=-100$. The branch cuts are indicated by white dashed lines and start at the white dots where there are logarithmic divergences. The blue triangles indicate the saddle-points where $\alpha$ takes a non-zero imaginary value while the green square indicates the real saddle-point. The blue lines show the three-piece contour ($I$, $II$, and $III$) for which the real saddle-point dominates in the path integral. {\bf Right:} We plot the value of $-\Re(\beta F_q^{(0)})$ along contour $II$ which shows that the real-saddle point indeed dominates in the path integral.
}
\end{figure}

  As an example, in Figure~\ref{fig:complex-alpha-plane} we show $-\Re(\beta F_q^{(0)})$, for $\beta=10$ and $\kappa = -100$ (with the real saddle having $\alpha_{\tilde j}>0$). $-\Re(\beta  F_q^{0})$ is periodic along imaginary strips of period $2\pi i/\beta$ and there are branch cuts along $\Im \alpha=  (2n+1)\pi/\beta$ with $n \in \mathbb Z$, starting at the divergence points of the logarithms in the second term of \eqref{eq:periodic}, indicated by the white dots in Figure~\ref{fig:complex-alpha-plane}. While these branch cuts appear in the exponent, the integrand itself is well-behaved, and, thus, the branch cuts do not affect the choice of contour. The saddle-point solutions for $\alpha$ which have a non-zero imaginary component lie along these branch cuts and are shown in Figure~\ref{fig:complex-alpha-plane} with blue triangles. The purely real saddle point is indicated by the green square. Since, as shown in Section~\ref{QED3All}, $- \beta \,\partial^2 F_q^{(0)}/ \partial^2 \alpha|_{\alpha_{\tilde j}} > 0  $, the standard prescription for the method of steepest-descent indicates that the contour should pass through the real saddle point along the imaginary direction.

 In order to close the contour we can pass arbitrarily close to the branch cuts along $\Im \alpha =\pm \pi/\beta$, thus forming the three-piece contour shown in figure~\ref{fig:complex-alpha-plane}, with $Z_q/Z_0 = I + II + III$. However, given this prescription, the real saddle point does not give the global maximum of the exponent along the whole contour. Nevertheless, the real saddle is a maximum along the vertical contour $II$. 
 
  Thus, what remains to be shown is that the holonomy integral over the horizontal pieces of the contour are negligible. Since the sum of the logarithms in \eqref{eq:periodic} is invariant under $\alpha\to\alpha+2\pi i/\beta$, one can express the sum of integrals over the two horizontal contours as 
\es{IplusIII}{
 I + III  \sim&\,  {Z_0}\int_I d\alpha\,  \exp\Big[N \Big(-2\hat \kappa q \beta \Re(\alpha)+ d_{q-1/2}\log(1+e^{\beta \alpha})\\
&\, +\sum_{j>q-1/2} d_j \log\left(2\cosh(\beta \lambda_j) - 2\cosh(\beta \Re \alpha)\right) \Big)\Big] \sin(2 \hat k q \pi )\,.
}
Thus, the quantization of the CS level $\hat k \in \mathbb Z$ implies that the contribution of the horizontal contours fully vanishes, and that the saddle along the vertical piece $II$ dominates the path integral.

\section{Derivation of integral kernels}
\label{Green}
In this appendix we give explicit expressions for the integral kernels that appear in the sub-leading computations, and compute the linear in $\beta$ contributions $\overline{ \bold K}_{q,\ell}$ that give rise to $\frac{\log\beta}{\beta}$ terms in the free energy for QED$_3$ and scalar QED$_3$. To compute these terms we will first compute the thermal Green's functions for scalars and fermions on $S^2\times S^1_\beta$ in the gauge background \eqref{SaddleAnsatz}. 

\subsection{QED$_3$}
\label{GreenF}

Since we are computing a fermion Green's function on a curved manifold, it is necessary to introduce a convention for the frame and gamma matrices. We use the frame obtained from conformally mapping the standard frame from $\R^3$.  In particular, if we define
 \es{xDef}{
   \vec{x}=e^\tau\left(\sin\theta\cos\phi\quad\sin\sin\phi\quad\cos\theta\right) \,,
 }
which is the standard definition of the position vector in $\R^3$ in spherical coordinates, with radial coordinate $e^\tau$, then the line element on $\R^3$ can be written as $d\vec{x}^2$.  The line element on $\R \times S^2$ can be written as $d\vec{x}^2 / e^{2 \tau}$, so it can be described by the frame
\es{frame}{
e^i=e^{-\tau}dx^i\,.
}
We take the gamma matrices $\gamma_i=\sigma_i$, where $\sigma_i$ are the Pauli matrices.

We begin with the matter component of the position space kernel \eqref{subQEDF} corresponding to the first term in \eqref{QED3K}.  In terms of the single fermion thermal Green's function $G_q(x,x')=\langle\psi(x)\psi^\dagger(x')\rangle_q$, it can be written as
\es{GreenKernel}{
K_{q,\text{mat}}^{\mu\nu}(x,x')=-\tr\left(\sigma^\mu G_q(x,x')\sigma^\nu G_q(x',x)\right)\,,
}
where $\sigma^\mu=e^{i\mu} \sigma_i$. The single fermion thermal Green's function satisfies
\es{GreenDefF}{
\left(i\slashed{D}+\slashed{\mathcal{A}}\right)G_q(x,x')=-\delta(x-x')\,,
}
and obeys anti-periodic boundary conditions in imaginary time.
Using the spinor monopole harmonics defined in \cite{Borokhov:2002ib,Pufu:2013vpa} we find that 
\es{NM1}{
\left(i\slashed{\nabla}+\slashed{\mathcal{A}}\right)\begin{pmatrix}
T_{qjm}(\theta,\phi)e^{-i \omega_n\tau}\\
S_{qjm}(\theta,\phi)e^{-i \omega_n\tau}\\
\end{pmatrix}
=&
\bold{N}_{q,j}\left(\omega_n-i\alpha+i\bold{M}_{q,j}\right)
\begin{pmatrix}
T_{qjm}(\theta,\phi)e^{-i \omega_n\tau}\\
S_{qjm}(\theta,\phi)e^{-i \omega_n\tau}\\
\end{pmatrix}\,,\\
\bold{M}_{q,j}\equiv\begin{pmatrix}
\frac{\lambda^2_j}{j+1/2}& -\frac{q\lambda_j}{j+1/2} \\
-\frac{q\lambda_j}{j+1/2} & -\frac{\lambda^2_j}{j+1/2} \\
\end{pmatrix}&\,,\qquad\bold{N}_{q,j}\equiv\begin{pmatrix}
-\frac{q}{j+1/2}&-\frac{\lambda_j}{j+1/2} \\
-\frac{\lambda_j}{j+1/2} &\frac{q}{j+1/2}\\
\end{pmatrix}\,.
}
For $j=q-1/2$ only $S_{qjm}(\theta,\phi)$ exists, so in all subsequent formulas we should consider $\bold{M}_{q,q-1/2},\bold{N}_{q,q-1/2}$ as scalars with values $0,1$, respectively. Using these formulas, we can then write the solution of \eqref{GreenDefF} as
 \es{TempGreen}{
&  G_q(x, x') = -\frac{1}{\beta}\sum_{n,j, m} \begin{pmatrix}T_{qjm}(\theta,\phi)&S_{qj m}(\theta,\phi)\end{pmatrix}\frac{e^{-i \omega_n (\tau - \tau')}}{\bold {N}_{q,j}(\omega_n - i \alpha+i\bold{M}_{q,j})}  \begin{pmatrix}T^\dagger_{qjm}(\theta',\phi')\\S^\dagger_{qj m}(\theta',\phi')\end{pmatrix}\\
 & =-\frac{1}{\beta}\sum_{n,j, m} \begin{pmatrix}T_{qjm}(\theta,\phi)&S_{qj m}(\theta,\phi)\end{pmatrix}\frac{(\omega_n-i\alpha)\bold{N}_{q,j}-\sigma_2\lambda_j}{(\omega_n - i \alpha)^2+\lambda_j^2}e^{-i \omega_n (\tau - \tau')}  \begin{pmatrix}T^\dagger_{qjm}(\theta',\phi')\\S^\dagger_{qj m}(\theta',\phi')\end{pmatrix}\,,
 } 
 where $\sigma_2$ is a Pauli matrix that for $j=q-1/2$ should be considered to be zero, $\lambda_j=\sqrt{(j+1/2)^2-q^2}$ are the usual fermionic eigenvalues, and recall that for fermions $\omega_n=\frac{(2n+1)\pi}{\beta}$. We now calculate
 \es{ToCalcT}{
   g_{q,j}(\tau,\tau') = e^{\alpha(\tau-\tau')}\frac{1}{\beta} \sum_{n\in\mathbb{Z} } \frac{(\omega_n-i\alpha)\bold{N}_{q,j}-\sigma_2\lambda_j}{(\omega_n - i \alpha)^2+\lambda_j^2}e^{-i (\omega_n-i\alpha) (\tau-\tau')}
 }
on the interval $\tau \in [-\frac{\beta}{2}, \frac{\beta}{2})$. We introduce the Poisson summation formula
\es{Poisson}{
\frac{1}{\beta}\sum_{n\in\mathbb{Z}}\hat f\left(\omega_n-i\alpha\right)=\sum_{k\in\mathbb{Z}}e^{k\alpha\beta+i k\pi}f(\beta k)\,,
}
where $\hat f$ is the Fourier transform of $f$. We
apply it to 
\es{PoissonFs}{
\hat f(x)=\frac{x\bold{N}_{q,j}-\sigma_2\lambda_j}{x^2+\lambda_j^2} e^{-ix(\tau-\tau')}\,,
}
which has the inverse Fourier transform
\es{InvFour}{
f(\beta k)&=\int_{-\infty}^\infty\frac{dx}{2\pi}e^{-ix(\beta k+\tau-\tau')}\frac{x\bold{N}_{q,j}-\sigma_2\lambda_j}{x^2+\lambda_j^2}\\
&=-\frac i2e^{-\lambda_{j}|\beta k+\tau-\tau'|}\left(-i\sigma_2+\bold{N}_{q,j}\sgn(\beta k+\tau-\tau')\right)\,.
}
Combining the above we find
\es{Poisson2}{
g_{q,j}(\tau,\tau')&=-\frac{i}{2}e^{\alpha(\tau-\tau')}\left[e^{-\lambda_{j}|\tau-\tau'|}\left(-i\sigma_2+\bold{N}_{q,j}\sgn(\tau-\tau')\right)\right.\\
&\left.-\frac{e^{-\lambda_{j}(\tau-\tau')}}{1+e^{\beta(-\alpha+\lambda_{j})}}\left(-i\sigma_2+\bold{N}_{q,j}\right)-\frac{e^{\lambda_{j}(\tau-\tau')}}{1+e^{\beta(\alpha+\lambda_{j})}}\left(-i\sigma_2-\bold{N}_{q,j}\right)\right]\,.\\
}
We now input the value $\alpha=-\sgn(\kappa-1/2)\left(\lambda_{j}+\beta^{-1}\log\frac{\xi_{\tilde j}}{1-\xi_{\tilde j}}\right)$ given in \eqref{QEDalph} and plug the Green's function defined by \eqref{TempGreen} and \eqref{Poisson2} into \eqref{GreenKernel} to find
\es{decay}{
K^{\mu\nu}_{q}(x,x')=&\sum_{j',j}e^{-(\lambda_j+\lambda_{j'})|\tau-\tau'|}A_{j,j'}^{\mu\nu}(\theta,\phi,\theta',\phi')+\sum_{j'<j}e^{-(\lambda_j-\lambda_{j'})|\tau-\tau'|}B_{j,j'}^{\mu\nu}(\theta,\phi,\theta',\phi')\\
&+\sum_{j'\leq j}e^{(\lambda_j-\lambda_{j'})|\tau-\tau'|}C_{j,j'}^{\mu\nu}(\theta,\phi,\theta',\phi')\\
&-{i\kappa\ov 2\pi}\,\delta(x,x') \,\epsilon^{\mu\nu\rho}\partial'_\rho+O(e^{-\beta\lambda_{q+1/2}})\,,
}
for some $\tau$- and $\beta$-independent functions $A_{j,j'}^{\mu\nu}$, $B_{j,j'}^{\mu\nu}$, and $C_{j,j'}^{\mu\nu}$.  The first term comes from the $e^{-\lambda_{j}|\tau-\tau'|}$ terms in \eqref{Poisson2} for each Green's function in \eqref{GreenKernel}. The second and third terms come from cross terms between $e^{-\lambda_{j}|\tau-\tau'|}$ and $e^{-\lambda_{j}(\tau-\tau')}$ type terms in \eqref{Poisson2}, where the third term includes those that are not exponentially decaying in $\tau$. The final term is due to the Chern-Simons kernel in \eqref{subQEDF}, which is not written in terms of Green's functions. The exponential correction comes from the smallest eigenvalues that appear in the Green's functions.

We can now take the Fourier transform 
\es{fourierKern}{
\bold{K}_{q,\ell}(\omega_n)=&\frac1\beta\int d^3xd^3x'\sqrt{g}\sqrt{g'} e^{i\omega_n(\tau-\tau')}  K^{\mu\nu}_{q}(x,x') \\
&\times\frac{1}{2\ell+1}\sum_{m=-\ell}^{\ell}  \begin{pmatrix}  \mathcal{E}^{\dagger}_{\mu,n\ell m}(x) \mathcal{E}_{\nu,n\ell m}(x')   &   \mathcal{B}^{\dagger}_{\mu,\ell m}(x)   \mathcal{E}_{\nu,n\ell m}(x') \\
\mathcal{E}^{\dagger}_{\mu,\ell m}(x) \mathcal{B}_{\nu,n\ell m}(x')   &   \mathcal{B}^{\dagger}_{\mu,n\ell m}(x)   \mathcal{B}_{\nu,n\ell m}(x') \end{pmatrix}\,,
}
where the vector spherical harmonics $ \mathcal{E}^\mu_{n\ell m}(x) $ and $ \mathcal{B}^\mu_{n\ell m}(x) $ are defined in \eqref{EBDef}. Using the symmetry of $S^2\times S^1_\beta$, it is clear that the integrand in \eqref{fourierKern} only depends on $x - x'$, so we can set $x' = 0$ and replace the integral $\int d^3x'$ by $4 \pi \beta$.   The Fourier transform of the  top line (that  decays exponentially in time) and the local Chern-Simons kernel in \eqref{decay} define $\widetilde{{\bf K}}_{q, \ell}(\omega_n)$ in \eqref{SmoothApprox}, which depends on temperature purely through $\omega_n$, and so can be analytically continued as $\omega_n\to\omega$. As explained in \eqref{Delta1Eq}, these terms contribute to the sub-leading energy, where the $O(e^{-\lambda_{q+1/2} \beta})$ correction comes from the smallest $\omega$ pole in $\widetilde{{\bf K}}_{q, \ell}(\omega)$.

The middle line of \eqref{decay} does not decay at infinity and turns out to be only nonzero for $j=j'=\tilde j$, in which case it is independent of $\tau$. The explicit form of $C_{\tilde j,\tilde j}^{\mu\nu}(x,x')$ is
\es{Cexplicit}{
C_{\tilde j,\tilde j}^{\mu\nu}(x,x')= \sum_{\ell=0}^{2\tilde j}\sum_{m=-\ell}^\ell \left[v^E_{\tilde j,q,\ell}\mathcal{E}^{\mu}_{0\ell m}(x)+v^B_{\tilde j,q,\ell}\mathcal{B}^{\mu}_{0\ell m}(x)\right]\left[v^E_{\tilde j,q,\ell}\mathcal{E}^{\nu}_{0\ell m}(x')+v^B_{\tilde j,q,\ell}\mathcal{B}^{\nu}_{0\ell m}(x')\right]  \,.
}
We can now take Fourier transform of this term to find the linear in $\beta$ term \eqref{DivInK}, where the lowest couple of values of $v^E_{\tilde j,q,\ell}$ and $v^B_{\tilde j,q,\ell}$ are
\es{QEDKvals}{
\tilde j=q-1/2:\qquad &v_{q-1/2,q,\ell}^E={\frac{\sgn(\kappa-1/2)(2q)!}{\sqrt{4\pi(2q-\ell-1)!(2q+\ell)!}}}\,,\qquad v_{q-1/2,q,\ell}^B=0\,,\\
\tilde j=q+1/2:\qquad &v_{q+1/2,q,\ell}^E=\frac{\sgn(\kappa-1/2)(2q+2)!(\ell^2+\ell-2-4q)}{\sqrt{16\pi(2q+1)^2(2q-\ell+1)!(2q+\ell+2)!}}\,,\\
&v_{q+1/2,q,\ell}^B=\frac{-i(2q+2)!\sqrt{\ell(\ell+1)}}{\sqrt{4\pi(2q+1)(2q-\ell+1)!(2q+\ell+2)!}}\,.
}

\subsection{Scalar QED$_3$}
\label{GreenB}

We begin by writing the matter component of the position space kernels \eqref{subQEDscal} in terms of the single scalar thermal Green's function $G_q(x,x')=\langle\phi(x)\phi^*(x')\rangle_q$ as
\es{wickCPN}{
{ K}_{q,\text{mat}}^{\mu\nu}(x,x')=&D^\mu G_q(x,x')D^{\nu} G_q(x',x)-G_q(x',x)D^{\mu}D^{\nu} G_q(x,x')\\
&+D^\mu G_q(x',x)D^{\nu} G_q(x,x')-G_q(x,x')D^{\mu}D^{\nu} G_q(x',x) \\
&+2g^{\mu\nu}\delta(x-x')G_q(x,x)\,,\\
{ K}_q^{\sigma\nu}(x,x')=&G_q(x,x')D^{\nu}G_q(x',x)- G_q(x',x)D^{\nu}G_q(x,x') \,,\\
{ K}_q^{\sigma\sigma}(x,x')=&G_q(x,x')G_q(x',x) \,,
}
where $D^\mu=\partial^\mu-i\mathcal{A}_q^\mu(x)$ and $D^{\nu}=\partial'^{\nu}+i\mathcal{A}_q^{\nu}(x')$ denote the gauge-covariant derivatives in the presence of the background gauge fields. The single scalar thermal Green's function satisfies
\es{GreenDefB}{
\left[-(\nabla_\mu-i\mathcal{A}_\mu)^2+\frac14+\mu\right]G_q(x,x')=\delta(x-x')\,,
}
and is periodic in imaginary time.
Using the scalar monopole harmonics $Y_{qjm}(\theta,\phi)$ introduced in \cite{Wu:1976ge,Wu:1977qk} we find that 
\es{NM1B}{
\left[-(\nabla_\mu-i\mathcal{A}_\mu)^2+\frac14+\mu\right]Y_{qjm}(\theta,\phi)
=
\left[(\omega_n-i\alpha)^2+\lambda_j^2\right]Y_{qjm}(\theta,\phi)\,.
}
We can then write the solution to \eqref{GreenDefB} as
 \es{TempGreenB}{
  G_q(x, x') &= \frac{1}{\beta} \sum_{n \in \Z} \sum_{j, m}\frac{e^{-i \omega_n (\tau - \tau')}}{(\omega_n-i\alpha)^2 +\lambda^2_{j}}\, Y_{qj m}(\theta,\phi)Y^*_{qj m}(\theta',\phi')\,,
 } 
 where $\lambda_j=\sqrt{(j+1/2)^2-q^2+\mu}$ are the usual scalar eigenvalues and recall that for scalars $\omega_n=\frac{2\pi n}{\beta}$. We now calculate 
 \es{ToCalcT2}{
   g_{q,j}(\tau,\tau') = e^{\alpha(\tau-\tau')}\frac{1}{\beta} \sum_{n \in \Z}  \frac{1}{(\omega_n - i \alpha)^2+\lambda_{j}^2}e^{-i (\omega_n-i\alpha) (\tau-\tau')}
 }
on the interval $\tau \in [-\frac{\beta}{2}, \frac{\beta}{2})$. We make use of the Poisson summation formula
\es{PoissonB}{
\frac{1}{\beta}\sum_{n\in\mathbb{Z}}\hat f\left(\omega_n-i\alpha\right)=\sum_{k\in\mathbb{Z}}e^{ k\alpha\beta}f(\beta k)
}
applied to 
\es{PoissonFsB}{
\hat f(x)=\frac{1}{x^2+\lambda^2_{j}}e^{-ix(\tau-\tau')}\,.
}
The inverse Fourier transform of $\hat f$ is
\es{InvFourB}{
f(\beta k)&=\int_{-\infty}^\infty\frac{dx}{2\pi}e^{-ix(\beta k+\tau-\tau')}\frac{1}{x^2+\lambda^2_{j}}=\frac{1}{2\lambda_{j}}e^{-\lambda_{j}|\beta k+\tau-\tau'|}\,.
}
Combining the above we find
 \es{ToCalcT3}{
   g_{q,j}(\tau,\tau') &= \frac{e^{\alpha(\tau-\tau')}}{2\lambda_{j}}\left[e^{-\lambda_{j}|\tau-\tau'|}+e^{-\lambda_{j}(\tau-\tau')}\sum_{k\geq1}e^{-k\beta(\lambda_{j}-\alpha)}+e^{\lambda_{j}(\tau-\tau')}\sum_{k\geq1}e^{-k\beta(\lambda_{j}+\alpha)}\right]\\
&= \frac{e^{\alpha(\tau-\tau')}}{2\lambda_{j}}\left[e^{-\lambda_{j}|\tau-\tau'|}+\frac{e^{-\lambda_{j}(\tau-\tau')}}{e^{\beta(\lambda_{j}-\alpha)}-1}+\frac{e^{\lambda_{j}(\tau-\tau')}}{e^{\beta(\lambda_{j}+\alpha)}-1}\right]\,.
 }
 We now input the value $\alpha=-\sgn(\kappa)\left(\lambda_{q}+\beta^{-1}\log\frac{\xi}{1+\xi}\right)$ given in \eqref{CPNalph}  and plug the Green's function defined by \eqref{TempGreenB} and \eqref{ToCalcT3} into \eqref{KernelsScalar} to find the position space kernels $K_q^{\mu\nu}(x,x')$, $K_q^{\sigma\nu}(x,x')$, and $K_q^{\sigma\sigma}(x,x')$. As in QED$_3$, these expressions contain terms that are exponentially decaying in $\tau$, as well as $\tau$-independent that occurs only for $j=\tilde j=q$ in \eqref{ToCalcT3} and take the form
\es{GreenKernel2scal}{
K_{q}^{\mu\nu}(x,x')&:\qquad \xi(1+\xi) \sum_{\ell=0}^{2\tilde j}\sum_{m=-\ell}^\ell \left[v^E_{q,\ell}\mathcal{E}^{\mu}_{0\ell m}(x)+v^B_{q,\ell}\mathcal{B}^{\mu}_{0\ell m}(x)\right]\left[v^E_{q,\ell}\mathcal{E}^{\nu}_{0\ell m}(x')+v^B_{q,\ell}\mathcal{B}^{\nu}_{0\ell m}(x')\right]  \,,\\
K_{q}^{\sigma\nu}(x,x')&:\qquad \xi(1+\xi) \sum_{\ell=0}^{2\tilde j}\sum_{m=-\ell}^\ell  v^\sigma_{q,\ell}Y_{\ell m}(x) \left[v^E_{q,\ell}\mathcal{E}^{\nu}_{0\ell m}(x')+v^B_{q,\ell}\mathcal{B}^{\nu}_{0\ell m}(x')\right]  \,,\\
K_{q}^{\sigma\sigma}(x,x')&:\qquad \xi(1+\xi) \sum_{\ell=0}^{2\tilde j}\sum_{m=-\ell}^\ell  \left( v^\sigma_{q,\ell} \right)^2 Y_{\ell m}(x) Y_{\ell m}(x')  \,.\\
}
 When we take the Fourier transform
\es{fourierKernscal}{
&\bold{K}_{q,\ell}(\omega_n)=\frac1\beta\int d^3xd^3x'\sqrt{g}\sqrt{g'} e^{i\omega_n(\tau-\tau')}\frac{1}{2\ell+1} \sum_{m=-\ell}^{\ell}  \\
& \begin{pmatrix} Y^{\dagger}_{\ell m}(x)K^{\sigma\sigma}_{q}(x,x')  {Y}_{\ell m}(x') & \mathcal{E}^{\dagger}_{\mu,n\ell m}(x)K^{\mu\sigma}_{q}(x,x')  Y_{\ell m}(x')   &   \mathcal{B}^{\dagger}_{\mu,n\ell m}(x)  K^{\mu\sigma}_{q}(x,x')  Y_{\ell m}(x') \\
Y^{\dagger}_{\ell m}(x)K^{\sigma\nu}_{q}(x,x')  \mathcal{E}_{\nu,n\ell m}(x') & \mathcal{E}^{\dagger}_{\mu,n\ell m}(x)K^{\mu\nu}_{q}(x,x')  \mathcal{E}_{\nu,n\ell m}(x')   &   \mathcal{B}^{\dagger}_{\mu,n\ell m}(x)  K^{\mu\nu}_{q}(x,x')  \mathcal{E}_{\nu,n\ell m}(x') \\
Y^{\dagger}_{\ell m}(x)K^{\sigma\nu}_{q}(x,x')  \mathcal{B}_{\nu,n\ell m}(x') &
\mathcal{E}^{\dagger}_{\mu,n\ell m}(x) K^{\mu\nu}_{q}(x,x')  \mathcal{B}_{\nu,n\ell m}(x')   &   \mathcal{B}^{\dagger}_{\mu,n\ell m}(x) K^{\mu\nu}_{q}(x,x')   \mathcal{B}_{\nu,n\ell m}(x') \end{pmatrix}\,,
}
the exponentially decaying terms contribute to the free energy at $\beta\to\infty$ as
\es{scalDecay}{
\Delta_q^{(1)}+O\le(e^{-\beta\left(\lambda_{q+1}-\lambda_q\right)}\ri)\,,
}
where $(\lambda_{q+1}-\lambda_q)$ is the lowest $\omega$ pole in the Fourier transform of these terms. The $\tau$-independent term yields the linear in $\beta$ terms in the kernel
\es{GreenKernel2scal2}{
\overline{\bold K}_{q,\ell}=\xi(1+\xi)  \begin{pmatrix}
   v^\sigma_{q, \ell} \\   v^E_{q, \ell} \\ v^B_{q, \ell} 
   \end{pmatrix} 
   \begin{pmatrix}
   v^\sigma_{q, \ell} & v^E_{q, \ell} & v^B_{q, \ell} 
   \end{pmatrix}\quad \text{for}\quad0\leq\ell\leq2\tilde j \,,
}
which determine $C_{q,\ell}$ in \eqref{F1smoothCPN} to be
 \es{DivInKscal}{
  C_{q,\ell}= \begin{pmatrix}
   v^\sigma_{q, \ell} &   v^E_{q, \ell} & v^B_{q, \ell} 
   \end{pmatrix} \widetilde{\bold K}^{-1}_{q,\ell}(0)
   \begin{pmatrix}
   v^\sigma_{q, \ell} \\  v^E_{q, \ell} \\ v^B_{q, \ell} 
   \end{pmatrix} \,,
 } 
where $\widetilde{\bold K}_{q,\ell}(\omega)$ is the $\beta$-independent part of the kernel and 
\es{scalQEDKvals}{
v_{q,\ell}^\sigma&=\frac{(2q)!}{\sqrt{4\pi(2q-\ell)!(2q+\ell+1)!}}\,,\\
v_{q,\ell}^E&=\frac{\sgn(\kappa)(2q+1)!}{\sqrt{4\pi(2q-\ell)!(2q+\ell+1)!}}\,,\\
v_{q,\ell}^B&=\frac{-i(2q)!\sqrt{\ell(\ell+1)}}{\sqrt{4\pi(2q-\ell)!(2q+\ell+1)!}}\,.
}

 \section{Further details in $\mathcal N=2$ SQED}
 \label{SQEDAPPENDIX}

 \subsection{Saddle-point values for $\sigma$ and $D$}
 \label{SQEDN2SigmaAndD}
   \begin{figure}[H]
    \centering
   \includegraphics[width=0.47\textwidth]{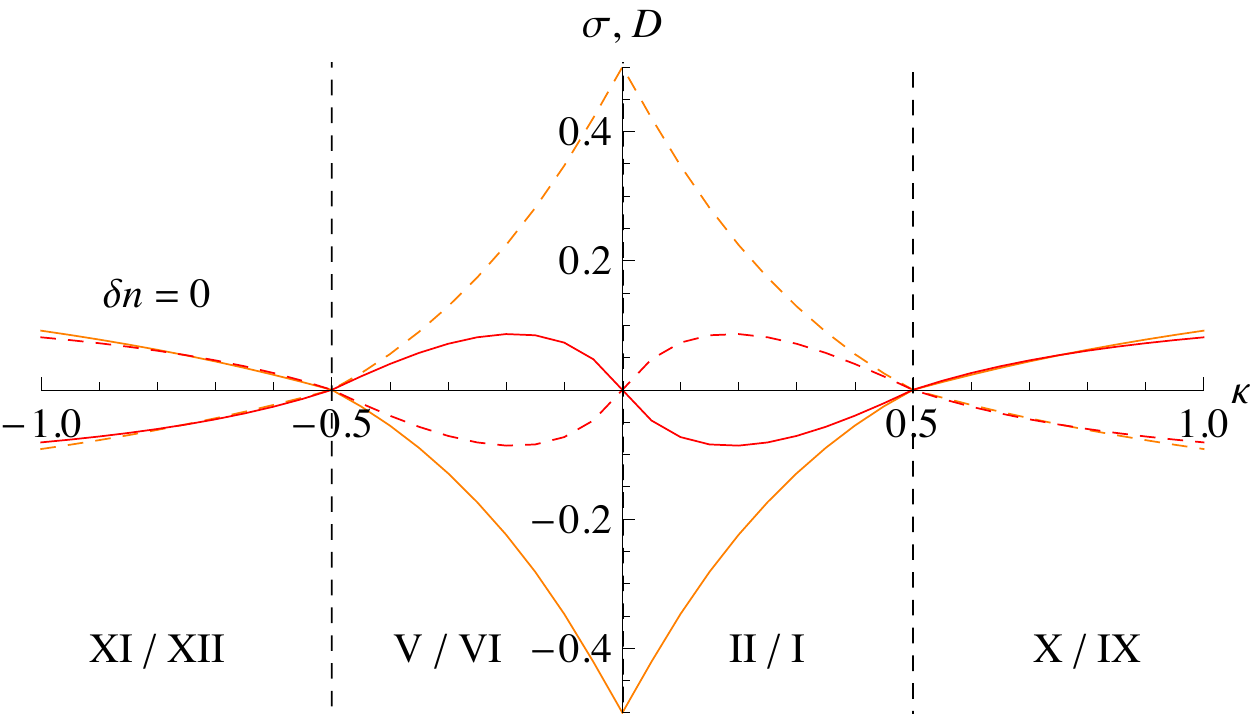}
   \includegraphics[width=0.47\textwidth]{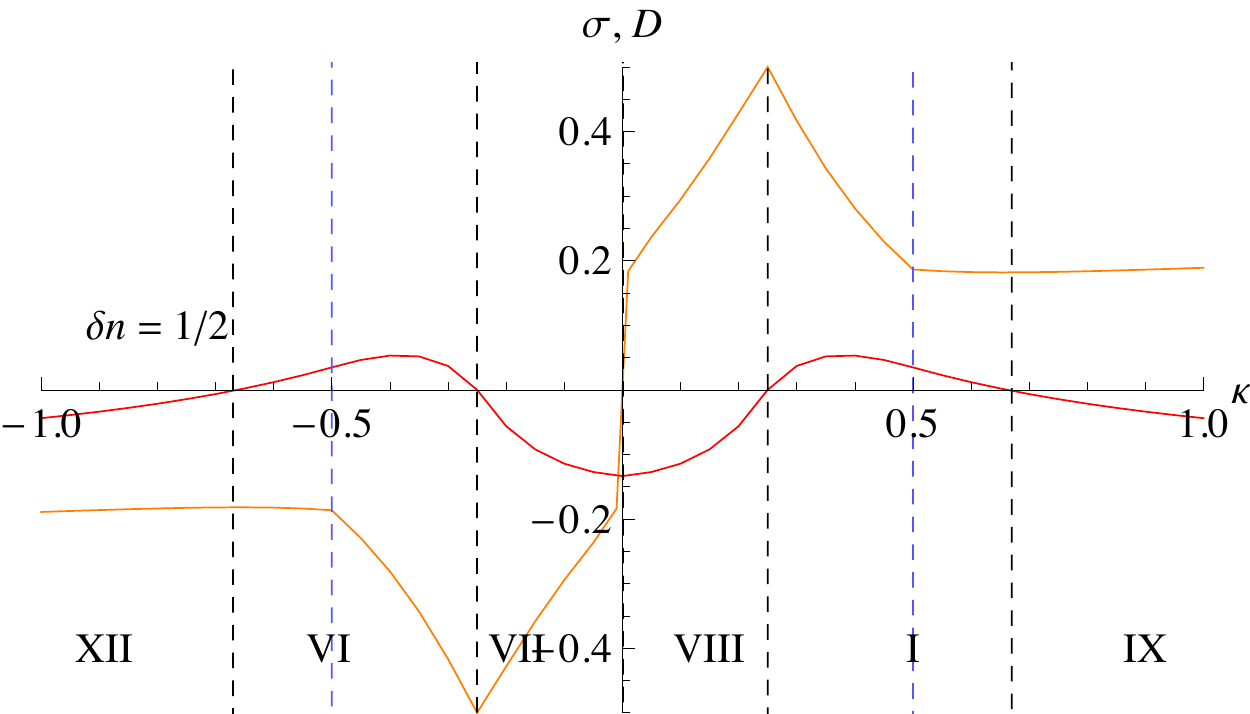}\\ \includegraphics[width=0.47\textwidth]{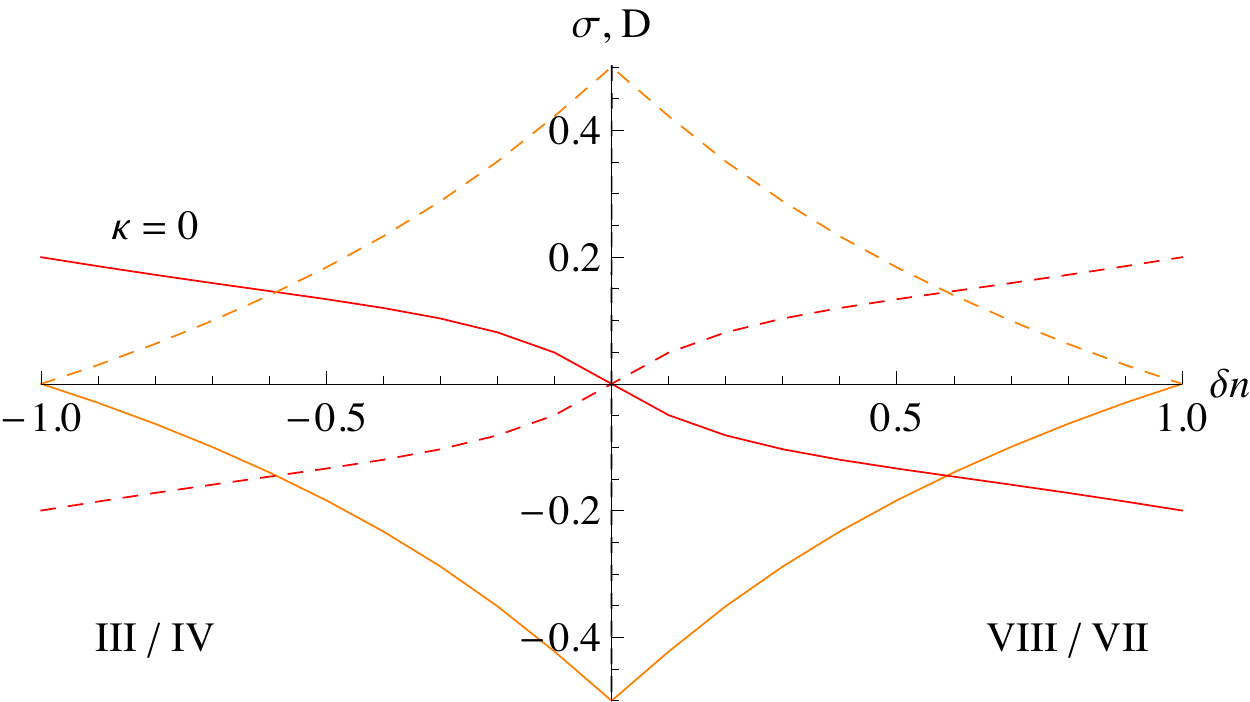}\includegraphics[width=0.47\textwidth]{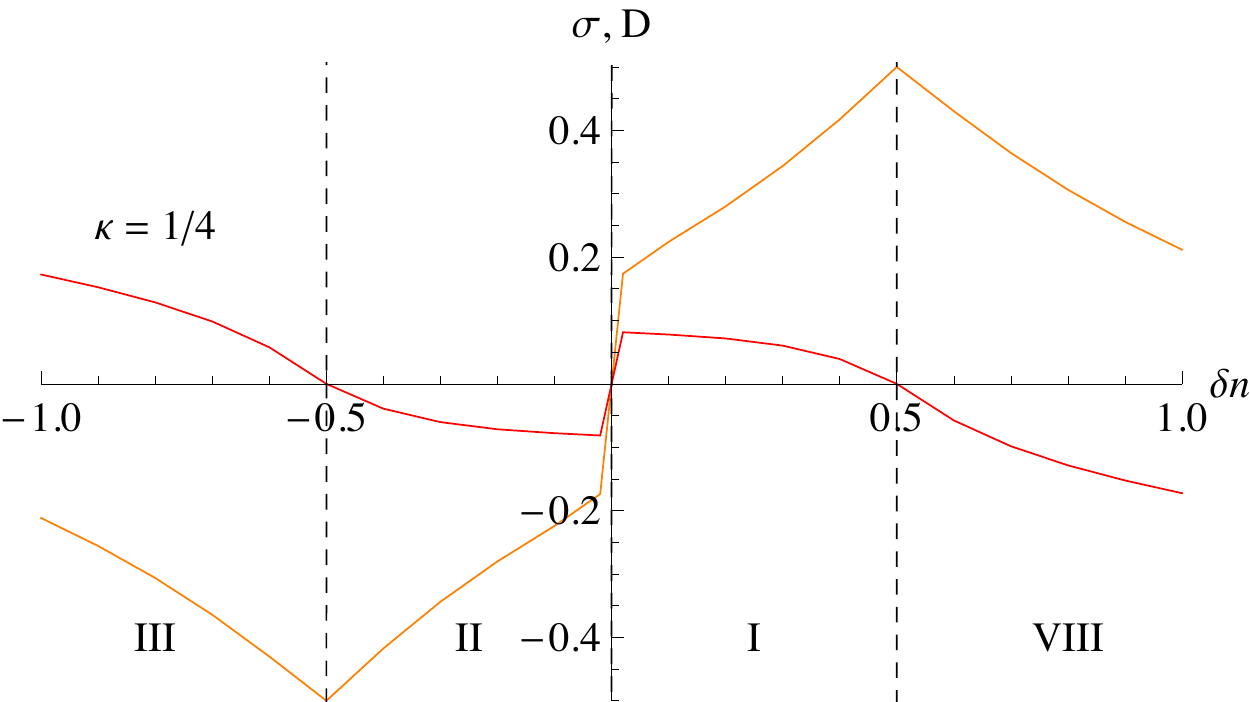}
    \caption{\label{fig:univ-bounds-sigmaAndD}  A horizontal (top) and vertical (bottom) bisection of the density plot in Figure~\ref{SQEDN2}. We show the values of $\sigma$ (orange) and $D$ (red) in the top for $\delta n = 0,\,1/2$, and in the bottom for  $\kappa = 0,1/4$.  For $\delta n = 0$ and $\kappa =0$ there are two saddle point with equal free energy; we plot one of the solutions for  $\sigma$ and $D$ using dashed curves. Note the several transition points which reproduce the signs of $\sigma$ and $D$ shown in the table in Figure~\ref{SQEDN2}.
    }\label{SQEDN2-otherdirection}
  \end{figure} 
 
 \subsection{Superconformal index of ${\cal N} = 2$ SQED}
\label{SQEDN2Index}

\subsubsection{Computation of the partition function}

A variant of the thermal partition function computation in Section~\ref{sec:Neq2} can be repeated for the superconformal index.  Since the superconformal index can be computed exactly using supersymmetric localization, it is possible to compare our large $N$ results with the supersymmetric localization ones.

In an ${\cal N} = 2$ SCFT one can define a superconformal index with respect to any of the four Poincar\'e supercharges. For concreteness, let us define the index with respect to the supercharge $Q$ with $U(1)$ R-charge $\tilde R = +1$ and eigenvalue $-1/2$ under the $j_3$ component of angular momentum.  The index is
\es{index}{
I=\Tr\left[(-1)^F e^{-\beta_1(\Delta-\tilde R-j_3)}e^{-\beta_2(\Delta+j_3)}\prod_n t_n^{f_n}\right]\,,
}
where the trace is taken over the radial quantization states, $\tilde R$ is the charge under the $U(1)_{\tilde R}$ symmetry that is part of the superconformal algebra (not to be confused with the $U(1)_R$ symmetry of the previous section), $j_3$ the charge under the Cartan of $SU(2)_\text{rot}$, $F = {2 j_3}$ is the fermion number, and $t_n$ are the fugacities of the various conserved charges  $f_n$ that correspond to global symmetries.  The superconformal algebra implies that $\Delta-\tilde R-j_3 = \{Q, Q^\dagger\}$, where $Q^\dagger$ is the conjugate of $Q$ in radial quantization.  Consequently,   
$I$ is independent of $\beta_1$.  Note that the computation of the index requires knowledge of the $U(1)_{\tilde R}$ charges of the various operators.  These $U(1)_{\tilde R}$ charges must be linear combinations of the other $U(1)$ charges:  $U(1)_\text{top}$, $U(1)_R$, and $U(1)_A$ introduced in Section~\ref{sec:Neq2}.  The precise linear combination can be found using $F$-maximization \cite{Jafferis:2011zi,Jafferis:2010un,Closset:2012vg,Pufu:2016zxm}.  At large $N$, however, $F$-maximization gives $\tilde R = R + O(1/N)$.

In the path integral formalism, the superconformal index \eqref{index} can be computed as the $S^2 \times S^1_\beta$ partition function of a theory similar to \eqref{SQEDN2Action}. To simplify the calculation, let us choose $\beta_1=\beta_2=\beta/2$, focus on the contribution to $I$ from operators with fixed monopole charge $q$, and set the rest of the global symmetry fugacities to $1$, so we are calculating
\es{index2}{
I_{q}=\Tr_q\left[(-1)^Fe^{-\beta(\Delta-R/2)}\right]\,.
} 
As in the case of the thermal free energy, let us calculate \eqref{index2} in the limit $\beta \to \infty$.  
After these simplifications, the $S^2 \times S^1_\beta$ partition function that calculates \eqref{index2} differs from \eqref{SQEDN2Action} in only two ways:  1) the fermions are periodic on $S^1_\beta$ in order to preserve supersymmetry (half the number of supercharges of the superconformal theory);  2) the action \eqref{SQEDN2Action} receives $1/r$ corrections that depend on $\tilde R$, where $r$ is the radius of $S^2$.  At large $N$, however, because $\tilde R = R + O(1/N)$ these $1/r$ corrections are further suppressed in $1/N$.  We can thus use the action \eqref{SQEDN2Action} to calculate the superconformal index at leading order in $1/N$.

Let us sketch the computation of the supersymmetric $S^2 \times S^1_\beta$ partition function corresponding to \eqref{index2} using our large $N$ method, and then compare it to the exact supersymmetric localization computation.  

As in previous sections, we integrate out the matter fields \eqref{SQEDN2Action} and expand around the large $N^\pm,\,k$ saddle point in the background \eqref{SaddleAnsatz} with extra auxiliary fields $\sigma,D$ to find $I_{q}=NI_{q}^{(0)}+O(N^{0})$ with 
\es{SQEDN2F}{
 I_{q}^{(0)}(\alpha, \sigma, D)&= \beta^{-1}\sum_\pm n^\pm \bigg[\sum_{j\geq q} d_j \log(2 \cosh(\beta \lambda_j^{\pm}) - 2 \cosh(\beta(\alpha\pm 1/4)) )  \\
 &- d_{q-1/2} \log\left |2\sinh\left(\beta(\alpha\mp(1/4+\sigma))/2\right)\right | \\
 &- \widehat{\sum}_{j> q-1/2} d_j \log(2 \cosh(\beta \hat{\lambda}_j) -2 \cosh(\beta (\alpha \mp 1/4))) \bigg] -2\kappa( q\alpha-D\sigma) \,,
} 
where $\lambda_j^\pm$, $\hat{\lambda}_j$, and $d_j$ are given in \eqref{eigsSQEDN2}.

The holonomy and the auxiliary fields are fixed by the saddle point equations
\es{SQEDN2SaddleEqsIndex}{
\frac{\partial I_{q}^{(0)}}{\partial \alpha}\Big\vert_{\alpha,D,\sigma}=\frac{\partial I_{q}^{(0)}}{\partial D}\Big\vert_{\alpha,D,\sigma}=\frac{\partial I_{q}^{(0)}}{\partial \sigma}\Big\vert_{\alpha,D,\sigma}=0\,.
}
The solution of these equations is
\es{SQEDN2Ansatz}{
 \sigma &= q \,, \qquad D = 0 \,, \\
\alpha&=\mp_\text{tot}\left[\lambda_q^{{\pm}}\big\vert_{D=0,\sigma=q}-1/4+\beta^{-1}\log\frac{\xi}{1+\xi}\right]+O(e^{-(\lambda_{q+1}^\pm-\lambda_{q}^\pm)\beta})\,,\quad \xi={q|2\kappa-\delta n |\ov n^{\mp_\text{tot}}}\,.
}
  We plug the solution back into \eqref{SQEDN2F} to find 
\es{SQEDN2Final2}{
I_{q}^{(0)}&=\Delta_{q}^{(0)}-\frac12R_{q}^{(0)}-\beta^{-1}S_{q}^{(0)}\,,
}
with 
 \es{DRSDef}{
 \Delta_{q}^{(0)}&=q/2+(1/2+q)n^{\mp_\text{tot}}\xi\,,\qquad
R_{q}^{(0)}=q/2+\frac12n^{\mp_\text{tot}}\xi\,,\\
S_{q}^{(0)}&=-n^{\mp_\text{tot}}\left(\xi\log\xi-(1+\xi)\log[1+\xi]\right)\,.
 }

As already mentioned, the supersymmetric $S^2 \times S^1_\beta$ partition function of the theory \eqref{SQEDN2Action} can also be computed using supersymmetric localization \cite{Imamura:2011su,Kim:2009wb}.\footnote{Using supersymmetric localization, it is also possible to compute the partition function of the theory corresponding to $\tilde R$ at any $N$.}  The computation proceeds as follows.  The first step is to add a positive-definite $Q$-exact term to the action with a large coefficient.  Due to supersymmetry, the partition function is independent of this deformation, and it can thus be evaluated in a saddle point approximation.  The saddles are \cite{Imamura:2011su,Kim:2009wb}
 \es{SaddleLoc}{
  A = {\cal A}\,, \qquad
   \sigma = q \,, \qquad D = 0 \,,
 } 
with all other fields vanishing.  Here, ${\cal A}$ is as in \eqref{SaddleAnsatz}, with $q$ and $\alpha$ arbitrary parameters.  The action~\eqref{SQEDN2Action} evaluated on this saddle equals $-2 \kappa q \alpha \beta$.  Around this saddle, one then has to compute a functional determinant of fluctuations of the fields in the vector multiplet and chiral multiplets.  When combined with the classical contribution, the partition function in the charge $q$ sector is then written as
 \es{ILoc}{
  Z_q = \int d\alpha\, e^{-I_\text{loc}(q, \alpha)} \,,
 }
(The full partition function is the sum $Z = \sum_q Z_q$.)  At large $N$, $I_\text{loc}(q, \alpha)$ is approximately equal to $N I_q^{(0)} (\alpha, q, 0)$, with the same $I_q^{(0)}$ as in \eqref{SQEDN2F}, the reason being that the chiral multiplet localizing term is precisely the same as the chiral multiplet kinetic term.  Thus, the localization formula takes the form
 \es{ILocAgain}{
  Z_q \approx \int d\alpha \, e^{- N I_q^{(0)}(\alpha, q, 0)} \,.
 }
Performing a saddle point approximation of this one-dimensional integral requires solving $\frac{\partial I_{q}^{(0)}}{\partial \alpha}\Big\vert_{D=0,\sigma=q} = 0$, which yields the value of $\alpha$ given in \eqref{SQEDN2Ansatz}.  With this value of $\alpha$, one can then write approximately $Z_q \approx e^{- N I_q^{(0)}}$, with $I_q^{(0)}$ given in \eqref{SQEDN2Final2}.  Thus, the result of the supersymmetric localization computation followed by a large $N$ saddle point approximation manifestly agrees with the result of our large $N$ saddle point approximation.

\subsubsection{Microscopic interpretation}

The large $\beta$ limit of the index \eqref{SQEDN2Final2} also has an interpretation as the contribution of a state in a Fock space.  The properties of the Fock vacuum and of the creation operators are the same as in Table~\ref{SQEDN2modes}.  In particular, plugging in $\sigma = q$ and $D=0$, we obtain:
\es{aux}{
\Delta_\text{bare} &= R_\text{bare}= \frac 12 Nq\,,\\
A_\text{bare}&= -Nq\,,\\
G&= 2qN(\kappa-\delta n/2)\,.
}
Note, however, that while the Fock space here is isomorphic to the one in the previous section, some of its properties, namely the energies of the vacuum and of the modes, are different.  Indeed, these energies depend on the saddle point values of $\sigma$ and $D$, which are now set to \eqref{SQEDN2Ansatz} whereas in the previous section they were found by solving \eqref{SQEDN2alph}.  

Because the Fock vacuum is not gauge invariant, it does not appear in the index.  The operator that the index captures in the limit $\beta \to \infty$ is the lowest dimension gauge-invariant BPS operator, i.e.~an operator obeying $\Delta = R + j_3$.  Such an operator can be constructed by acting on the Fock vacuum $\abs{G}$ times with the creation operator $a^{(\mp_\text{tot})i,\dagger}_{qq}$, because only the creation operators $a^{(\pm)i,\dagger}_{j m}$ and $\hat b^{(\pm)\dagger}_{j m, i}$ obey the condition $\Delta = R + j_3$, and $a^{(\mp_\text{tot})i,\dagger}_{qq}$ has lowest energy and a gauge charge of the right sign.  The quantum numbers of the gauge-invariant BPS monopole constructed this way are: 
\es{localBPSb}{
\Delta_{q}^{\text{BPS}}=&\,\Delta_\text{bare}+\left( q + \frac 12 \right)\abs{G}\,,\\
j_{q}^{\text{BPS}}=&\,q\abs{G}\,,\\
R_{q}^{\text{BPS}}=&\,\Delta_\text{bare}+ \frac 12 \abs{G}\,,\\
A_{q}^{\text{BPS}}=&\,A_\text{bare}+\abs{G}\,,\\
SU(N^{\pm_\text{tot}})\times SU(N^{\mp_\text{tot}}) \text{ irrep: }&\Big(\bold 1\,,\,\underbrace{\begin{ytableau}
 {} &{}&{}&{} \\
\end{ytableau}}_{\abs{G} }\Big)\,.
}

One can check that the expressions for $\Delta^\text{BPS}_q$ and $R^\text{BPS}_q$ in \eqref{localBPSb} agree with those given in \eqref{DRSDef} after plugging in the value of $\xi$ from \eqref{SQEDN2Ansatz}.   The entropy $S_{q}^{(0)}$ computed in \eqref{DRSDef} also matches the large $N$ expansion of the logarithm of the dimension of the $SU(N)$ irrep of the dressed BPS monopole:
\es{SQEDboxes}{
\log\left[\frac{(N^{\mp_\text{tot}}(1+\xi)-1)!}{(N^{\mp_\text{tot}}-1)!(N^{\mp_\text{tot}}\xi)!}\right]=NS_{q}^{(0)}-\frac12\log N^{\mp_\text{tot}} + O\left(1\right)\,.
}
Note that unlike the thermal free energies computed in the main text, for which the entropy included a variety of flavor irreps, here the index is expected to pick out a unique flavor irrep.   Thus, one expects that the entropy at constant order in $N$ should not have an associated $\log \beta$ term.

We can understand now why the BPS monopoles are the lowest energy states in the monopole sector only if they coincide with the bare monopole, as seen in Figure~\ref{SQEDN2} and already remarked there. Above, we have found that the BPS monopoles are obtained by dressing the bare monopole with scalar modes. We can always do energetically better than that by at least partially dressing with the lowest fermion modes, hence the BPS monopoles are only minimal energy states when there is no dressing.

\bibliographystyle{ssg}
\bibliography{monopole}
\end{document}